\documentclass[aps,prb,showpacs,a4paper,superscriptaddress,groupedaddress]{revtex4}
\usepackage[english]{babel}
\usepackage[utf8]{inputenc}
\usepackage{amsmath}
\usepackage{graphicx}
\usepackage[toc,page]{appendix}
\usepackage{multirow}
\usepackage{nonfloat}
\usepackage[colorinlistoftodos]{todonotes}
\usepackage{verbatim}

\begin{document}
\title{The impact of microcavity wire width on polariton soliton existence and multistability}

\author{G. Slavcheva}\email{g.slavcheva@bath.ac.uk}
\affiliation{Department of Physics, University of Bath, Bath, BA2 7AY, United Kingdom}
\author{M. V. Koleva}
\affiliation{Department of Physics, University of Oxford, Oxford, OX1 3RH, United Kingdom}
\author{A. Pimenov}
\affiliation{Weierstrass Institute, Mohrenstrasse 39, D-10117 Berlin, Germany}

\date{\today}

\begin{abstract}
We have developed a model of the nonlinear polariton dynamics in realistic 3D non-planar microcavity wires in the driven-dissipative regime. We find that the typical microcavity optical bistability evolves into multistability upon variation of the model parameters. The origin of the multistability is discussed in detail. We apply linear perturbation analysis to modulational instabilities, and identify conditions for localisation of composite multi-mode polariton solitons in the triggered parametric oscillator regime. Further, we demonstrate stable polariton soliton propagation in tilted and tapered waveguides, and determine maximum tilt angles for which solitons still exist. Additionally, we study soliton amplitude and velocity dependence on the wire width, with a view to engineering quantum photonic devices.
\end{abstract}

\maketitle
\section{Introduction}
\label{sec1}
Semiconductor quantum-well (QW) microcavities are 1D photonic crystal structures, specifically designed to control light-matter interactions. The strong coupling cavity-emitter regime of operation is realised when the QW exciton-cavity photon interaction exceeds any dissipative rates in the system. The eigenmodes of the system are mixed, entangled light-matter states that can be viewed as photons `dressed' with the medium polarisation (exciton); these give rise to bosonic quasiparticles known as microcavity exciton-polaritons. Owing to their photon component, exciton polaritons are extremely light particles. As a result of their excitonic component, however, they exhibit strong repulsive inter-particle interactions, leading to strong nonlinearities nearly four orders of magnitude higher than in typical nonlinear solid-state optical media \cite{Snoke&Sanvitto}. The nonlinearities arise primarily from parametric scattering of exciton-polaritons, driven by a Coulomb exchange interaction between polariton-excitonic constituents, with additional, smaller contributions originating from phase space filling \cite{Ciuti, Whittaker, Kwong}.

Nonlinear self-localisation and coherent propagation phenomena with exciton polaritons in planar microcavities have been extensively studied in recent years. Formation of moving 2D self-localised, non-equilibrium polariton droplets travelling without loss at high speeds ($\sim 1 \%$ of the speed of light) has been experimentally demonstrated in coherently pumped semiconductor microcavities operating in the strong-coupling regime \cite{Amo1}. These polaritons display collective dynamics consistent with superfluidity. When studying polariton flow around a defect, it has been demonstrated that at high flow velocities, the perturbation induced by the defect gives rise to the turbulent emission of quantised vortices, and to the nucleation of oblique, dark `quantum hydrodynamic' solitons \cite{Amo2}. Accurate tracking in space and time of long-life ($100-200\,\,\mathrm{ps}$) polaritons reveals long-range ballistic propagation and coherent flow over macroscopic distances -- from hundreds of $\mathrm{\mu m}$ to millimetres within the cavity \cite{Snoke1,Snoke2,Snoke3}.

It is well known that a high-density and low-temperature gas of microcavity polaritons exhibits effects pertinent to cold-atom Bose-Einstein condensates \cite{Deng}. However, unlike atomic condensates -- for which bright solitons exist only for attractive interactions, while repulsive interactions give rise to dark solitons -- bright polariton solitons supported by repulsive exciton-exciton interactions have been theoretically predicted \cite{Egorov&Dmitry} and experimentally demonstrated \cite{Sich_NaturePhotonics} for pump momenta beyond the inflection point of the lower polariton branch, where the polariton effective mass becomes negative. Self-localisation (in a direction collinear with the pump wave vector) occurs when the dispersion induced by the negative polariton effective mass compensates the repulsive polariton-polariton interaction.

Coherent polariton propagation has also been experimentally demonstrated in laterally confined microcavity wires, fabricated by lateral etching of planar semiconductor microcavities \cite{Wertz_APL}. Condensed polaritons spread over the whole wire, with the higher-energy excitons remaining at the excitation spot \cite{Wertz_Nature}.

Polariton waves can be confined in structures with sub-$\mu m$ size. This opens up possibilities for fabrication of polaritonic integrated circuits based on structured semiconductor microcavities on a chip. Polaritonic structured media are under active development, with arrays of microcavity wires currently being fabricated to act as waveguides for polariton waves. Structured microcavities are a particularly promising integration platform due to the large polaritonic nonlinearity, broad transparency window, mature fabrication technology and the possibility of monolithic integration with semiconductor diode lasers and VCSELs. In this respect, new theories and numerical methods are needed to model the nonlinear polariton dynamics in non-planar waveguides.

We have recently developed a driven-dissipative, scalar (spinless) Gross-Pitaveskii model of the nonlinear polariton dynamics in realistic microcavity wires. In addition, we have numerically demonstrated multi-mode polariton solitons and a peculiar type of spatial multistability \cite{our_OL}. Under suitable conditions, different modes within the polariton soliton wave packet interact among themselves in such a way as to give rise to a self-localisation mechanism that prevents the pulse from broadening. Self-localisation phenomena in multi-mode systems arise due to counter-balancing nonlinearity with a combination of dispersive effects: material (polariton) dispersion due to a frequency-dependent dielectric response, cavity modes dispersion, and variation of the group velocity of each mode. In a recent work \cite{PRB2016}, using the coupled-mode expansion technique we showed that a new component of the nonlinearity has to be taken into account -- namely, the intermodal nonlinear coupling. The latter arises from two distinct nonlinear phenomena: intermodal cross-phase modulation (through Kerr nonlinearity) and degenerate four-wave mixing (through polariton parametric scattering).

In this work, we provide more details of our general mean-field, driven-dissipative Gross-Pitaevskii model. We apply it to the problem of polariton soliton propagation in microcavity wires with different widths -- with a view to designing integrated polaritonic devices. Using the model, we study the soliton existence domain, multistability and group velocity dependence on the wire width, and identify optimum geometry parameters for soliton formation. We provide a qualitative explanation of the non-monotonic dependence of the maximum soliton amplitude and group velocity dependence on the wire width. Additionally, we investigate the dependence of the multistability on the cavity-emitter system's dissipation parameter, which includes exciton reservoir dephasing and cavity photon decay. Furthermore, we demonstrate numerically a new coherent propagation phenomenon of a radiating polariton soliton exhibiting periodic collapses and revivals, and identify regimes for existence of this new class of polariton solitons in a realistic microcavity wire. Finally, we simulate polariton soliton propagation in tilted and tapered microcavity wires, and determine the maximum tilt angle for which the soliton persists.

\section{Theoretical Model}
\label{sec2}
\begin{figure}
\vspace{10pt}
\begin{center}
\resizebox{.9\textwidth}{!}{%
\includegraphics[height=3cm]{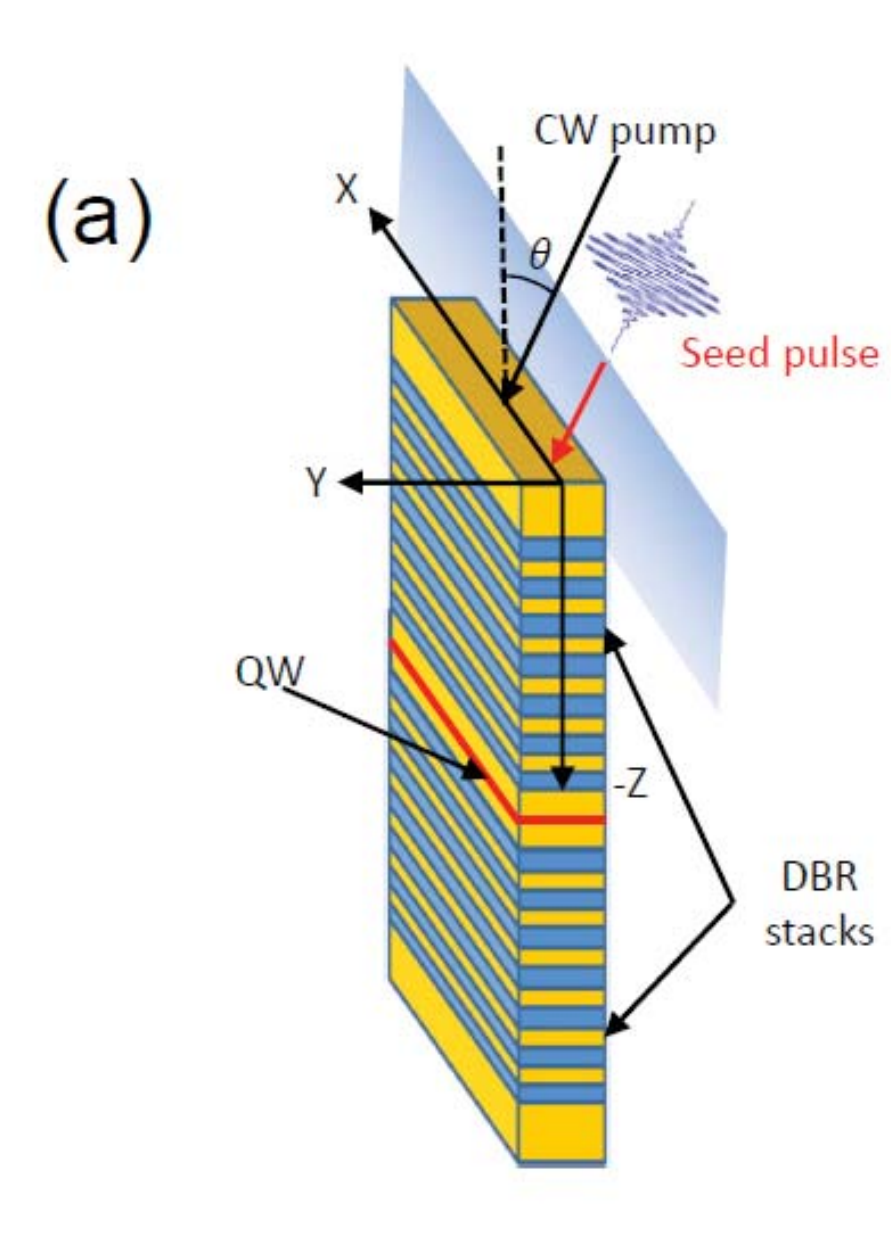}
\quad
\includegraphics[height=3cm]{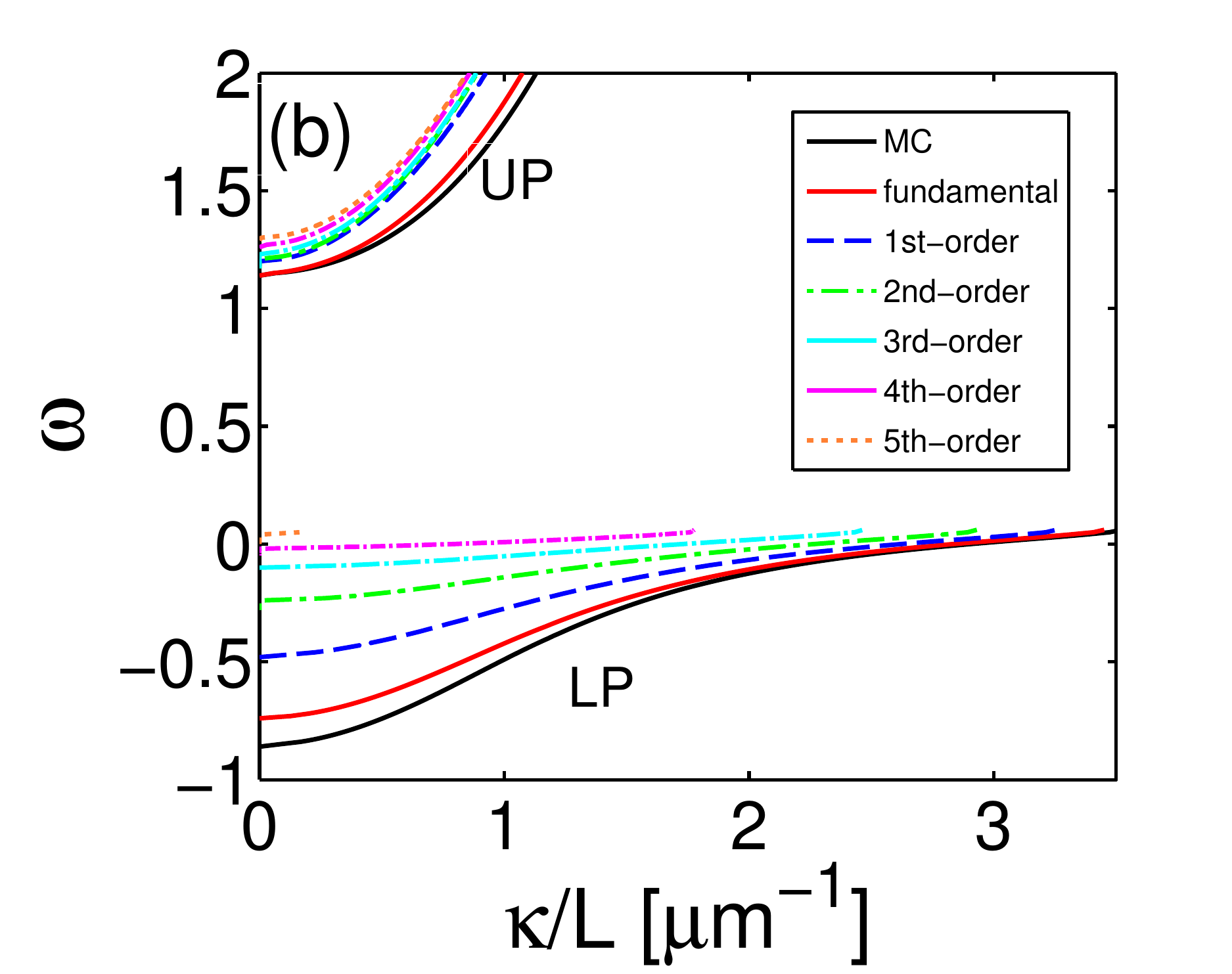}
}
\end{center}
\begin{center}
\resizebox{.9\textwidth}{!}{%
\includegraphics[height=3.5cm]{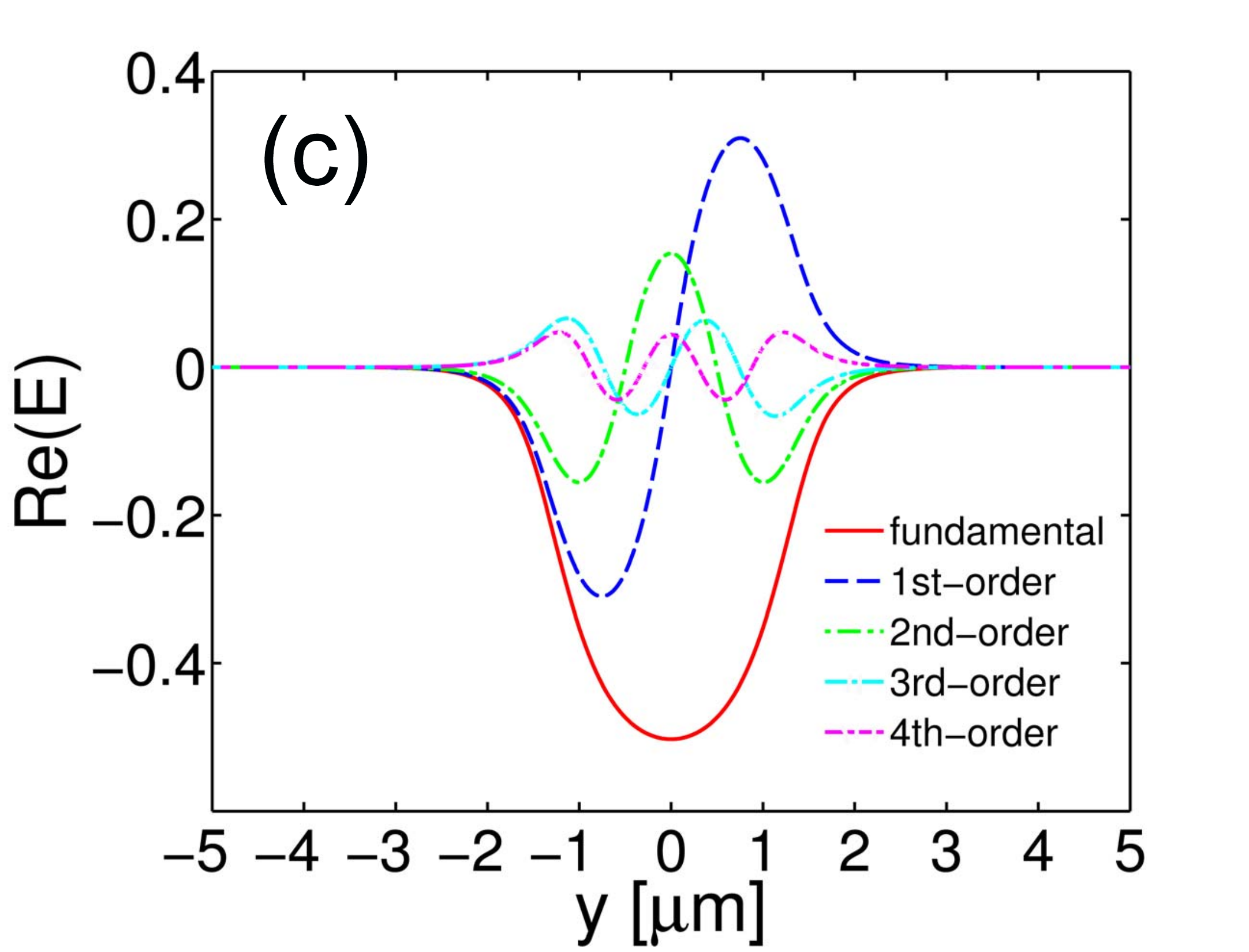}
\quad
\includegraphics[height=3.5cm]{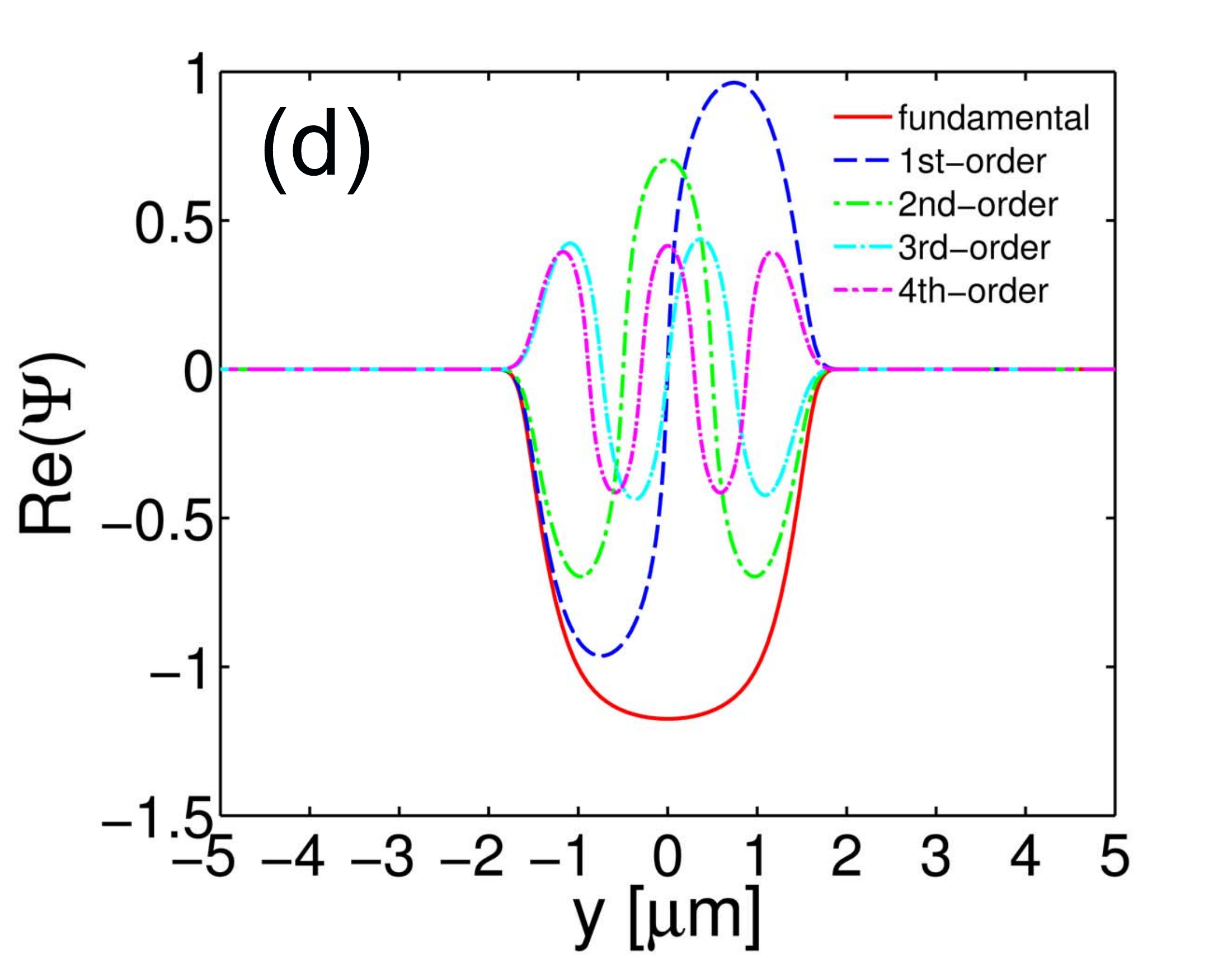}
}
\end{center}
\caption[Fig 1] {(Color online) (a) Scheme of the microcavity polaritonic wire structure and the excitation geometry. The CW pump and the seed pulse are incident along the $x$ direction at angle $\theta$, and have no components in the $y$ direction. (b) Linear polariton dispersion for the fundamental and next five higher-order modes at zero detuning ($\Delta=0$), wire width $w=3 \,\,\mathrm{\mu m}$ and a pump incidence angle $\theta=20^{\circ}$. Both the lower (LP) and upper (UP) polariton branches are shown. The planar microcavity dispersion is displayed with solid black lines for comparison. Note that $\omega$ is normalised with respect to the Rabi frequency. Amplitudes of the corresponding linear (c) photonic ($E$) and (d) excitonic ($\Psi$) electric-field modes.}
\label{fig:excitation_structure_scheme}
\end{figure}

We consider a microcavity wire, fabricated by laterally etching a planar GaAs/AlGaAs microcavity, of the type described in \cite{Wertz_APL}. The corresponding geometry is schematically shown in Fig.~\ref{fig:excitation_structure_scheme} (a). The wire is pumped by a CW laser linearly polarised in the $y$ direction, such that it supports quasi-TE modes in the wire. The pump is inclined along the $x$ direction at an angle $\theta$ contained within the $x$-$z$ plane, so that it has a non-zero momentum along the wire. Similar to the case of a planar microcavity \cite{Sich_NaturePhotonics}, solitons propagating along the wire are triggered by a short seed pulse also incident at $\theta$, having the same linear polarisation as the CW pump.

We describe the dynamics of the system by a scalar mean-field, driven-dissipative Gross-Pitaevskii model \cite{our_OL}:

\begin{eqnarray}
\hspace*{-1cm} \partial _t {E} - i\left( {\partial _x^2  + \partial _y^2 } \right){E} + \left[ {\gamma _c  - i\delta _c -i\Delta  - iU(y)} \right]{E}
&= i\Omega _R(y)\Psi+ E_p e^{i\kappa x}\;,\qquad
\label{eqE}\\
\hspace*{1.5cm} \partial _t \Psi  + \left( {\gamma _e  - i\delta _e -i\Delta} \right)\Psi  + i\left| \Psi  \right|^2 \Psi &= i\Omega _R(y) {E}\;, \qquad
\label{eqPsi}
\end{eqnarray}
where $E_p$ is the normalized pump amplitude, $g$ is the strength of exciton-exciton interaction \cite{Sich_NaturePhotonics}, $\omega_R$ is the Rabi frequency in a planar homogeneous cavity, time is measured in units of $T=1/\omega_R$, and $E$ and $\Psi$ are the averages of, respectively, the photon and exciton creation or annihilation operators. The normalisation is such that $(\omega_R/g)\vert E \vert^2$ and $(\omega_R/g)|\Psi|^2$ are the photon and exciton numbers per unit area, respectively. The unit length, $L = \sqrt{\hbar/(2m_c \omega _R)}$, is determined by the effective cavity photon mass, $m_c$. $\delta_e$, $\delta_c$, and the pump frequency, $\Delta$, are detunings from a reference frequency, $\hbar \omega_0=1.55\,\,\mathrm{eV}$ ($\lambda_0=800 \,\,\mathrm{nm}$), normalised with respect to the Rabi frequency. The reference frequency is blue-shifted with respect to the resonance frequency of the lower polariton branch for a given angle of incidence of the pump; the blue-shift is important because only then is excitation of a bistable mode possible over a finite range of $E_p$ \cite{PRB2016}. $\gamma_c$ and $\gamma_e$ are cavity and exciton decay rates.

The lateral confinement in the cavity plane (along the $y$ axis) is described by an effective potential $U(y)$ in the photonic component, and a spatially confined normalised coupling, $\Omega_R(y)$:
\begin{eqnarray}
\hspace*{3cm} U(y)&=U_{bg}\left[1-e^{-\left(2y/w\right)^8}\right]\;,
\label{eqPot}\\
\hspace*{3cm} \Omega_R(y)&=e^{-\left(2y/w\right)^8}\;,
\label{eqPot2}
\end{eqnarray}
where $w$ is the dimensionless wire width and $U_{bg}$ is the background confinement potential. As demonstrated in our previous work \cite{our_OL}, the above super-Gaussian potentials give an accurate description of the dispersion of the lowest six guided modes of the wire, which includes the fundamental mode.

\section{Numerical Results}
\label{sec3}
\subsection{Free-polariton mode dispersion and model parameters}
\label{Free-polariton}
The free-polariton dispersion is calculated from our mean-field model (Eqs.~(\ref{eqE}) and (\ref{eqPsi})). By setting $E_p=0$ in Eq.~(\ref{eqE}) and neglecting the nonlinear term in Eq. (\ref{eqPsi}), using the ansatz $E=A(y)e^{i\kappa x-i\omega t}$, $\Psi=B(y)e^{i\kappa x-i\omega t}$, the dispersion of the linear modes, $\omega(\kappa)$, is given by the following eigenvalue problem:
\begin{eqnarray}
\hspace*{1cm} \kappa^2 A=\left[\omega+\delta_c+U-\frac{\Omega_R^2}{\omega+\delta_e+i\gamma_e}+\partial^2_y-i\gamma_c\right]A\;.
\end{eqnarray}
The resulting linear polariton dispersion for the fundamental and the next 5 modes is shown in Fig.~\ref{fig:excitation_structure_scheme} (b), along with the single-mode planar microcavity dispersion (black solid lines) for a microcavity wire width of $3 \,\,\mathrm{\mu m}$ and a pump tilt angle of $20 \mathrm{^{\circ}}$ at zero pump detuning ($\Delta=0$). The planar microcavity can be regarded as the limiting case when the width, $w \rightarrow \infty$. Thus, it is clear from the figure that, for the pump energy that we are using (blue-shifted from the LP resonance frequency), the planar microcavity dispersion in the reciprocal momentum space is steeper than that of the microcavity wire. The real part of the corresponding linear $E$-field and $\Psi$-field fundamental and higher-order mode profiles are shown in Fig.~\ref{fig:excitation_structure_scheme} (c) and (d), respectively.

Using the finite-element electromagnetic-field solver, COMSOL, we obtain the model free parameters -- namely the Rabi frequency $\omega_R$, the spatial scale $L$, and the effective potential depth $U_{bg}$ -- by fitting the free-polariton dispersion with the numerically computed dispersion of a realistic microcavity wire structure, as described in \cite{our_OL}.

For the chosen value of the normalised quantum-well oscillator strength, $H=0.015$, the Rabi splitting in the fundamental mode is $2\hbar \omega_R\approx 11 \,\,\mathrm{meV}$ at normal incidence ($\kappa=0$), which is comparable to the one in planar cavities. Using this value, the model parameters in Eqs. (\ref{eqE}) and (\ref{eqPsi}) are: $T_0\approx 0.12\,\,$ps, $\delta_c=-0.191$, $\delta_e=-0.1$, and $\gamma_c=\gamma_e=0.04$, corresponding to exciton and cavity linewidths, $\Gamma_e=\Gamma_c=0.22\,\,\mathrm{meV}$ \cite{Sich_NaturePhotonics}.

\subsection{Stationary nonlinear modes}

The stationary solutions of the nonlinear system, Eqs.~(\ref{eqE}) and (\ref{eqPsi}), in the presence of a spatially homogeneous monochromatic pump are obtained by setting all time derivatives to zero, restoring the pump and the nonlinearity, and seeking stationary solutions to the homogeneous system of the form $\left\{ {E,\Psi} \right\} = \left\{ {A\left( y \right)e^{i\kappa x} ,B\left( y \right)e^{i\kappa x} } \right\}$. The system then reduces to:

\begin{eqnarray}
\left[\delta_c+\Delta+U-\kappa^2+\partial^2_y+i\gamma_c\right]A+\Omega_R B&=iE_p\;,
\label{nonlinear_modes}
\\
\hspace*{1cm} \left(\delta_e+\Delta+i\gamma_e\right)B-|B|^2B
+\Omega_R A&=0\;.
\label{nonlinear_modes2}
\end{eqnarray}
After discretising the $y$ co-ordinate, this system transforms into a set of coupled nonlinear algebraic equations, which is solved numerically for $E$ and $\Psi$, e.g. using Newton-Raphson iterations for each fixed set of pump parameters $E_p$ and $\kappa$.

In what follows, all calculations are carried out for a microcavity polaritonic wire of width $w=3 \,\mathrm{\mu m}$ at pump incidence angle $\theta=20^\circ$ and model parameters as set out at the end of Sec. \ref{Free-polariton}. We choose to plot the $\Psi$-field power, $\int {\left| \Psi  \right|} ^2 dy$, integrated in a transverse direction to the channel, in order to avoid any spurious non-monotonic behaviour which arises if the maximum or the channel midpoint value of $\left| {E} \right|^2$ or $\left| {\Psi} \right|^2$ are plotted.

\begin{figure*}
\vspace{10pt}
\begin{center}
\resizebox{.9\textwidth}{!}{%
\includegraphics[height=3cm]{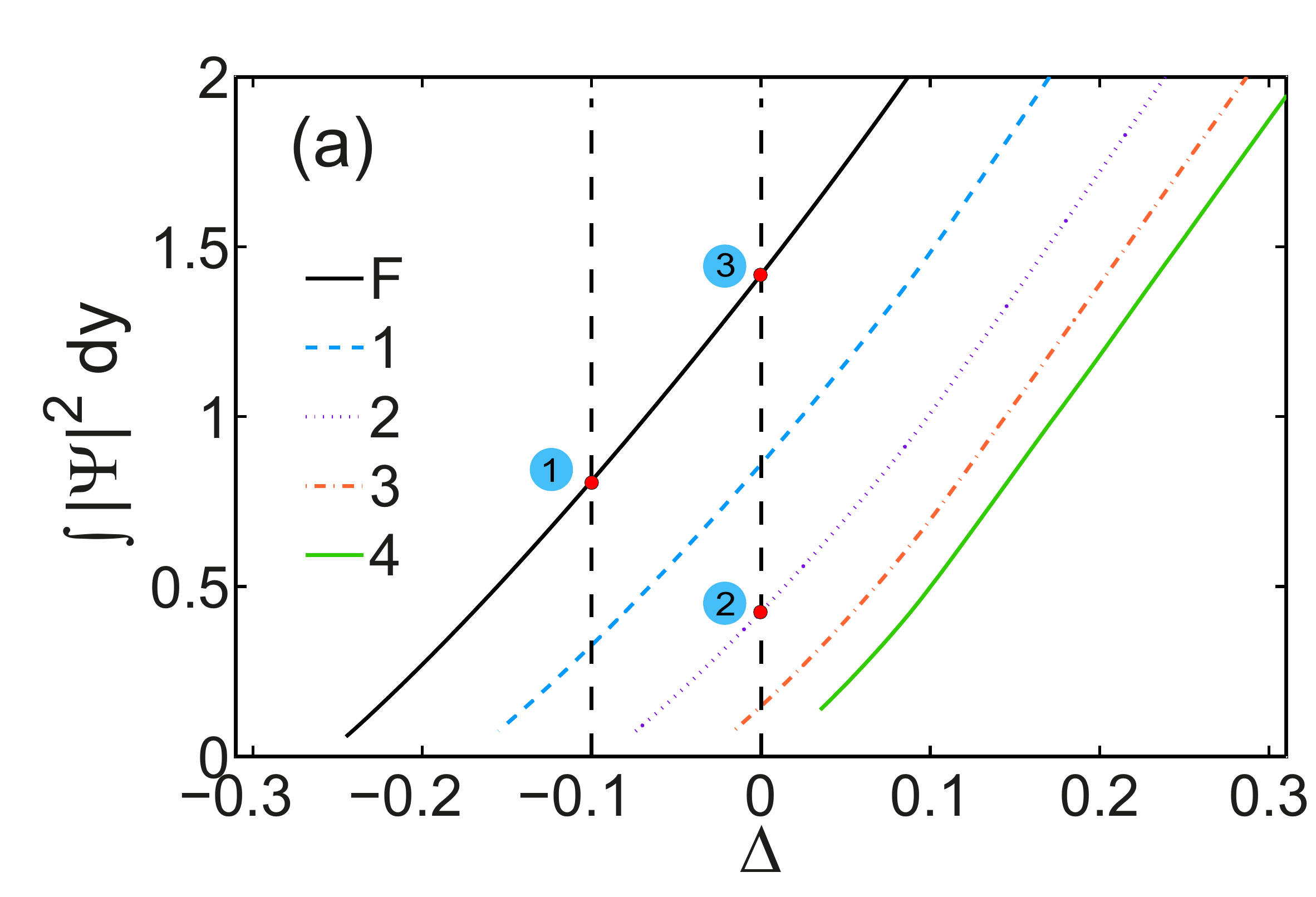}
\quad
\includegraphics[height=3cm]{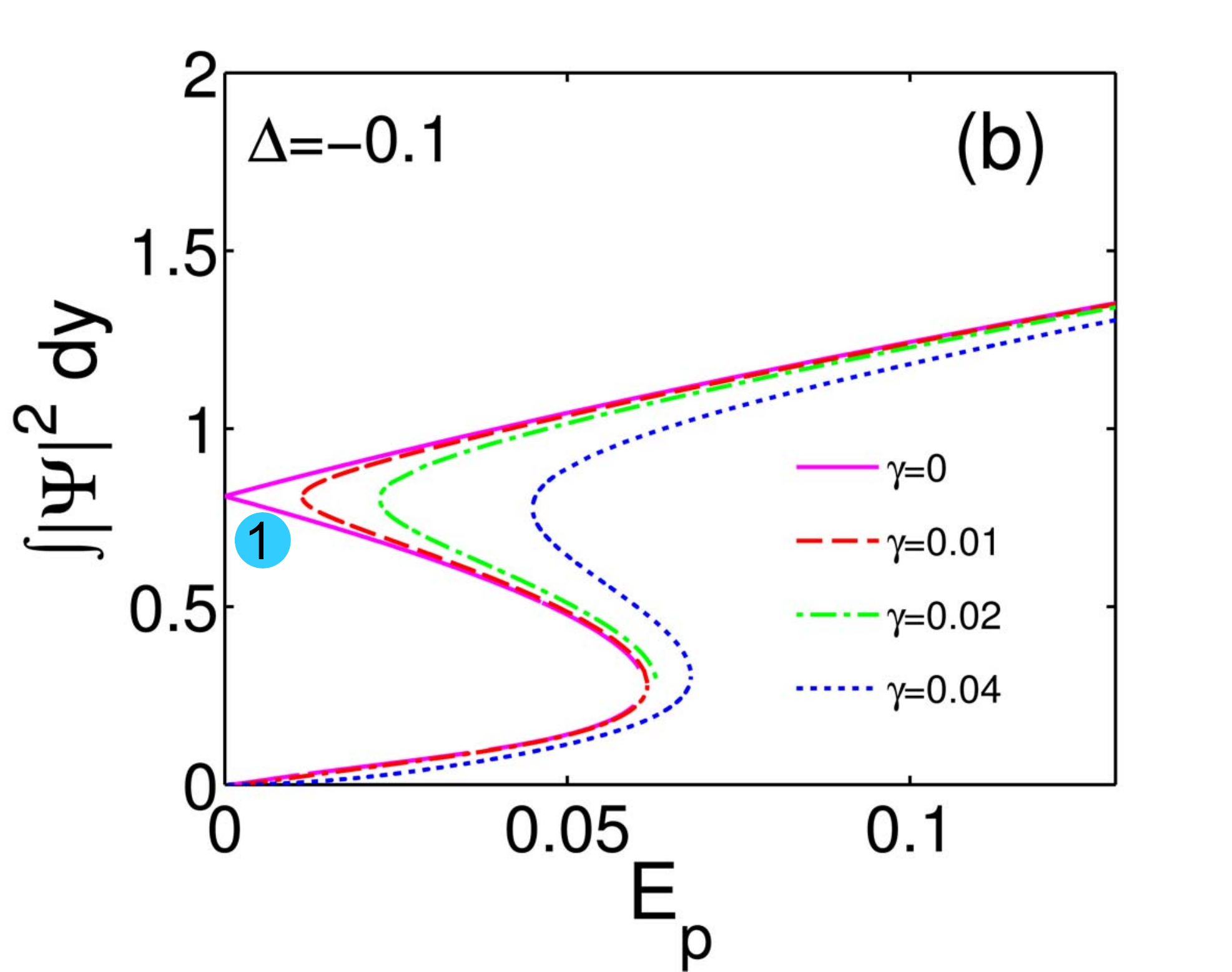}
}
\end{center}

\begin{center}
\resizebox{.9\textwidth}{!}{%
\includegraphics[height=3cm]{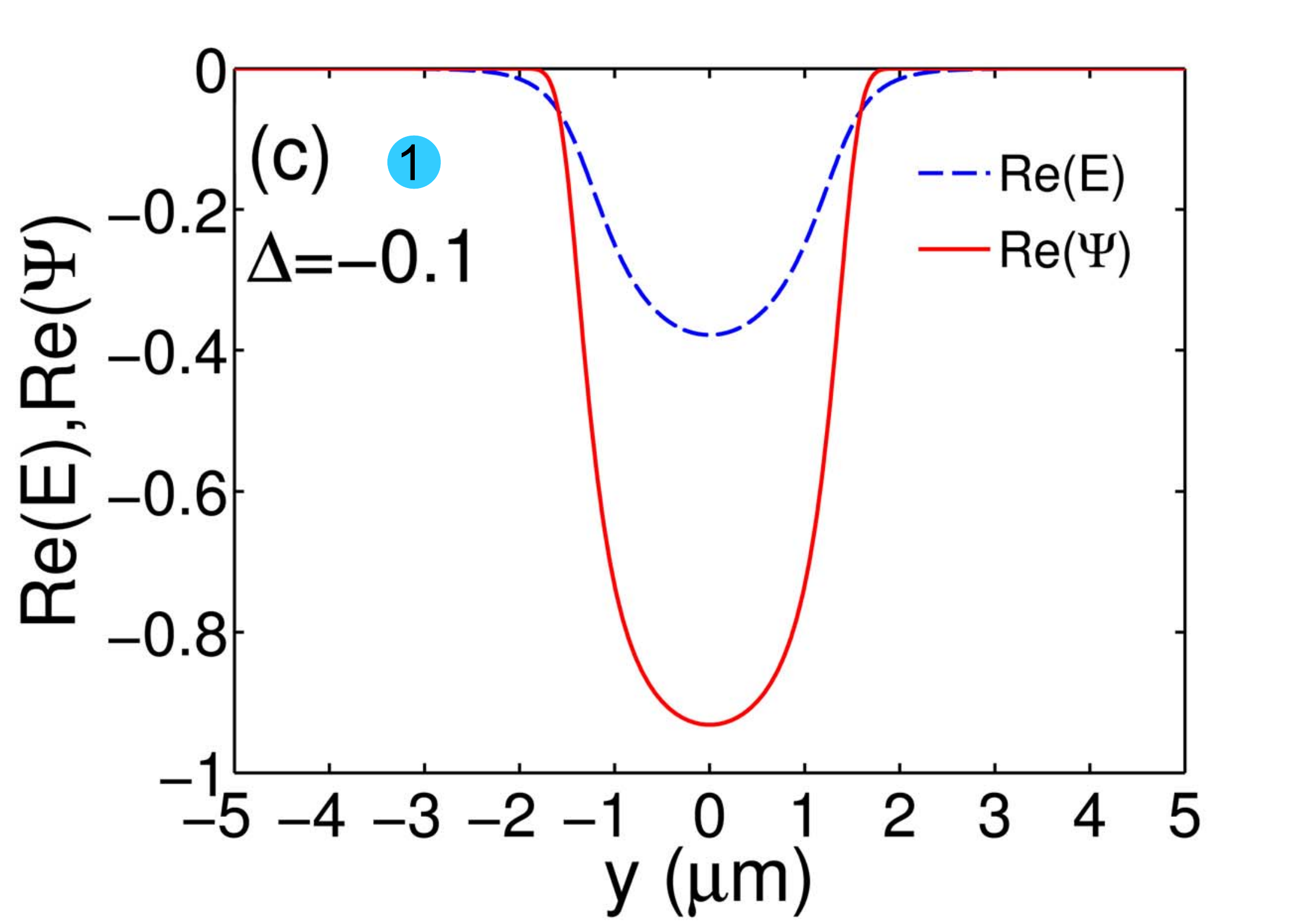}
\quad
\includegraphics[height=3cm]{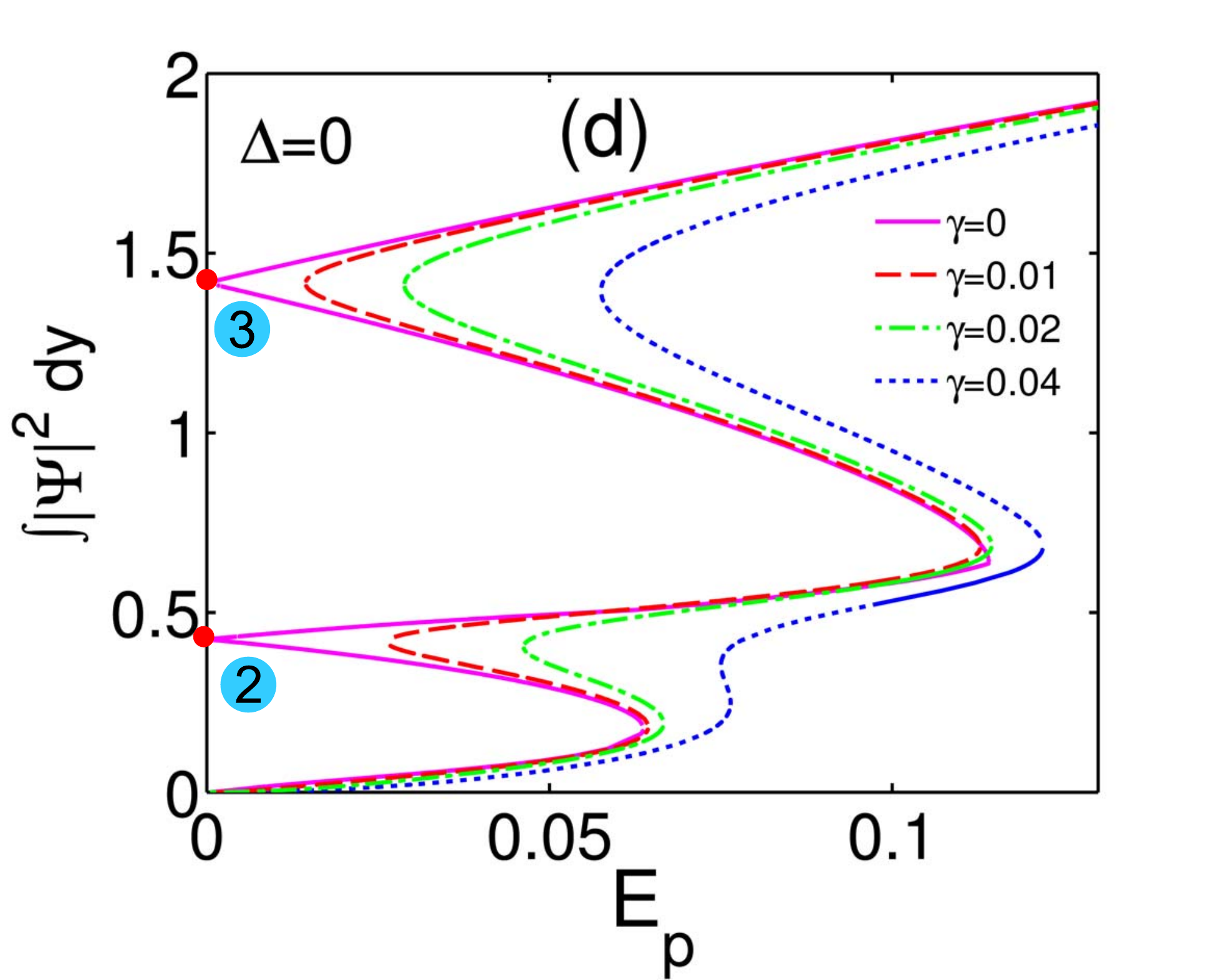}
}
\end{center}

\begin{center}
\resizebox{.9\textwidth}{!}{%
\includegraphics[height=3cm]{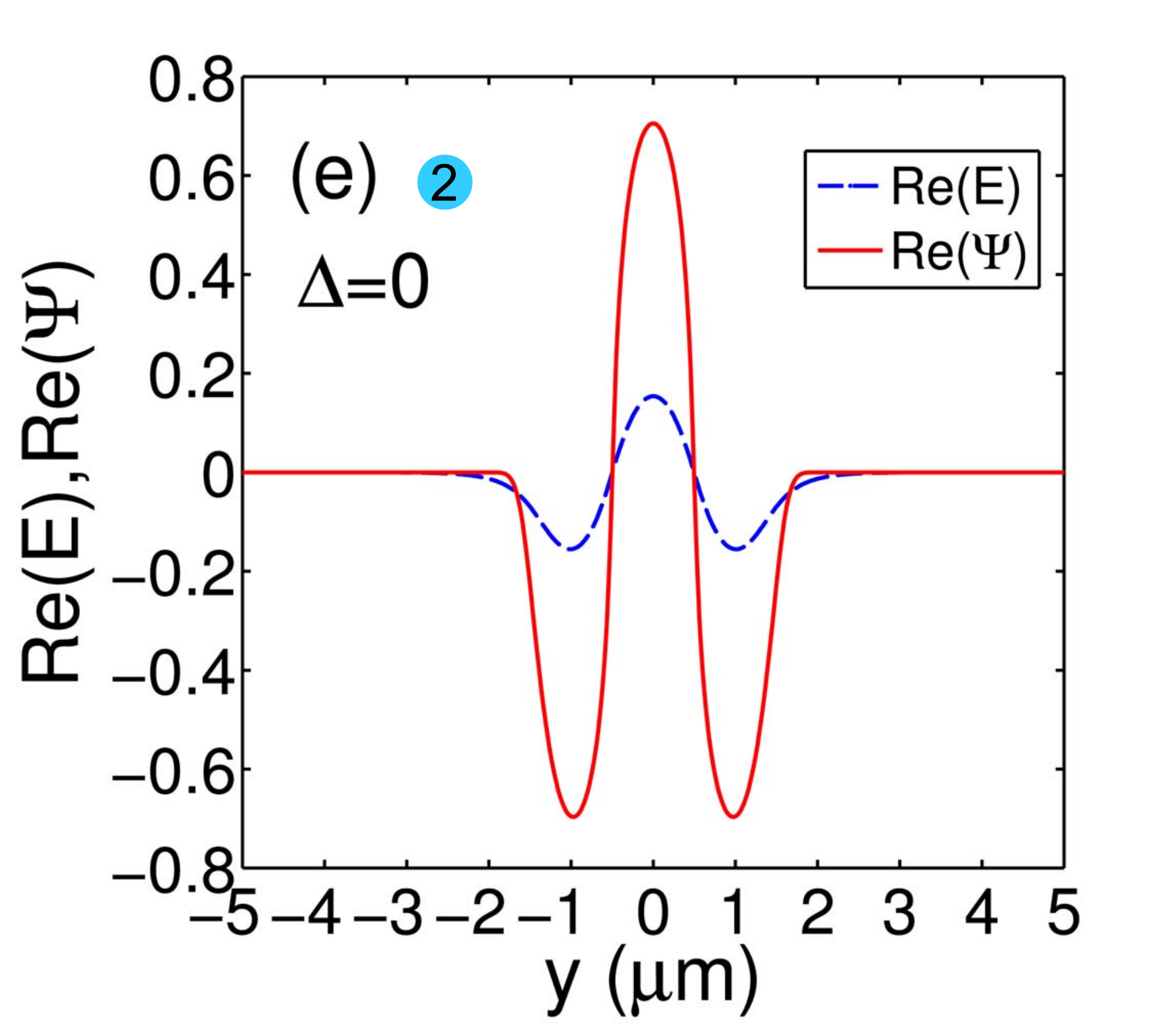}
\quad
\includegraphics[height=3cm]{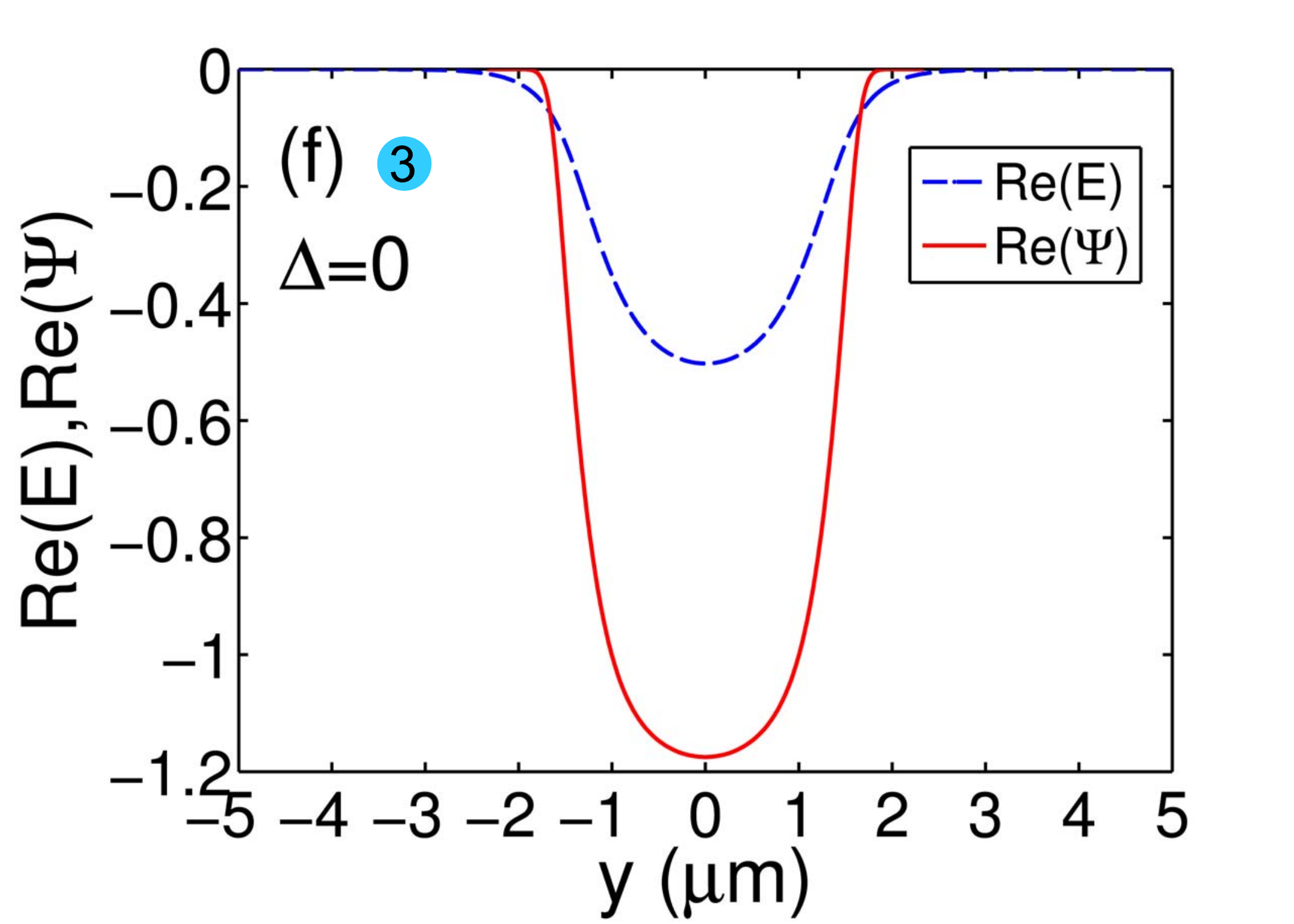}
}
\end{center}

\caption[Fig 2] {(Color online) (a) Integrated power of the excitonic ($\Psi$) field at pump amplitude $E_p=0$ and $\gamma_c=\gamma_e=0$ for five nonlinear modes (F, fundamental; 1, first order; 2, second order; 3, third order; 4, fourth order) as a function of $\Delta$, the detuning from the reference pump frequency, $\omega_0$. Vertical dashed lines help to determine which modes are present for different detunings. (b) $\Psi$-field power as a function of $E_p$ for different dissipation parameters, $\gamma_e=\gamma_c=\gamma$, and $\Delta=-0.1$. A bistability curve is observed. Note that we take the pump to be spatially symmetric. The single folding point (denoted `1') corresponds to the fundamental mode (see panel (a) and text for more details). (c) Profile of the real parts of the photonic ($E$) electric field and $\Psi$-field of the fundamental polariton mode, corresponding to point 1 in (b). (d) $\Psi$-field power as a function of $E_p$ for different dissipation parameters, $\gamma_e=\gamma_c=\gamma$, and $\Delta=0$. A multistability curve is observed. The multistability arises from the fact that, for this detuning, two rather than one polaritonic modes are supported (see panel (a) and text for more details). The multistability curve touches the $y$ axis at two folding points, denoted `2' and `3'. The bi- and multistability in (b) and (d) becomes more pronounced as $\gamma$ is decreased -- from $0.04$ to $0$. (e), (f) Real part of the $E$-field and $\Psi$-field of the nonlinear polariton mode profiles at $\Delta=0$ at the folding points 2 and 3 in (d), respectively. Point 2 corresponds to a second-order mode, while point 3 corresponds to a fundamental mode.}
\label{fig:multistability_origin}

\end{figure*}

The integrated power of the $\Psi$-field, plotted at zero pump amplitude ($E_p=0$) and no dissipation ($\gamma_e=\gamma_c=0$) for different detuning ($\Delta$) is shown in Fig.~\ref{fig:multistability_origin} (a). The number of confined modes for the specific potential considered is $5$ (including the fundamental mode). Scanning through a range of pump amplitudes (Fig.~\ref{fig:multistability_origin} (b, d)) shows that, depending on the detuning, a bi- or multistability family of curves is obtained for a range of dissipation parameters, $\gamma$. We cover the range $0\leq \gamma \leq 0.04$, as $0.04$ (corresponding to $\Gamma_e=\Gamma_c=0.22\,\mathrm{meV}$) is the most realistic value of $\gamma$ \cite{Sich_NaturePhotonics}, and $0$ is the limiting case of absence of dissipation. Note that, for both detunings considered, the bi- and multistability behaviour becomes increasingly pronounced as the dissipation is reduced. The curve in (d) clearly exhibits three distinct branches and two loops, as opposed to the commonly found single bistability loop observed in (b).

The multiple loops found is a new result specific to 1D microcavity wires, and is markedly distinct from the simple bistability observed in planar microcavities. Here below we analyse the origin of the multistability.

Each point of the bi- or multistability curve corresponds to a polariton ($E$, $\Psi$) mode. For the case $\Delta=-0.1$, only the first two (fundamental and first-order) modes of the family could potentially be excited (see left vertical dashed line in Fig.~\ref{fig:multistability_origin} (a) intersecting two modes). However, since the pump is spatially symmetric, the first-order mode does not arise, and therefore only the fundamental polariton mode (shown in Fig.~\ref{fig:multistability_origin} (c)) is observed in both the lower and upper bistability branches in Fig.~\ref{fig:multistability_origin} (b) -- including at the turning point, denoted `1'.

At $\Delta=0$, four modes could be excited (see right vertical line in Fig.~\ref{fig:multistability_origin} (a), intersecting four nonlinear modes). Note that, again, due to the even parity of the pump, only even modes survive -- in this case, the fundamental and second order. The real and imaginary parts of the modes corresponding to each turning point (denoted `2' and `3') are plotted in Fig.~\ref{fig:multistability_origin} (e) and (f), respectively. By comparing Fig.~\ref{fig:multistability_origin} (d) with the intersection points in (a) for $\Delta=0$, one can clearly see that these turning points correspond to a second-order and a fundamental mode, respectively. From this it could be inferred that the appearance of additional turning points may herald the appearance of additional stable states.

As we explain above, the predicted multistability is an interplay between the symmetric Gaussian pump and the number of modes supported by the microcavity wire potential. The number of supported modes can be controlled by selecting the confinement potential depth ($U_{bg}$) and width of the wire ($w$). The exact combination of modes can then be fine-tuned using the pump detuning.

\subsection{Modulational instability}

Once a stationary mode $\{A,B\}$ is found, we can analyze its stability by introducing a small perturbation:

\begin{eqnarray}
E=A(y)+\alpha[\epsilon_f(y)e^{iqx-i\delta t+\lambda t}+\epsilon_b^*(y)e^{-iqx+i\delta t+\lambda t}]e^{i\kappa x}\;,\\
\Psi=B(y)+\alpha[p_f(y)e^{iqx-i\delta t+\lambda t}+p_b^*(y)e^{-iqx+i\delta t+\lambda t}]e^{i\kappa x}\;,
\end{eqnarray}
where $\delta$ and $\lambda$ are real, $\alpha$ is a scaling parameter, $q$ is the perturbation wave vector, and the subscripts $f$ and $b$ stand for `forward' and `backward'-propagating waves, respectively.

\begin{figure*}
\centering
\vspace{10pt}
\resizebox{\columnwidth}{!}{
\includegraphics[width=0.33\textwidth]{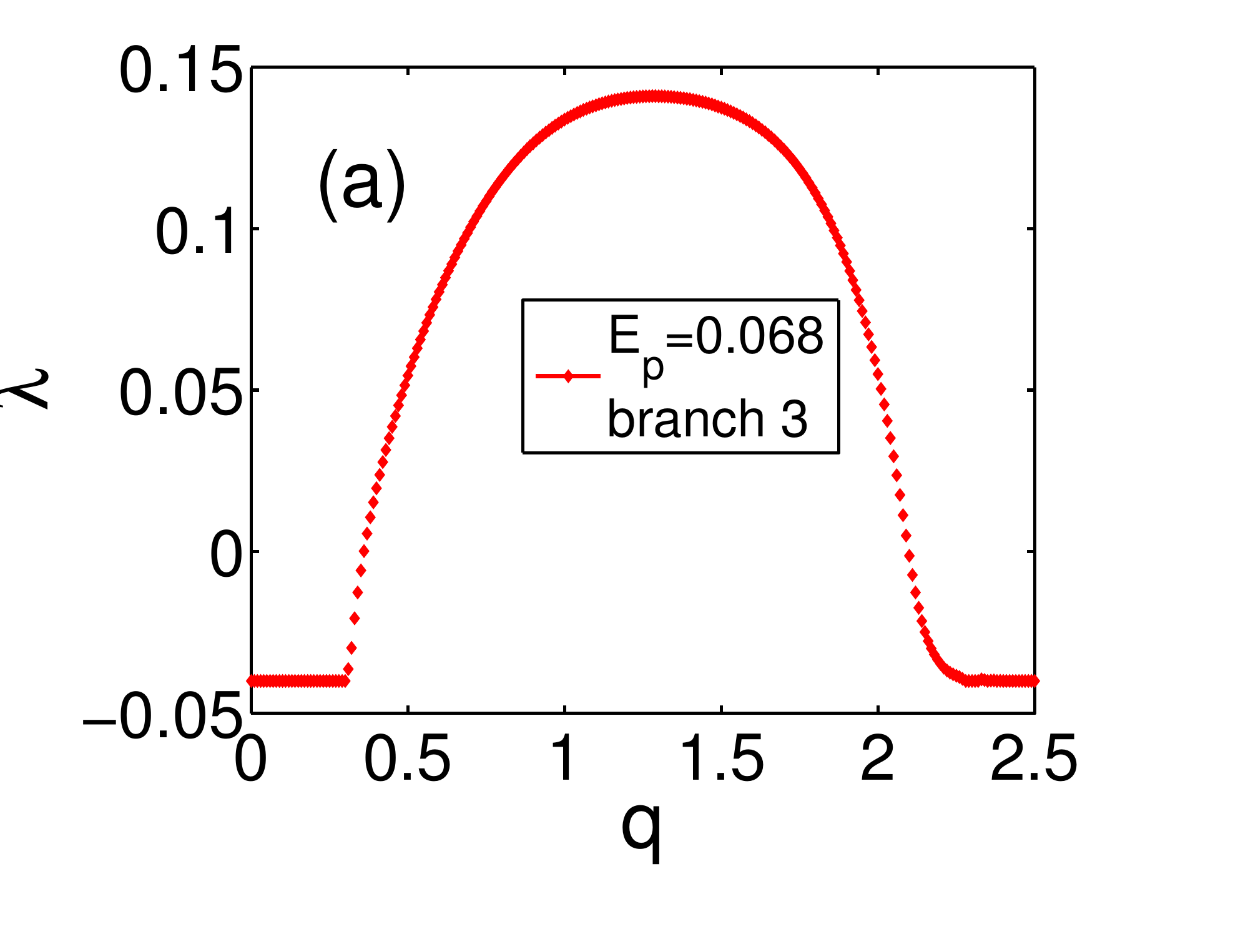}
\includegraphics[width=0.33\textwidth]{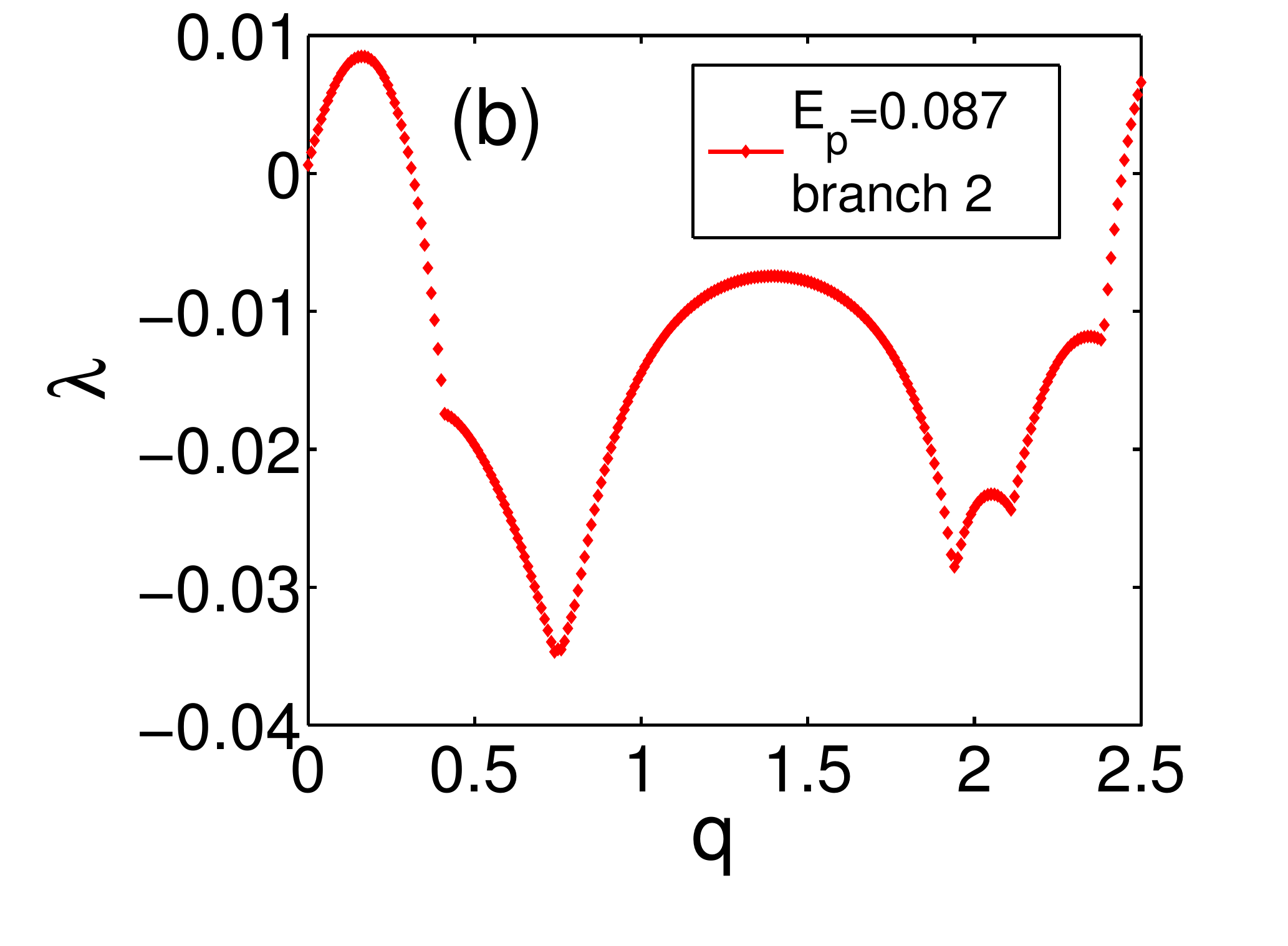}
\includegraphics[width=0.30\textwidth]{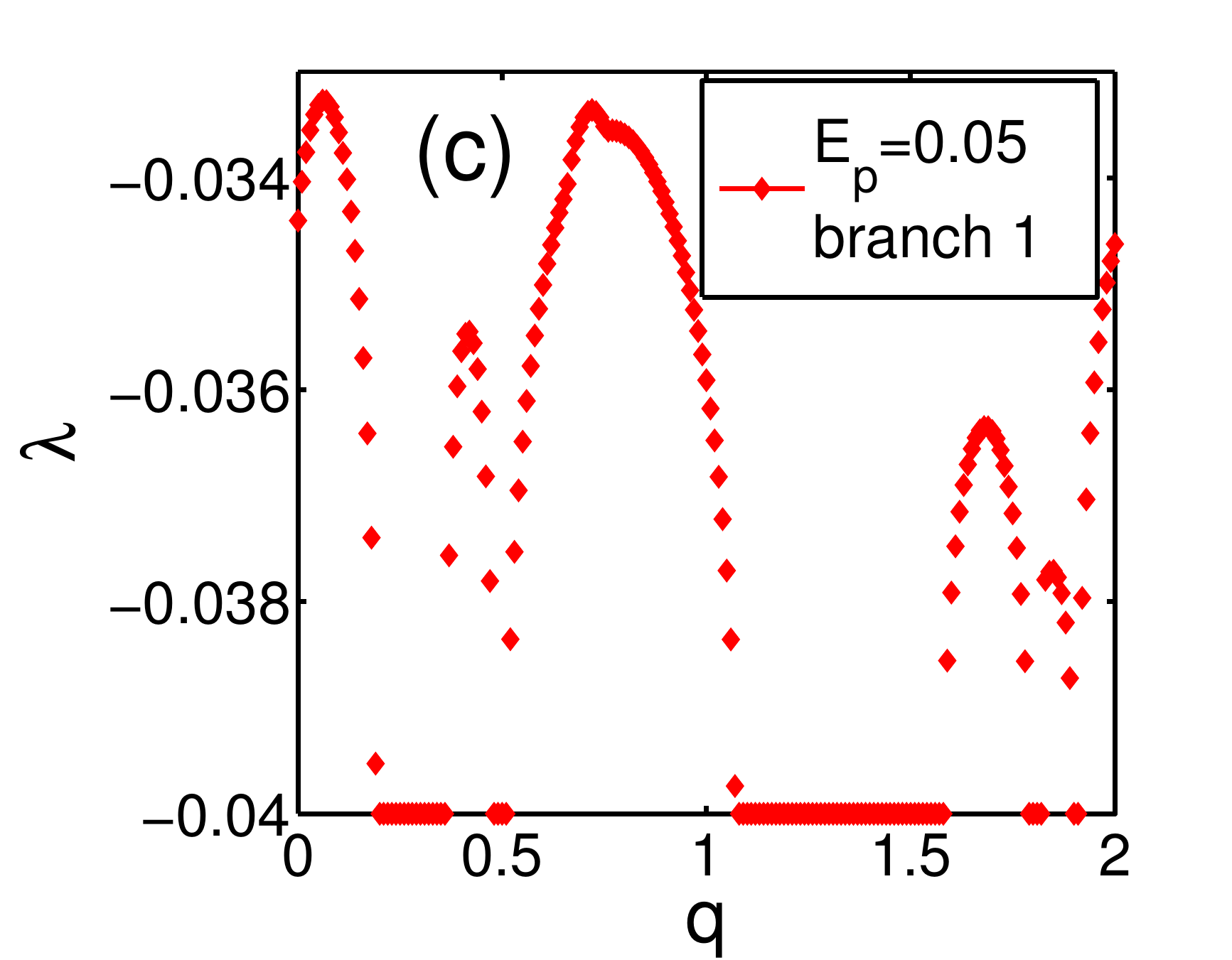}
}
\caption[Fig 3] {(Color online) Maximum imaginary part ($\lambda$) of the eigenvalue spectrum of the matrix $\hat M$, as a function of the perturbation wave vector, $q$, for a stationary nonlinear solution. If branches 1, 2 and 3 signify consecutive sections of the curve in Fig.~\ref{fig:multistability_origin} with positive slope, beginning to count from the abscissa upwards, then the results shown on this figure are representative of (a) branch 3 at pump amplitude, $E_p=0.068$; (b) branch 2 at $E_p=0.087$; (c) branch 1 at $E_p=0.05$, for $\gamma=0.04$.}
\label{fig:eigenvalue_spectra}
\end{figure*}

Assuming small amplitude perturbations $\varepsilon _{f,b}$ and $p_{f,b}$, neglecting the second-order terms $O\left( {\varepsilon _{_{f,b} }^2 } \right)$ and $O\left( {p_{_{f,b} }^2 } \right)$ and introducing $\vec{x}=[\epsilon_f,\epsilon_b,p_f,p_b]^T$, the system can be recast as the eigenvalue problem:

\begin{eqnarray}
(\delta+i\lambda)\vec{x}=\hat{M}\vec{x}\;,\hspace{10 mm}
\hat{M}=\left[
\begin{array}{cccc}
-\mathcal{L}_f & 0 & -\Omega_R & 0 \\
0 & \mathcal{L}_{b}^* & 0 & \Omega_R  \\
-\Omega_R & 0 & -\mathcal{P} & B^2 \\
0 & \Omega_R & -(B^*)^2 & \mathcal{P}^*
\label{eigenvalue_problem}
\end{array}
\right]\;,
\\
\hspace*{-1cm} \mathcal{L}_{f,b}=\partial^2_y-(\kappa\pm q)^2+\delta_c+\Delta+U+i\gamma_c,\hspace{5 mm}
\mathcal{P}=\delta_e+\Delta-2|B|^2+i\gamma_e\;.
\label{Operators}
\end{eqnarray}
We find the eigenvectors $\vec x$ and the complex eigenvalues ($\delta+i\lambda$) of the large sparse $4N \times 4M$ matrix, $\hat M$, each matrix element of which is a $N \times M$ block matrix, $N$ and $M$ being the number of grid points along the $x$ and $y$ directions (typically, $N=M=2048$). We designate the names `branch 1', `branch 2' and `branch 3' within the curves shown in Fig.~\ref{fig:multistability_origin} (d) for the consecutive sections with a positive gradient, starting to count from the bottom of the graph. We then plot the maximum of the eigenvalues' imaginary part, $\lambda$, against a small range of values of the $q$ wave vector corresponding to each of these branches (Fig.~\ref{fig:eigenvalue_spectra}), choosing $\Delta=0$.

\begin{figure*}
\centering
\vspace{10pt}
\includegraphics[width=0.5\textwidth]{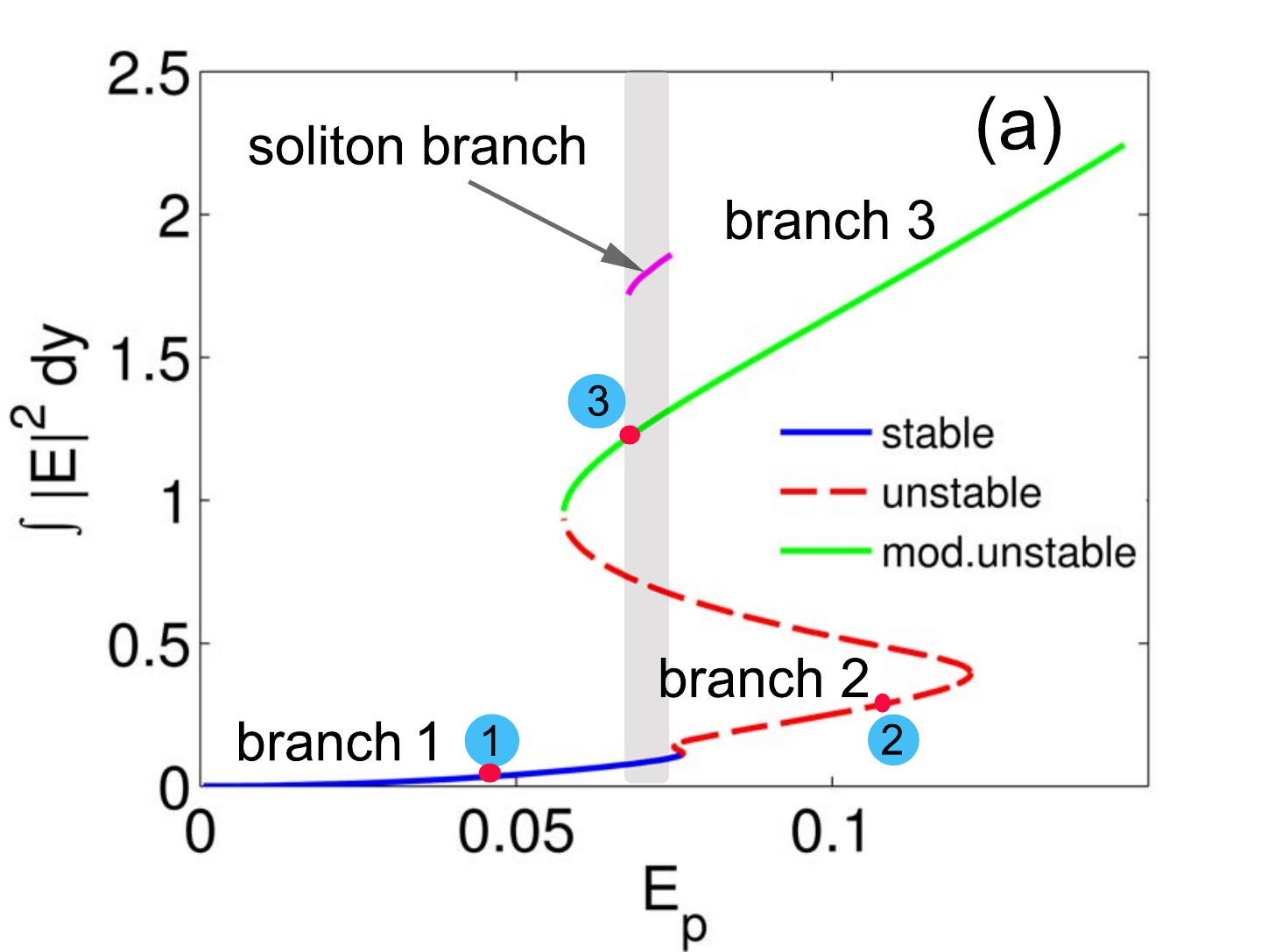}
\\
{\fontfamily{phv}\selectfont (b)}
\resizebox{\columnwidth}{!}{
\includegraphics[width=0.5\textwidth]{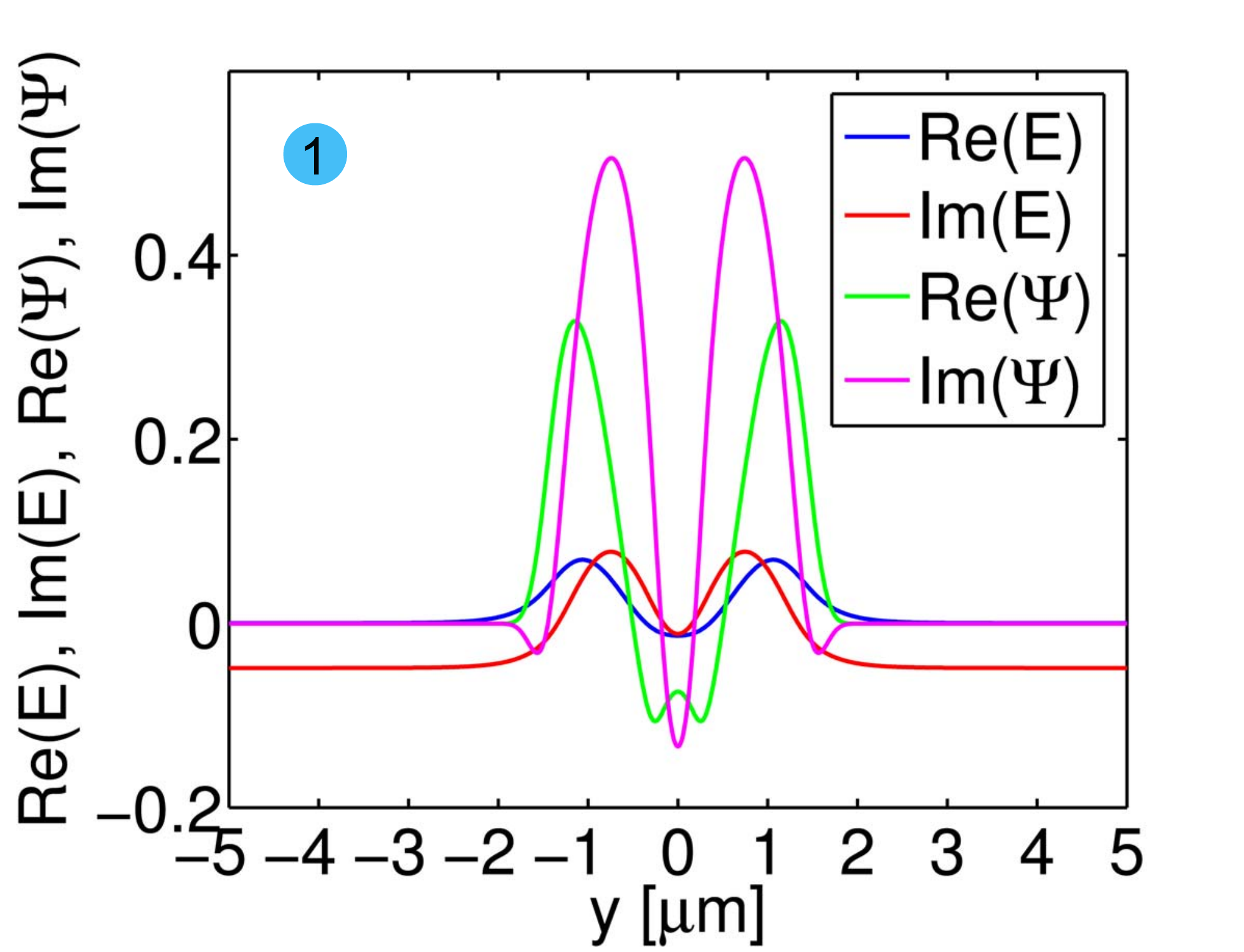}
\includegraphics[width=0.5\textwidth]{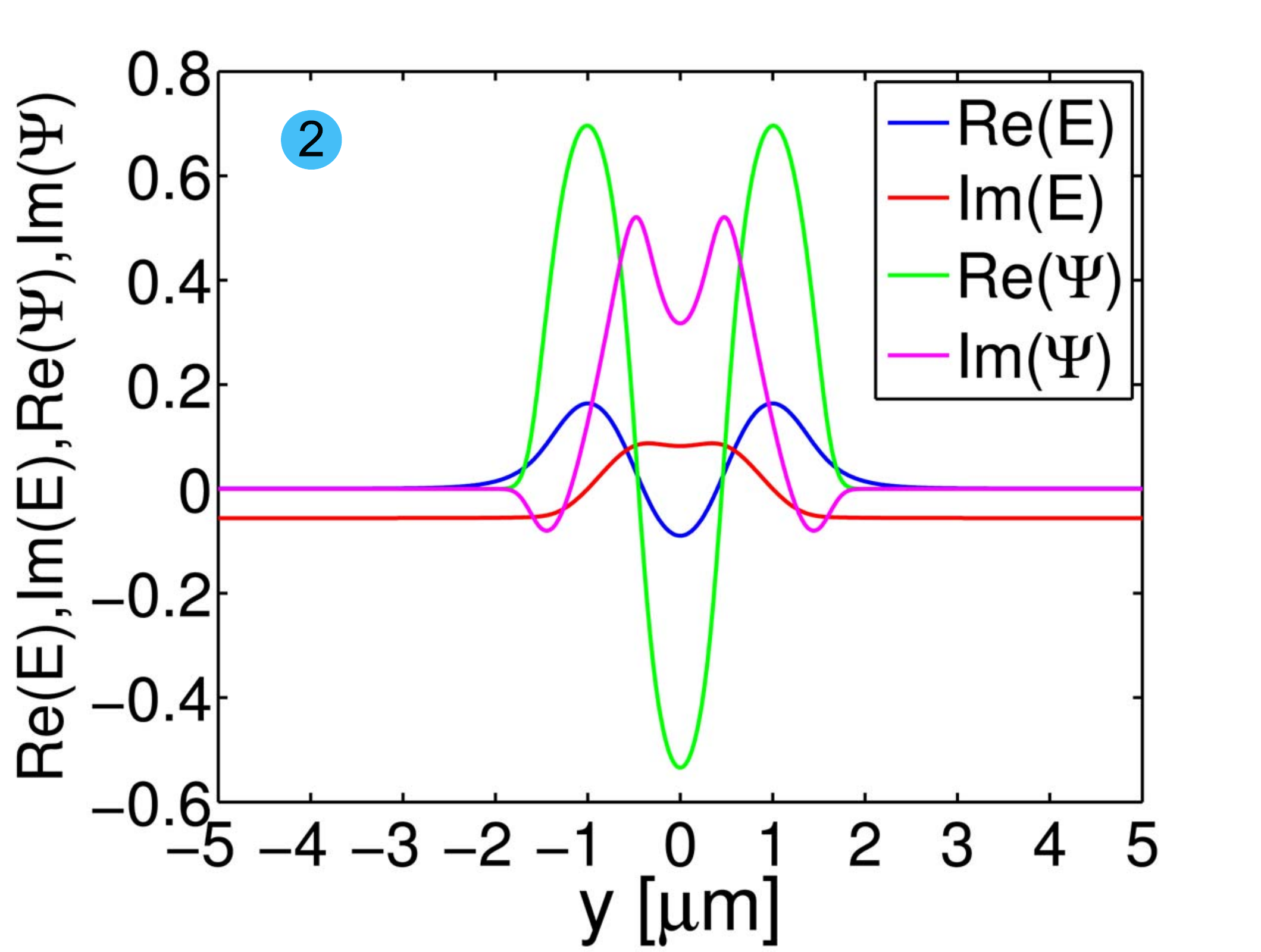}
\includegraphics[width=0.48\textwidth]{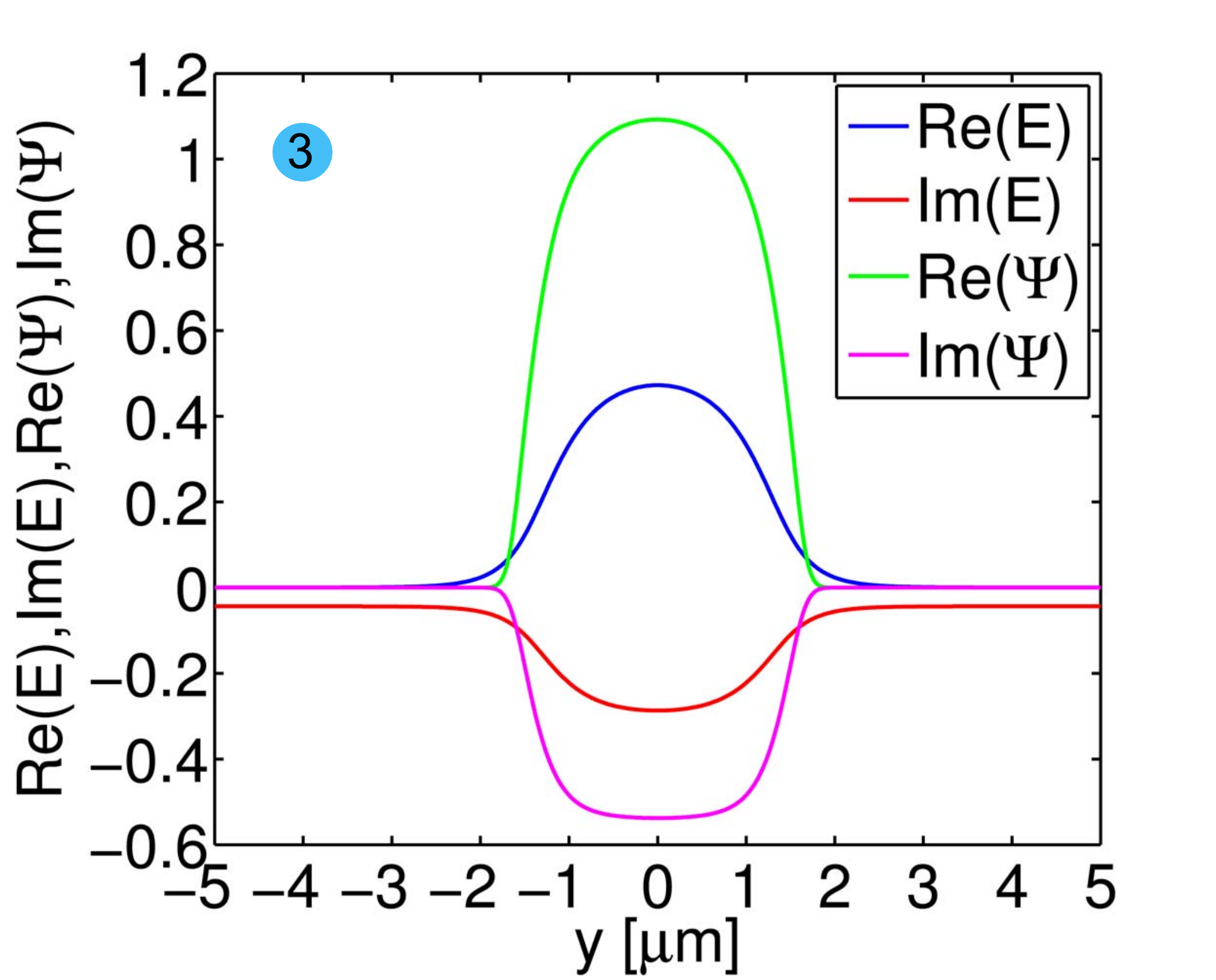}
}
\caption[Fig 4] {(Color online) (a) Multistability curve of the power of the photon's electric field ($E$), integrated along the $y$ direction, as a function of pump amplitude, $E_p$, at $\gamma=0.04$ and $\Delta=0$ for the stationary solutions of a microcavity polaritonic wire of width $w=3 \,\mathrm{\mu m}$ and pump incidence angle $\theta=20^\circ$. Each point on the curve corresponds to a specific polariton mode, shown in (b) (bottom row of this figure). Three distinct branches are discernible, depending on the type of solution obtained: stable (blue), modulationally unstable (green), and unstable (dashed red). The curve in magenta above the topmost branch represents the soliton branch, and was calculated numerically by the 2D Newton method. The shaded area indicates the region in which solitons exist. (b) Real and imaginary parts of the nonlinear photon ($E$) and exciton ($\Psi$) field modes at the points indicated in (a). Note that: we have a fourth-order mode at point 1 (on branch 1); a second-order mode at point 2 (on branch 2); a fundamental mode at point 3 (on branch 3).}
\label{fig:multistability_stationary_sols}
\end{figure*}

In Fig.~\ref{fig:multistability_stationary_sols} (a), we plot the equivalent of the multistability curve from Fig.~\ref{fig:multistability_origin} (d) at $\gamma=0.04$ for the integrated $E$-field power as a function $E_p$. The stability of each point on the curve is worked out as follows. If, for a given $E_p$, the relevant curve in Fig.~\ref{fig:eigenvalue_spectra} contains at least one point for which $\lambda>0$ (as long as it is not the one at $q=0$), then the corresponding point on the multistability curve is deemed to be `modulationally unstable'. By inspection of Fig.~\ref{fig:eigenvalue_spectra} (a), this is the case for branch 3, and so the manifold of all points belonging to this branch is modulationally unstable (colour-coded in green in Fig.~\ref{fig:multistability_stationary_sols} (a)). This type of instability may nevertheless lead to the formation of stable localised entities, such as solitons. If, on the other hand, the curve lies below zero (Fig.~\ref{fig:multistability_stationary_sols} (c)), the corresponding point on the multistability curve is classified as stable (such as is the case for branch 1). If the curve lies entirely above zero, or if at least $\lambda(q=0)>0$ (Fig.~\ref{fig:multistability_stationary_sols} (b)), the corresponding point is classified as `unstable' (branch 2). The modes (real and imaginary parts of the $E$- and $\Psi$-fields) corresponding to each of the multistability curve branches, are shown in Fig.~\ref{fig:multistability_stationary_sols} (b), bottom row.

\begin{figure*}
\centering
\vspace{10pt}
\includegraphics[width=0.43\textwidth]{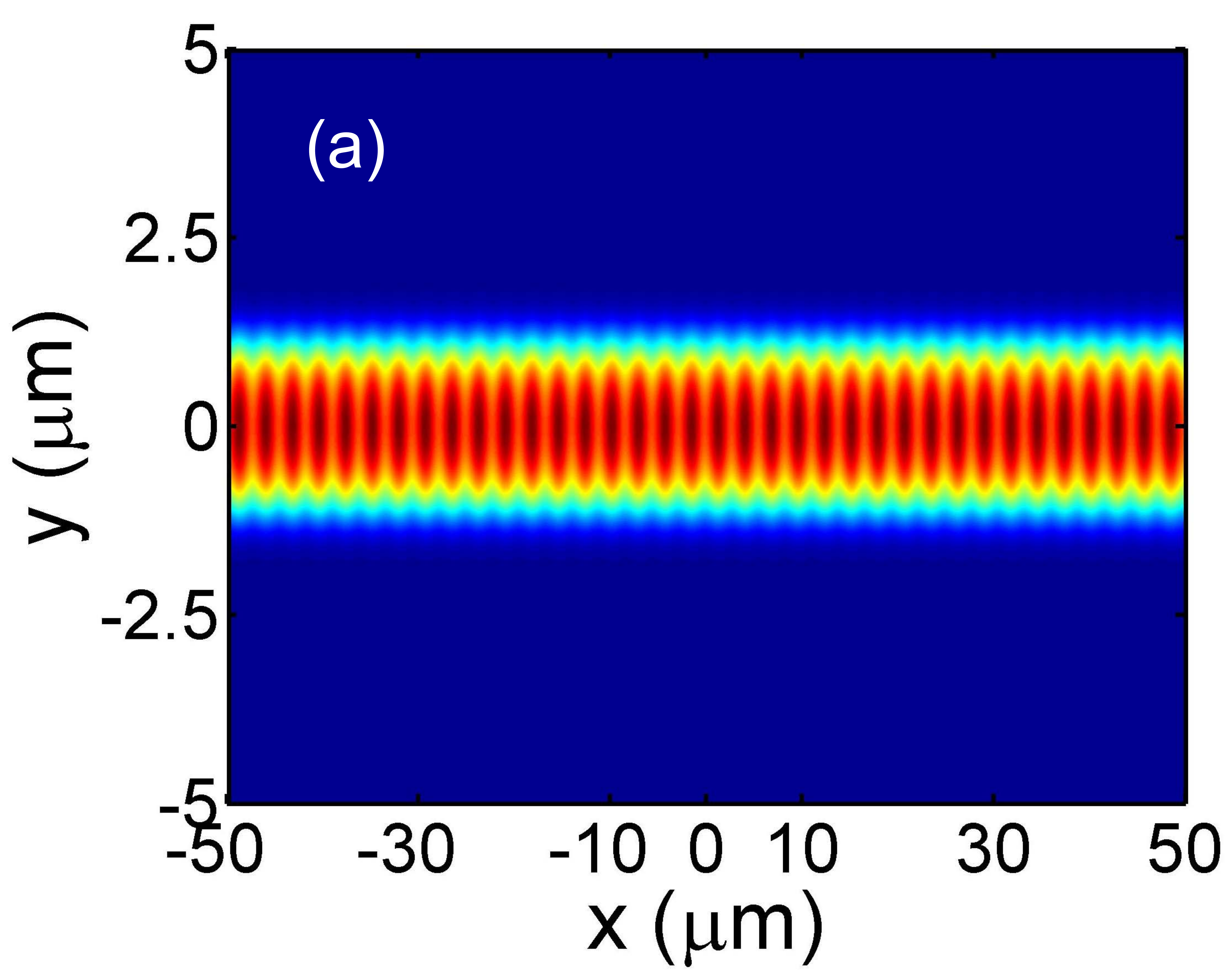}
\includegraphics[width=0.43\textwidth]{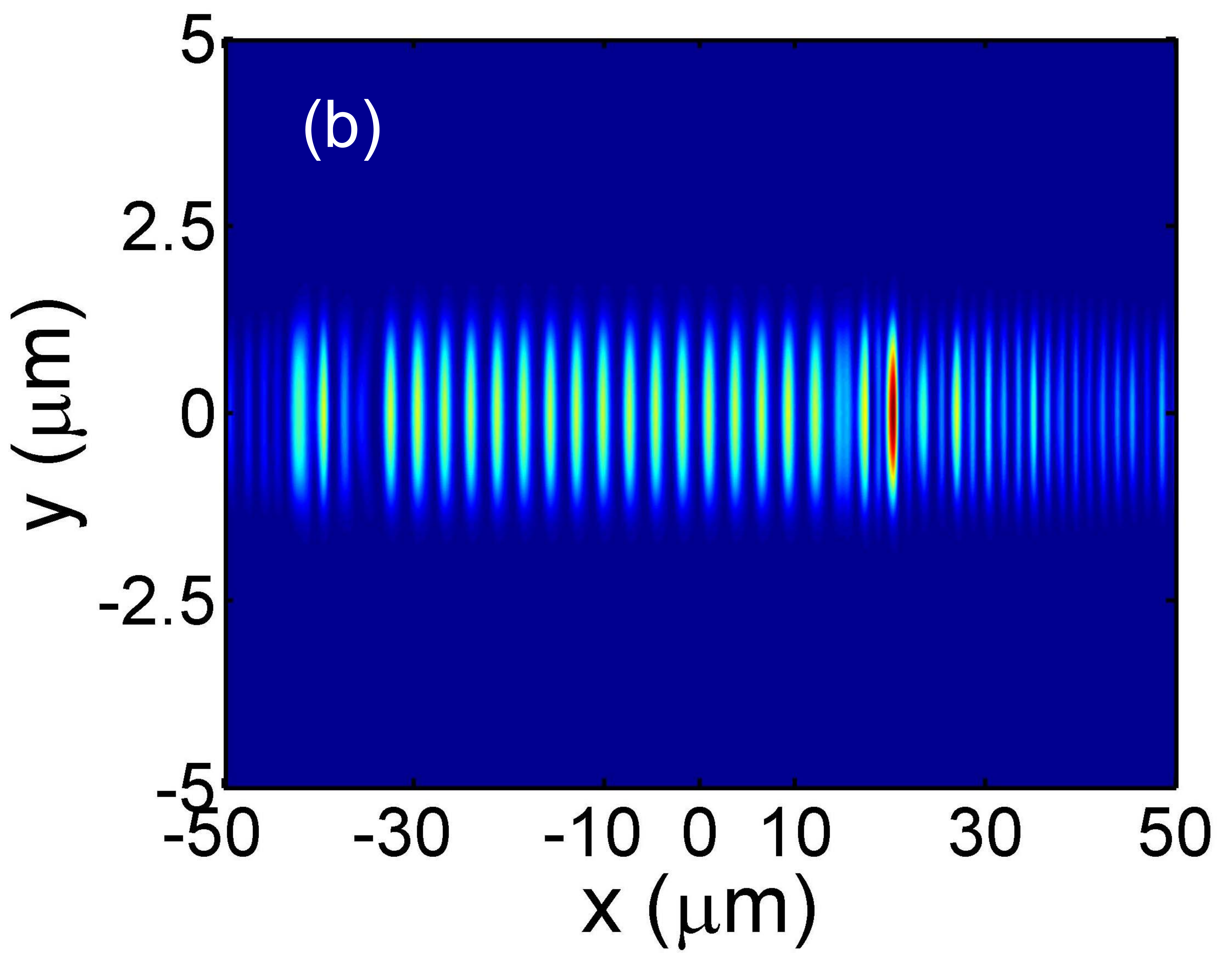}
\caption[Fig 5] {(Color online) Dynamical check of the modulational instability of the stationary solution, by means of solving Eqs.~(\ref{eqE}) and (\ref{eqPsi}) in the time domain using the split-step method. Snapshots of the time evolution of the perturbed mode's photonic electric-field intensity at (a) $t= 0.4 \,\mathrm{ps}$ and (b) $t=20 \,\mathrm{ps}$, after excitation with a Gaussian-shaped seed pulse of duration $2$ ps at time $t=0$. The perturbation wave vector is $q=1.2$, corresponding to the local maximum of growth rate in Fig.~\ref{fig:eigenvalue_spectra} (a) for $E_p=0.068$, lying on the upper branch (3) of the multistability curve in Fig.~\ref{fig:multistability_stationary_sols} (a). Clearly, the modulational instability on the third branch leads to localisation and concomitant formation of a soliton.}
\label{fig:bistability_dyn_check1}
\end{figure*}

\begin{figure*}
\centering
\vspace{10pt}
\includegraphics[width=0.43\textwidth]{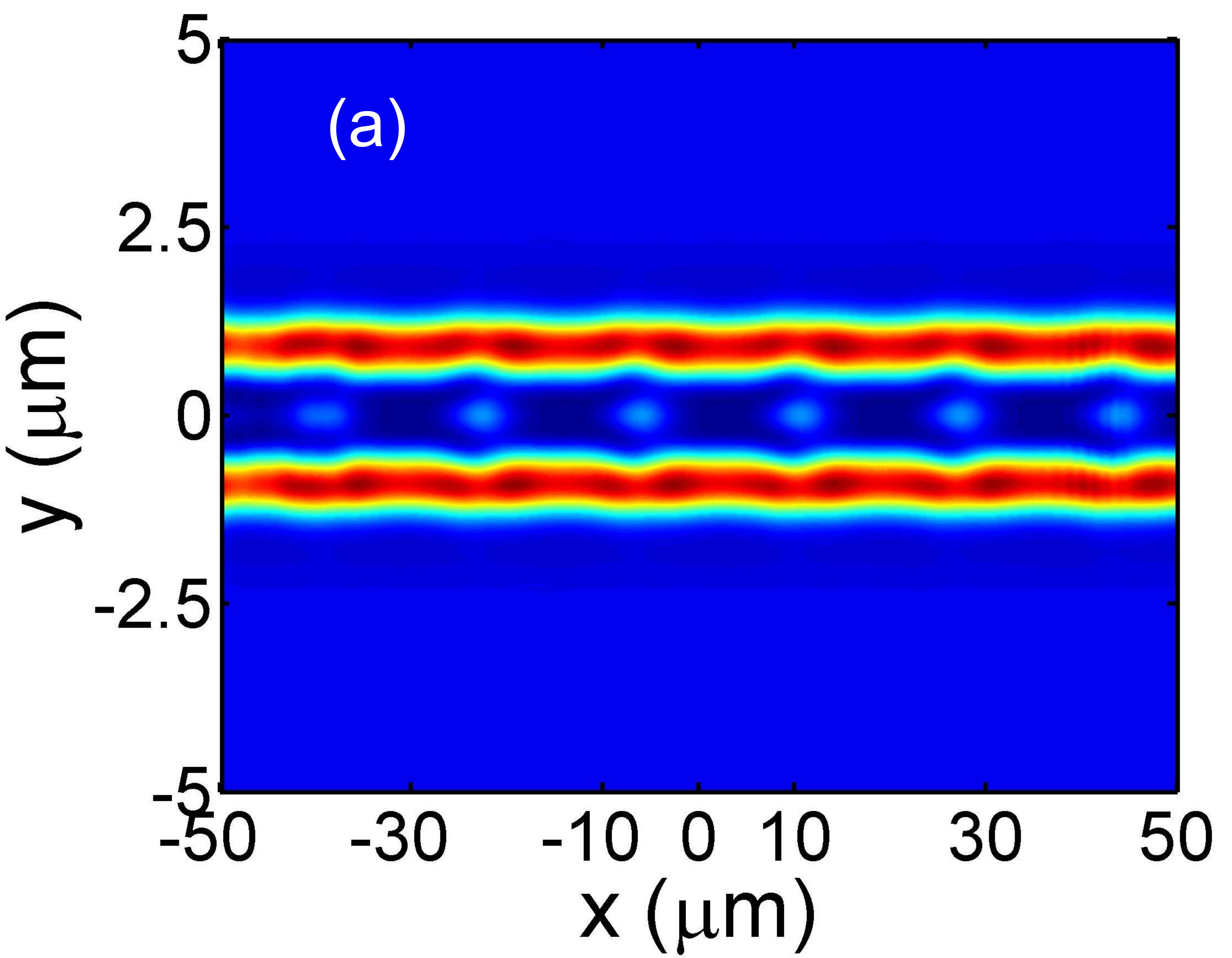}
\includegraphics[width=0.43\textwidth]{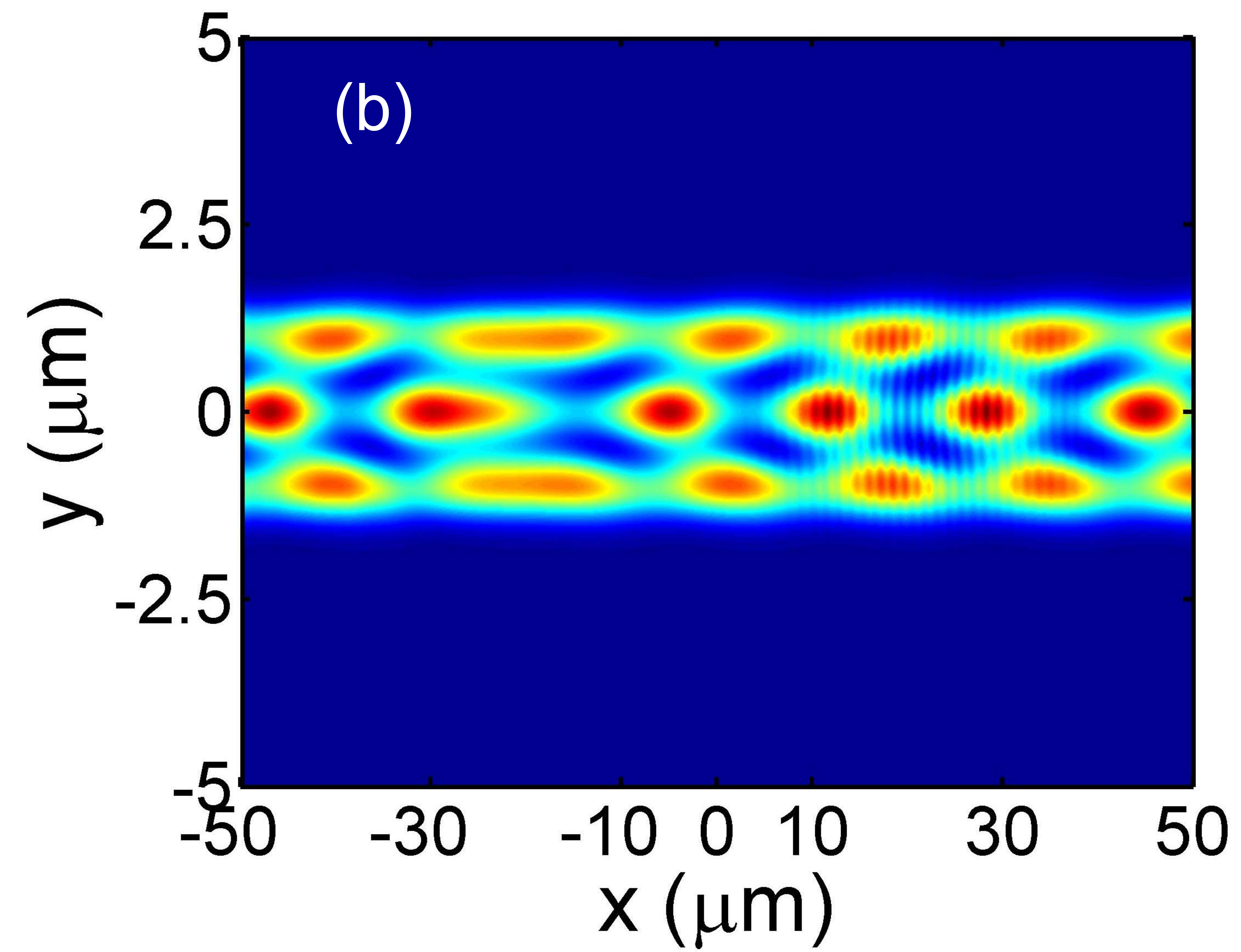}
\caption[Fig 6] {(Color online) Dynamical check of the instability of the stationary solution, by means of solving Eqs.~(\ref{eqE}) and (\ref{eqPsi}) in the time domain using the split-step method. Snapshots of the time evolution of the perturbed mode's photonic electric-field intensity at (a) $t=0.4 \,\mathrm{ps}$ and (b) $t=28.4 \,\mathrm{ps}$, after excitation with a Gaussian-shaped seed pulse of duration $2$ ps at time $t=0$. The perturbation wave vector is $q=0.2$, corresponding to the local maximum of growth rate in Fig.~\ref{fig:eigenvalue_spectra} (b) for the pump amplitude, $E_p=0.087$, lying on branch 2 of the multistability curve in Fig.~\ref{fig:multistability_stationary_sols} (a). Perturbation of the mode on the second branch clearly leads to filamentation, rather than localisation. Therefore, the mode is unstable.}
\label{fig:bistability_dyn_check2}
\end{figure*}

To verify the above linear stability analysis, we perform a dynamical check by solving Eqs.~(\ref{eqE}) and (\ref{eqPsi}) in the time domain by the split-step method. First, we propagate the perturbed mode at $E_p=0.068$ and $q=1.2$ as in Fig.~\ref{fig:eigenvalue_spectra} (a). Snapshots at $t=0.4$ and $20 \,\mathrm{ps}$ of the time evolution of the perturbed solution are shown in Fig.~\ref{fig:bistability_dyn_check1} (a) and (b), respectively. In agreement with the stability analysis above, this perturbed solution is modulationally unstable, and exhibits localisation and soliton-like features at later times (Fig.~\ref{fig:bistability_dyn_check1} (b)). By contrast -- and still in agreement with the above analysis -- the perturbed solution for $E_p=0.087$ and $q=0.2$ (corresponding to Fig.~\ref{fig:eigenvalue_spectra} (b)), shown at $t=0.4$ and $28.4 \,\mathrm{ps}$ in Fig.~\ref{fig:bistability_dyn_check2} (a) and (b), respectively, is unstable. In this case, the time evolution exhibits filamentation, rather than localisation and soliton-like features.
The stationary nonlinear mode profiles and perturbation eigenvectors computed from Eqs.~\ref{eigenvalue_problem}-\ref{Operators} are given in \cite{Supplementary_material} for both cases considered above.

The magenta curve above the third (topmost) branch in Fig.~\ref{fig:multistability_stationary_sols} (a) is the soliton branch. It is represented by the integrated $E$-field power along the transverse cross-section through the soliton maximum, which was computed by the 2D Newton method. The shaded area indicates the region in which solitons exist.

\subsection{Dynamical modelling of soliton formation}
\label{sssec:3.4}
\begin{figure*}
\centering
\vspace{10pt}
\includegraphics[width=0.43\textwidth]{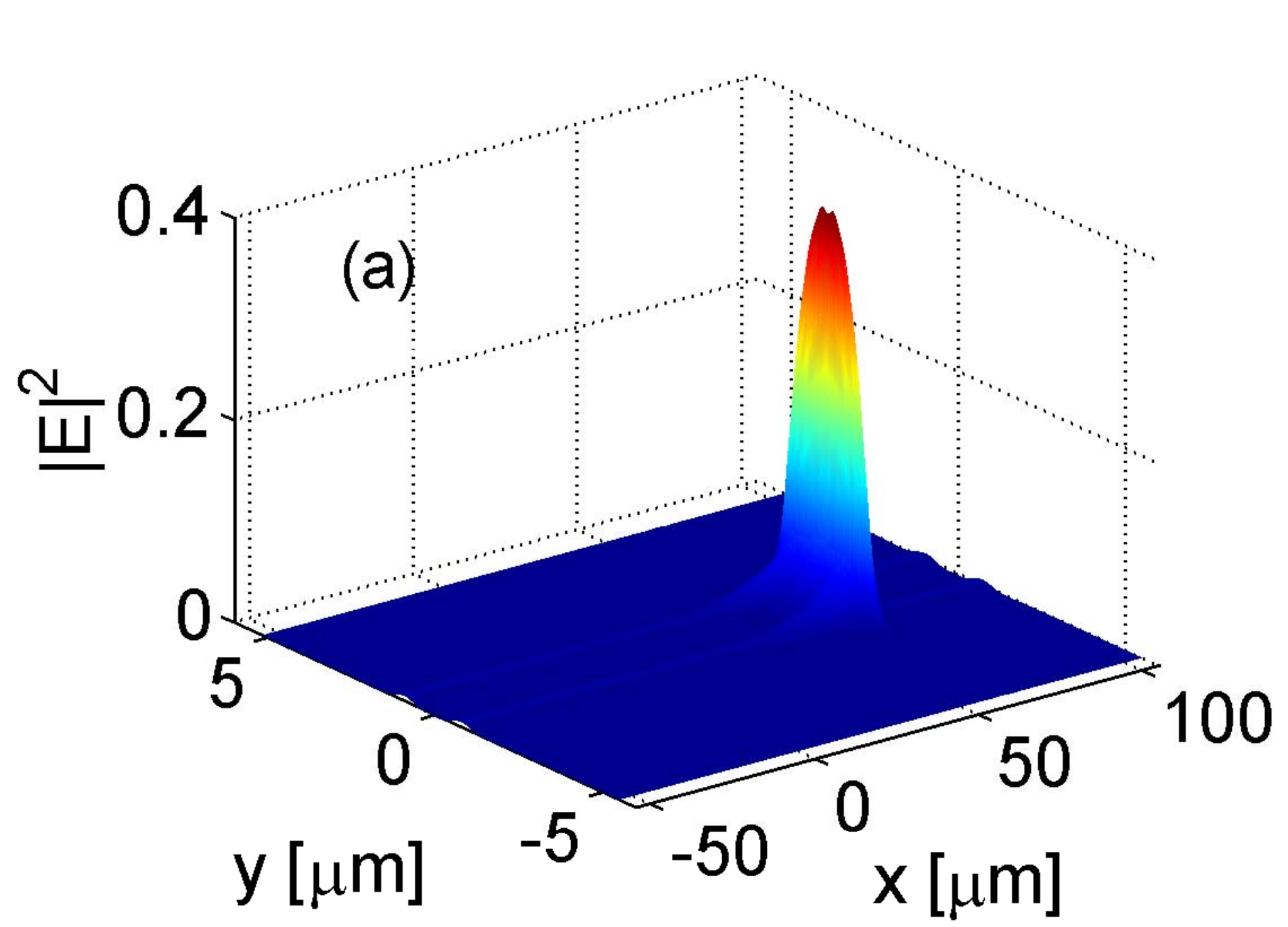}
\\
\includegraphics[width=0.43\textwidth]{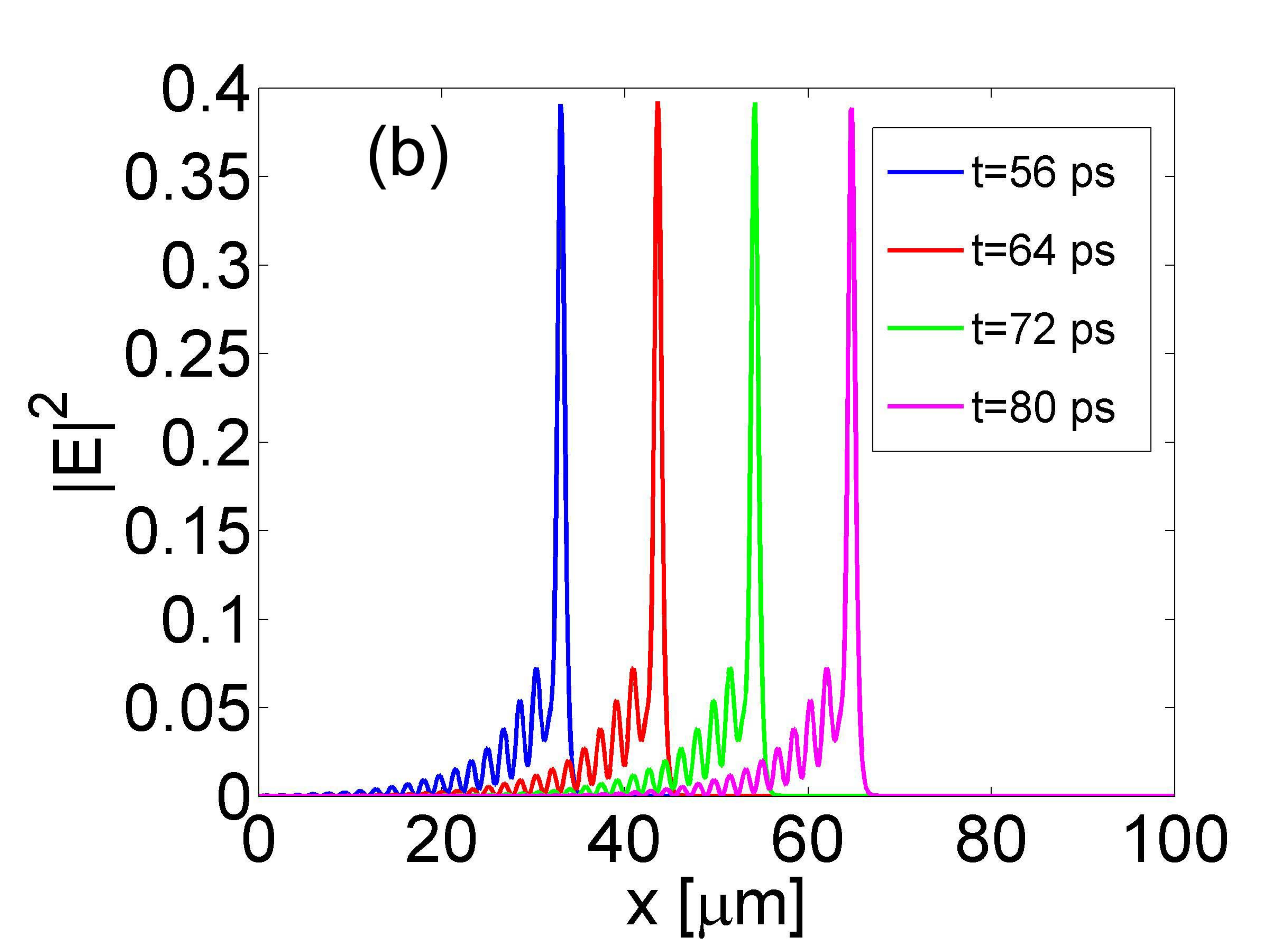}
\caption[Fig 7] {(Color online) Dynamical modelling of soliton propagation by means of the split-step method. A seed pulse of duration $\tau=2\,\mathrm{ps}$ launched at $t=0$ catalyses the formation of the soliton. (a) Snapshot at $t=80 \,\mathrm{ps}$ of the photonic electric-field intensity of a single-soliton solution at $\Delta=0$ and $\gamma=0.04$ for pump amplitude $Ep=0.075$. (b) Cross-section of the above along the middle of the microcavity wire (i.e. for $y=0$), showing the intensity profile of the soliton at four different time points: $t=56, 64, 72$ and $80 \,\mathrm{ps}$.}
\label{fig:soliton_1_2_3D}
\end{figure*}

We choose the pump amplitude within the interval of existence of solitons (see Fig.~\ref{fig:multistability_stationary_sols} (a), shaded area) and initialise the system with the stable mode from the lowest branch. The soliton formation is triggered at time $t=0$ by a short ($\tau=2 \,\mathrm{ps}$) Gaussian seed pulse with intensity FWHM of $3\,\,\mathrm{\mu m}$. The pulse is collinear with the pump and provides perturbation to the stationary solution. We solve the system of equations, Eqs.~(\ref{eqE}) and (\ref{eqPsi}), in the time domain by the split-step method. Following a transient process of initial pulse reshaping (data not shown), a soliton is formed. The stable soliton solution at $\Delta=0$ and $\gamma=0.04$ is displayed in Fig.~\ref{fig:soliton_1_2_3D} (a) for $E_p=0.075$. Longitudinal cross-sections of the soliton profile in (a) along the microcavity wire ($y=0$) are shown for different simulation times in Fig.~\ref{fig:soliton_1_2_3D} (b). From these, it is clear that the soliton shape is preserved during propagation along the wire (see also a movie of the soliton propagation in \cite{Supplementary_material}).

\begin{figure*}
\centering
\vspace{10pt}
\resizebox{\columnwidth}{!}{
\includegraphics[width=0.43\textwidth]{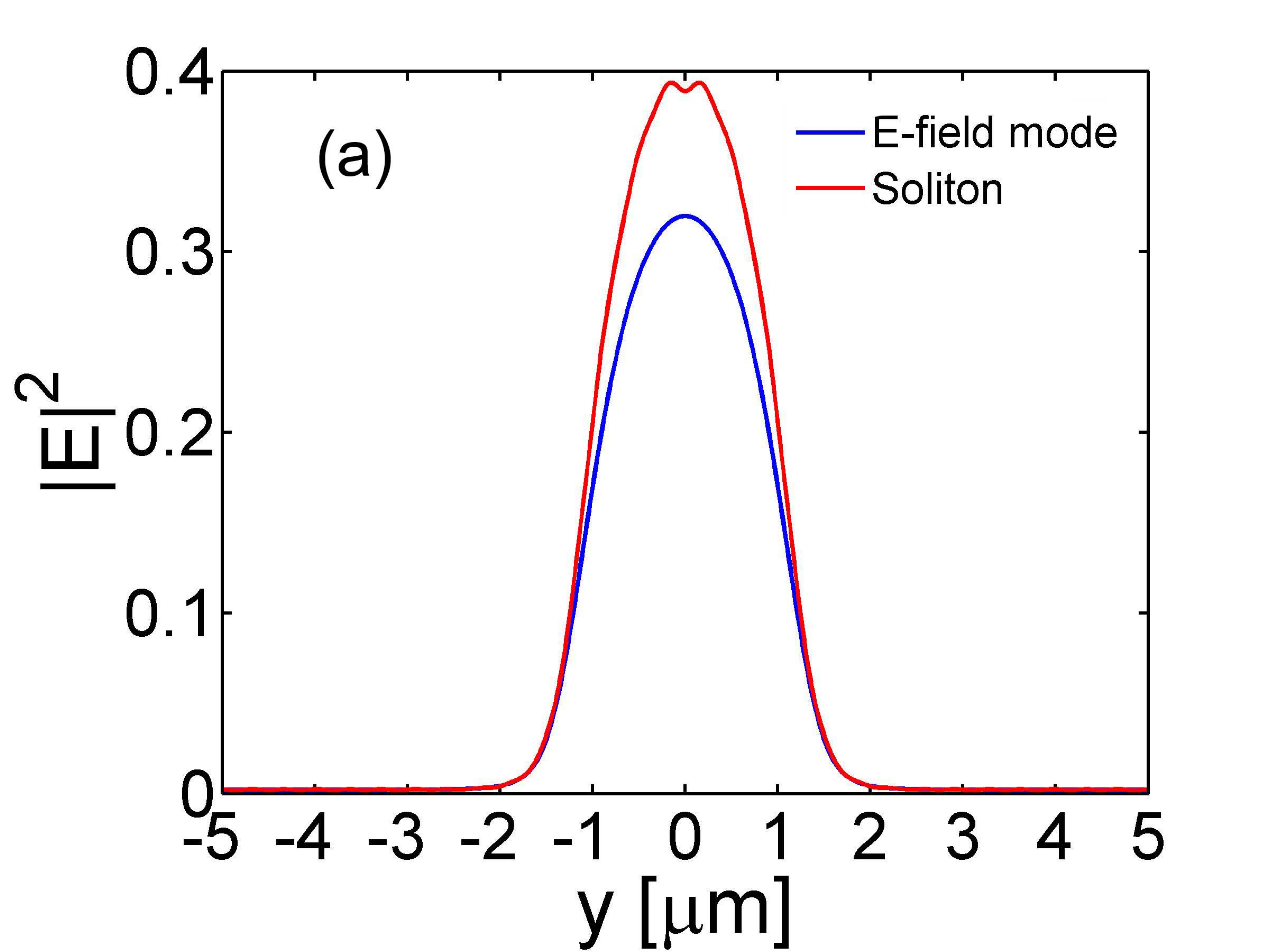}
\includegraphics[width=0.43\textwidth]{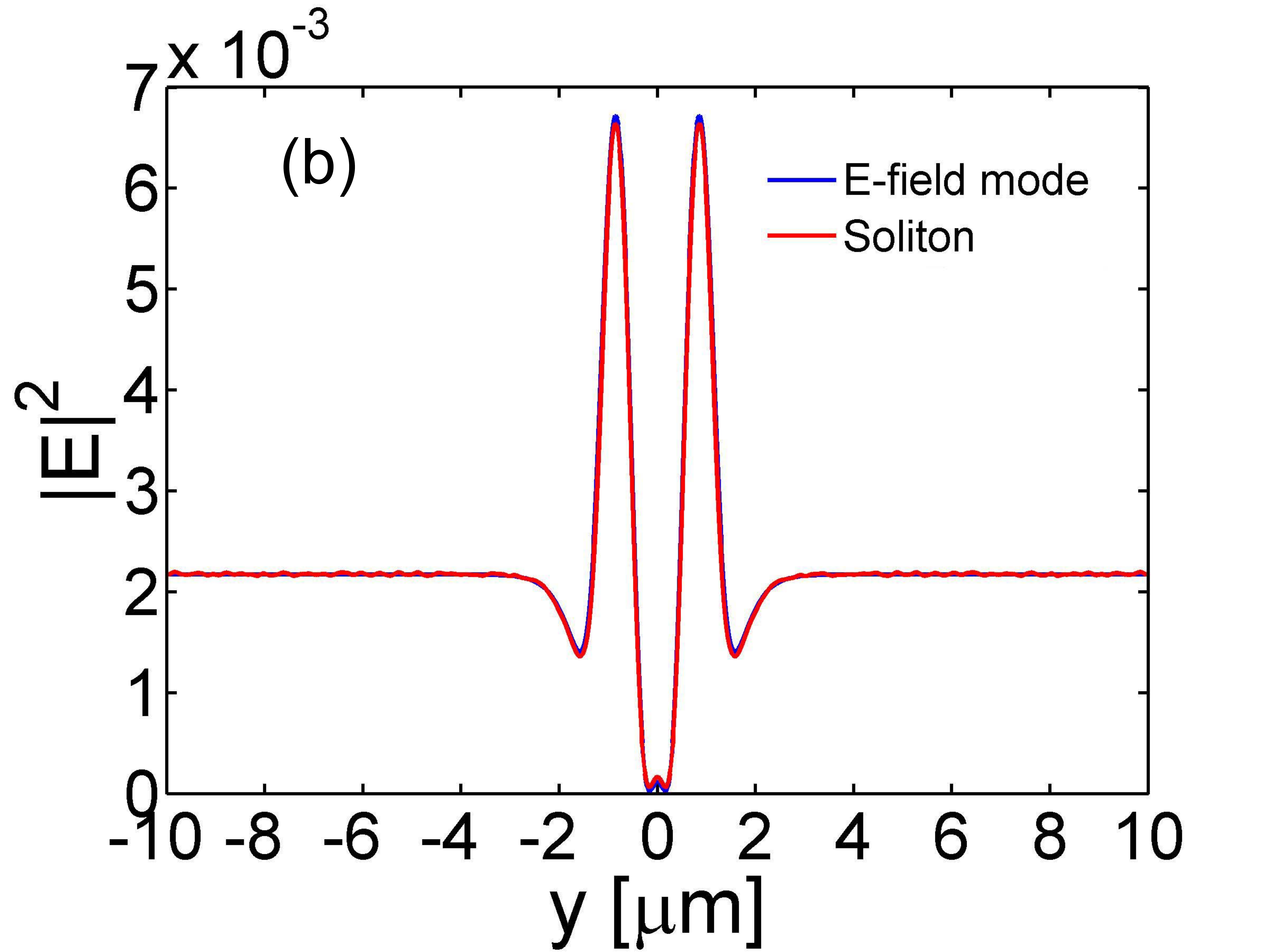}
}
\caption[Fig 8] {(Color online) Transverse cross-section through: (a) the soliton maximum (red curve) and the stationary mode (blue curve) on branch 3 of the multistability curve (see Fig.~\ref{fig:multistability_stationary_sols} (a)); (b) the soliton tail (red curve) and the stationary mode (blue curve) on branch 1. $w=3 \mu m$, $\theta=20^{\circ}$, $E_p=0.075$, $\Delta=0$, and $\gamma=0.04$.}
\label{fig:soliton_modes_cross_section}
\end{figure*}

In Fig.~\ref{fig:soliton_modes_cross_section} (a), a transverse cross-section of the $E$-field power, $\vert E\vert^2$, through the soliton maximum and the stationary nonlinear mode of branch 3 for $E_p=0.075$ are plotted on the same graph. The soliton is clearly more intense in the vicinity of that point than the mode of branch 3. In Fig.~\ref{fig:soliton_modes_cross_section} (b), a transverse cross-section of the $E$-field power in the soliton tail, along with the nonlinear mode corresponding to branch 1 with $E_p=0.075$ are shown. Whereas the soliton and nonlinear mode differ significantly in the peak, they appear quite similar in magnitude in the tail.

It is worth noting that for $E_p=0.075$, the soliton temporal width (at FWHM), $T_p=0.727 \,\mathrm{ps}$, is nearly one-half of the soliton temporal width in a planar microcavity ($ \sim1.25 \,\mathrm{ps}$) \cite{Sich_NaturePhotonics}, and the soliton is squeezed along the microcavity wire.

\subsection{Radiating solitons: collapses and revivals}
\label{sssec:3.5}
In this section, we study multistability and soliton formation in a microcavity wire with the same width and incidence angle as before (i.e., $w=3\,\mathrm{\mu m}$, $\theta=20 ^{\circ}$), whilst keeping $\Delta=0$ and selecting $\gamma=0.01$. These conditions correspond to the multistability curve denoted by the red dashed line in Fig.~\ref{fig:multistability_origin} (d).

\begin{figure*}
\centering
\vspace{10pt}
\includegraphics[width=0.43\textwidth]{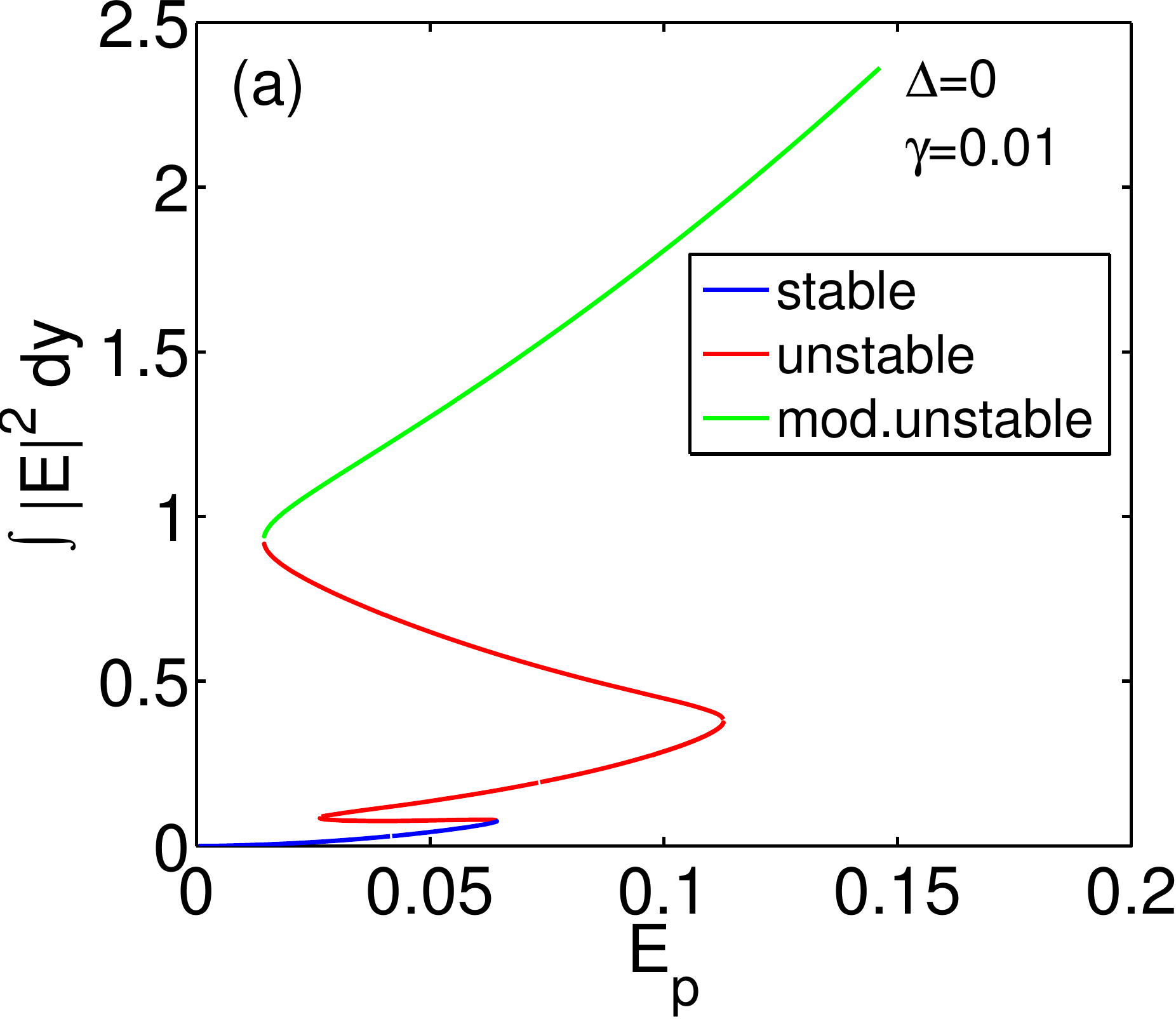}
\includegraphics[width=0.43\textwidth]{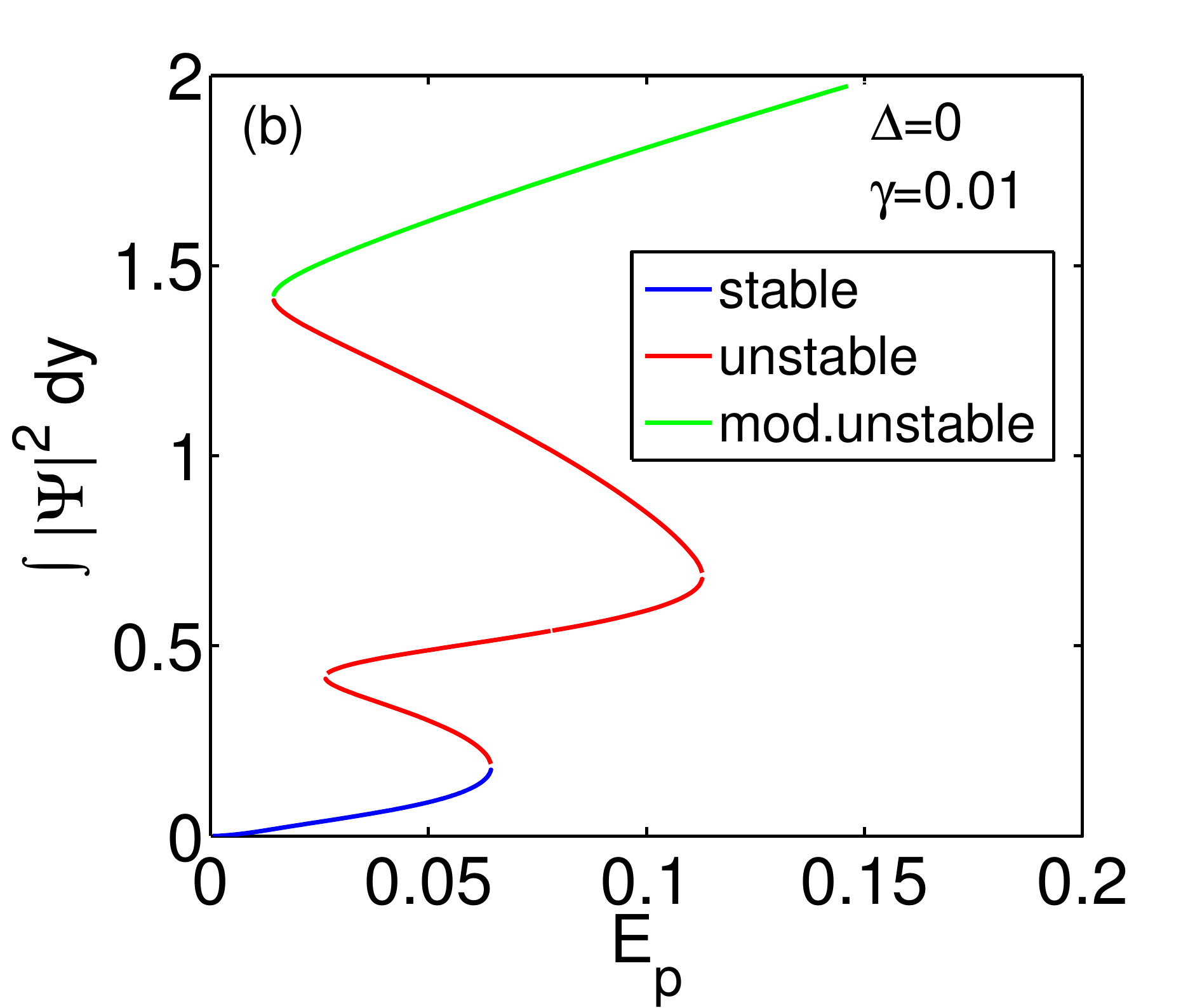}
\includegraphics[width=0.7\textwidth]{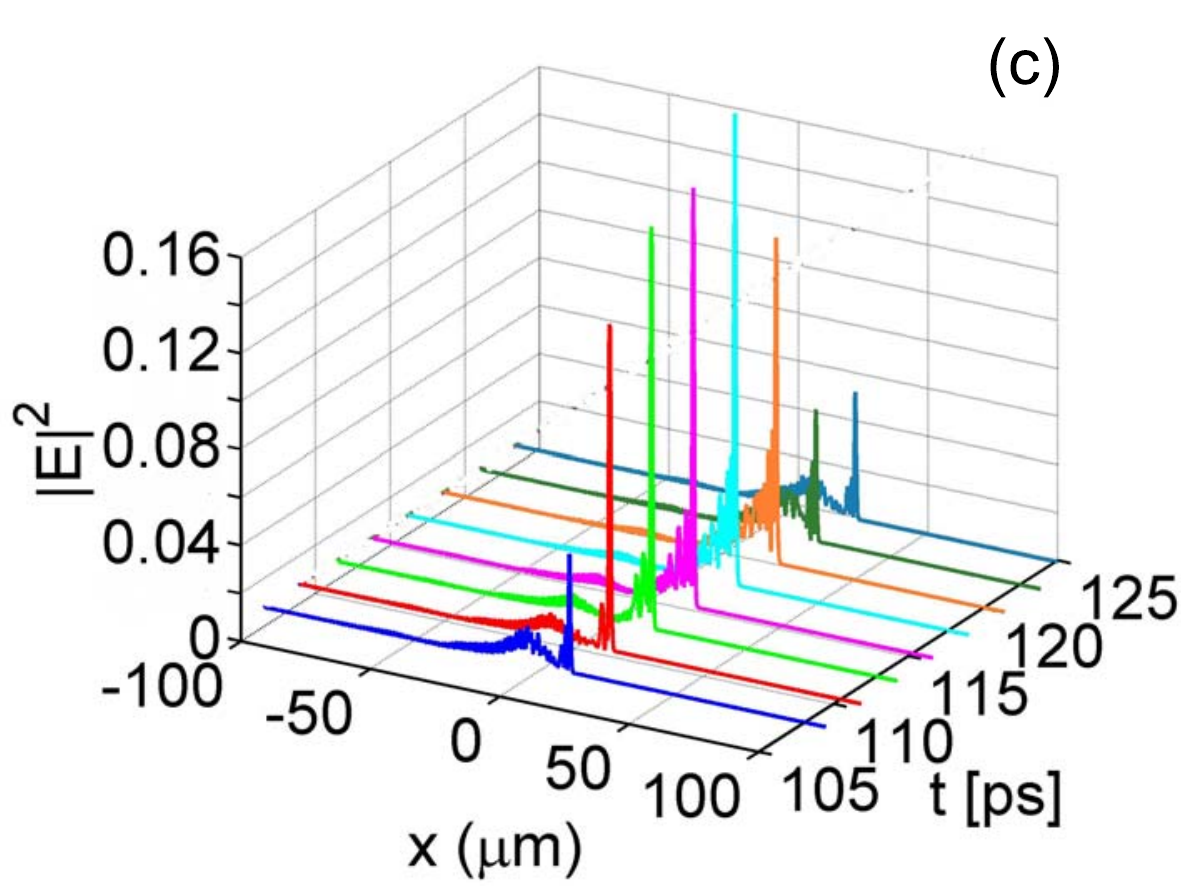}
\caption[Fig 9] {(Color online) Integrated electric-field intensity of the (a) photon ($E$) and (b) exciton ($\Psi$) component of the soliton, and its dependence on the pump amplitude, $E_p$, for $\Delta=0$ and $\gamma=0.01$. Two bistability loops are evident. Sections of the multistability curve corresponding to stable light states are depicted in blue, unstable in red, and modulationally unstable in green. (c) Time evolution of the radiating soliton propagating along the microcavity wire. Snapshots of the $E$-field intensity profile along the middle of the wire (i.e., along the $x$ axis) at $t=108, 110.4, 112.8, 115.2, 117.6, 120, 122.4$, and $124.8 \,\mathrm{ps}$. In this case, $E_p=0.0341$. Radiating solitons are observed only over a very narrow interval of pump amplitudes (approximately $[0.025, 0.035]$).}
\label{fig:multistability_rad_soliton}
\end{figure*}

The multistability curves for the $E$-field and $\Psi$-field integrated mode powers, obtained from Eqs.~(\ref{nonlinear_modes}) and (\ref{nonlinear_modes2}), are shown in Fig.~\ref{fig:multistability_rad_soliton} (a) and (b), respectively. After performing a linear stability analysis, we identified the sections of the curve corresponding to stable, modulationally unstable and unstable nonlinear homogeneous solutions. However, in this case we could not find solitons of the type that we found for the case of $\gamma=0.04$. Instead, we found an interesting type of radiating soliton in the vicinity of $E_p=0.0341$, which recovers after emitting radiation in a cyclical manner (see movie of the propagation of the radiating soliton in \cite{Supplementary_material}).

Regular snapshots of the time evolution of this radiating soliton is displayed in Fig.~\ref{fig:multistability_rad_soliton} (c). The  resonant radiation, manifested in the appearance of a second peak in the trailing edge of the pulse, is likely due to higher-order dispersion effects, and leads to pulse damping. As a result, the soliton dynamics is described by a series of collapses, followed by revivals. After a pulse break-up and emission (see pulse profile at $107.5$ ps, Fig.~\ref{fig:multistability_rad_soliton} (c)), the soliton is recovered (pulse profile at $125$ ps). The revivals and collapses take place at regular intervals. An estimate for the time interval between soliton collapses, $t_c$, can be derived from comparing the first and the last frames in Fig.~\ref{fig:multistability_rad_soliton} (c), giving $t_c \approx 10$--$15 \,\,\mathrm{ps}$. Remarkably, the observed periodic dynamics is strongly dependent on the initial conditions, and in particular on the pump amplitude. Only excitations in a very narrow interval around the pump amplitude (approximately $[0.025, 0.035]$, determined from our simulations) result in this peculiar dynamics.

This periodic behaviour is reminiscent of the collapses and revivals of semiclassical Rabi oscillations occurring when resonant coherent pulses propagate in and interact with a resonantly absorbing/amplifying medium. On the condition that the pulse duration is smaller than the relaxation times in matter, the resonant coherent pulse effectively becomes a polariton soliton, whose pulse area is conserved during propagation owing to the Pulse Area Theorem \cite{McCall&Hahn}. In the present case, the Pulse Area Theorem is indeed satisfied, since $\gamma=0.01$ corresponds to a relaxation time $T(=T_1=T_2) \approx 12 \,\mathrm{ps}$, which is greater than the pulse duration ($2$ ps).

It should be noted, however, that the above analogy with Rabi oscillation collapses and revivals is incomplete, since the revival component of the phenomenon cannot be explained classically (i.e., without introducing photon field quantisation) \cite{Scully&Zubairy}. The semiclassical Gross-Pitaevskii mean-field model we have presented in this paper could only explain destructive interference effects, such as pulse collapses. We therefore find numerically that it must be the interference of the multiple discrete nonlinear modes within the microcavity wire that results in a semiclassical coherent effect mimicking the quantum dynamics of collapses and revivals. This interesting coherent propagation phenomenon deserves further investigation and should be a subject of future studies.

\subsection{Impact of wire width on the linear polariton energy and mode group velocity dispersion}
\label{sssec:3.6}
\begin{figure*}
\centering
\vspace{10pt}
\includegraphics[width=0.43\textwidth]{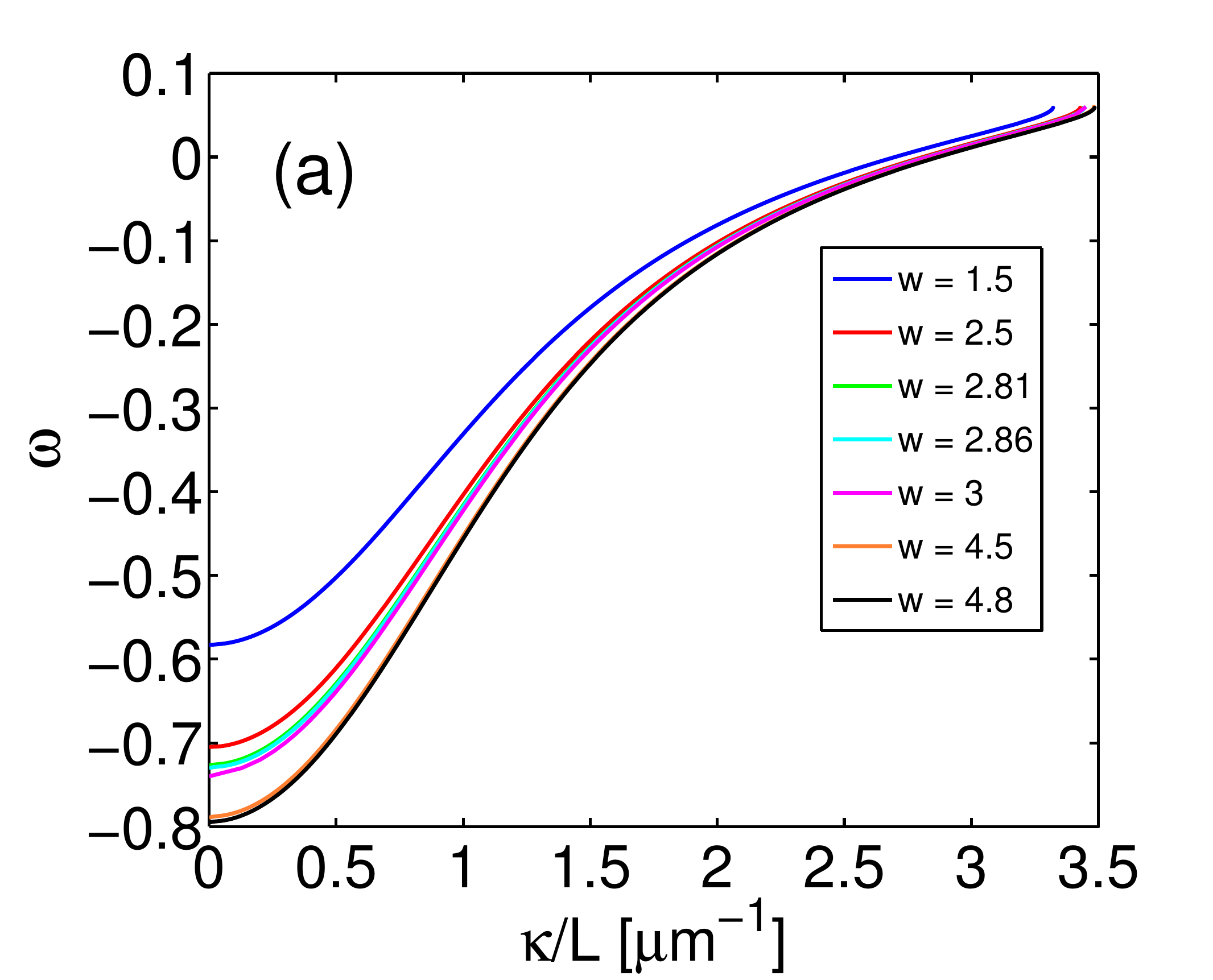}
\includegraphics[width=0.43\textwidth]{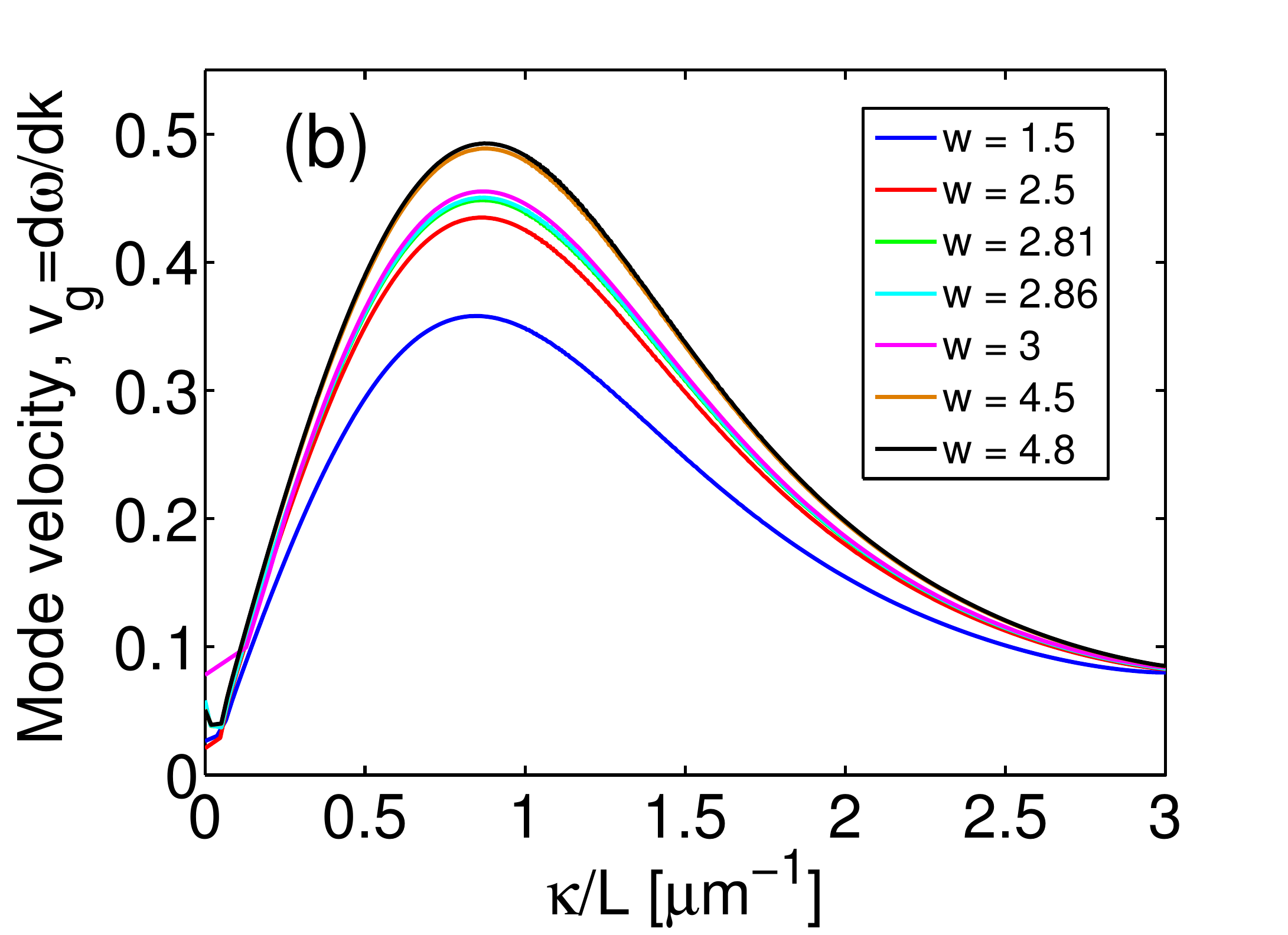}
\caption[Fig 10] {(Color online) (a) Linear fundamental polariton mode dispersion, $\omega(\kappa)$, for a range of microcavity wire widths. (b) Corresponding fundamental mode group velocity dispersion.}
\label{fig:mode_velocity_dispersion}
\end{figure*}

In this section, we investigate the dependence of the linear free polariton mode dispersion on the channel width. The linear energy dispersion, $\omega(\kappa)$, of the fundamental mode is shown in Fig.~\ref{fig:mode_velocity_dispersion} (a) for different microcavity widths. Similar to Fig.~\ref{fig:excitation_structure_scheme} (b), the dispersion becomes steeper with increasing wire width. In agreement with this trend, the group velocity (Fig.~\ref{fig:mode_velocity_dispersion} (b), being the slope of the energy dispersion curves at each point) monotonously increases with increasing wire width. The upper bound is the group velocity corresponding to a single-mode planar microcavity.

Let us compare first the soliton velocity with the one for non-interacting polaritons (i.e., absence of a pump and nonlinearity) at a channel width $w=3 \,\mathrm{\mu m}$. The group velocity of the fundamental free polariton mode is calculated as the slope of the fundamental mode linear dispersion curve at pump wave vector $\kappa/L=2.6862 \,\mathrm{\mu m^{-1}}$, corresponding to an angle of incidence $\theta=20^\circ$. This gives $v_g^{(0)}\approx 1.472 \,\mathrm{\mu m/ps}$. This value is very similar to the soliton velocity calculated by the 2D Newton-Raphson method for $E_p=0.075$ ($v_{g,sol}\approx 1.3 \,\mathrm{\mu m/ps}$), as well as to the one that can be inferred from the dynamical simulation through comparing consecutive time frames ($v_g^{(0)}\approx 1.1875\,\mathrm{\mu m/ps}$). The free polariton group velocity monotonically decreases with the mode number (see slopes for lower polariton branch in Fig.~\ref{fig:excitation_structure_scheme} (b)). At pump wave vector $\kappa/L=2.6862 \,\mathrm{\mu m^{-1}}$, the first-order mode group velocity is $v_g^{(1)}=0.96 \,\mathrm{\mu m/ps}$; for the second-order mode, $v_g^{(2)}=0.549 \,\mathrm{\mu m/ps}$, or nearly one-half of the soliton velocity. For the 3rd and 4th-order modes, it decreases even further: $v_g^{(3)}=0.313 \,\mathrm{\mu m/ps}$ and $v_g^{(4)}=0.227 \,\mathrm{\mu m/ps}$, respectively. Thus, we find that the soliton group velocity obtainable with this microcavity wire geometry (Fig.~\ref{fig:excitation_structure_scheme} (a)) is always lower than that of a soliton propagating in a single-mode, planar microcavity $(1.68\,\,\mathrm{\mu m/ps})$ \cite{Sich_NaturePhotonics}.

\subsection{Soliton branches at different channel widths}
\begin{figure*}
\centering
\includegraphics[width=0.43\textwidth]{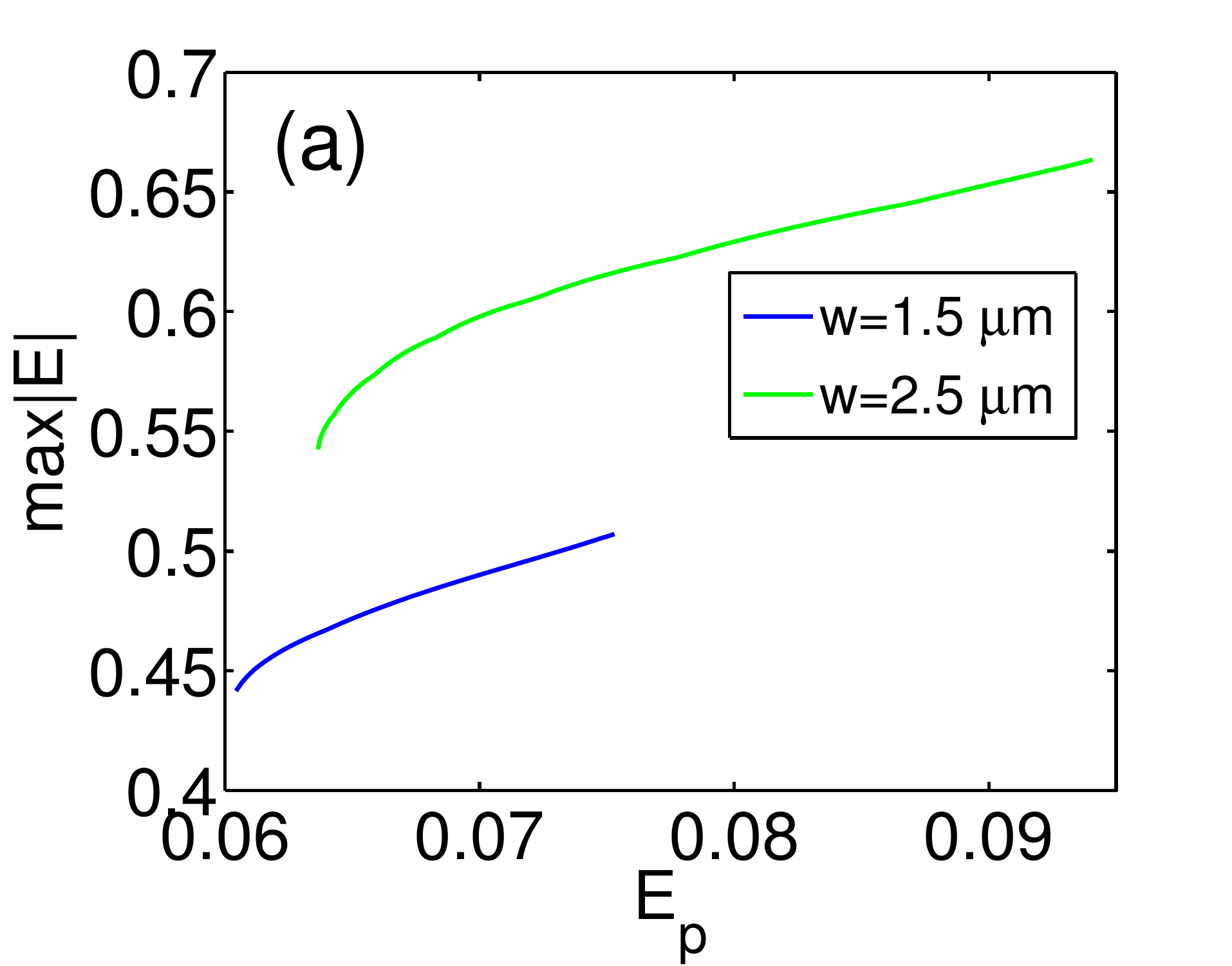}
\includegraphics[width=0.43\textwidth]{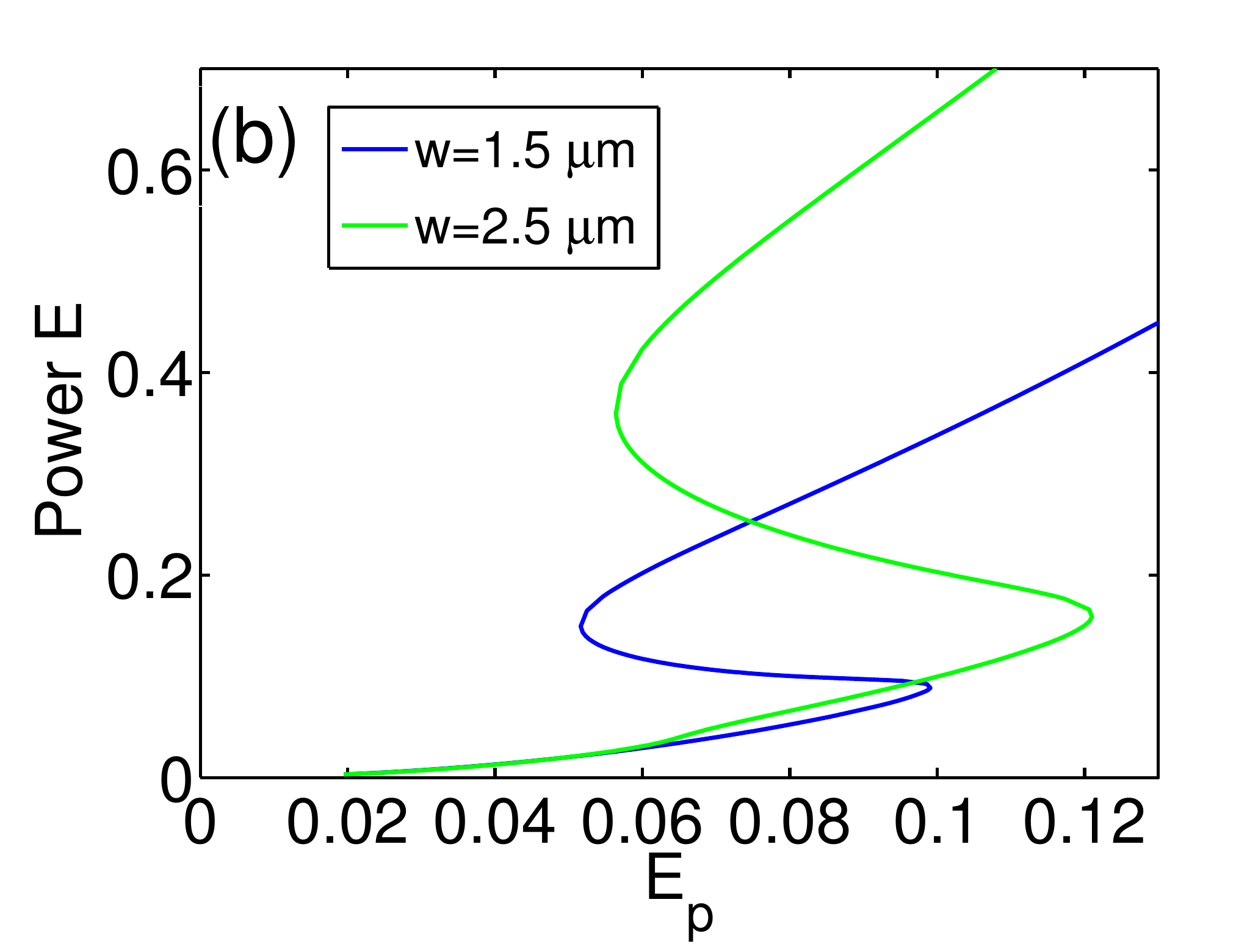}
\\
\includegraphics[width=0.43\textwidth]{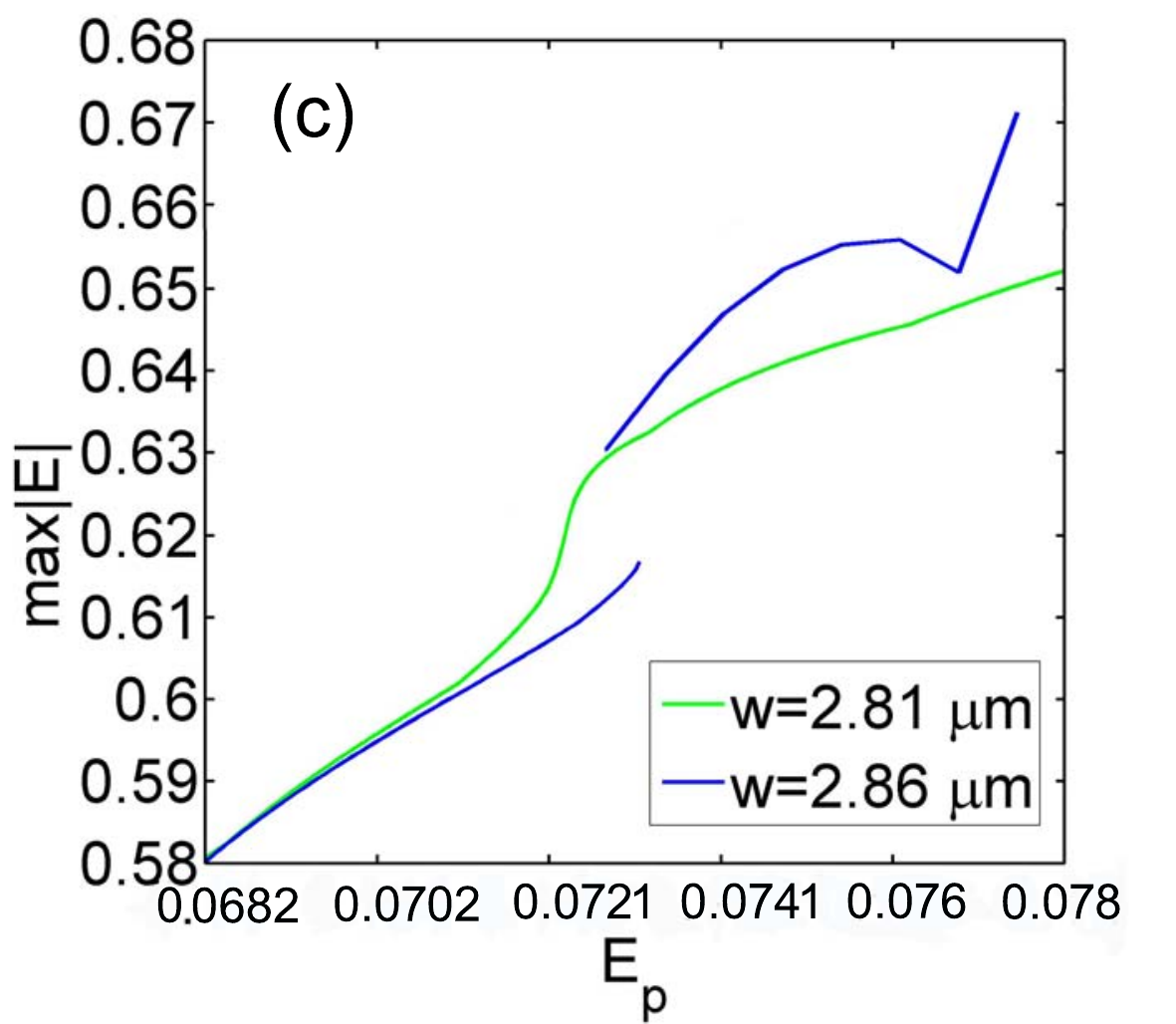}
\includegraphics[width=0.50\textwidth]{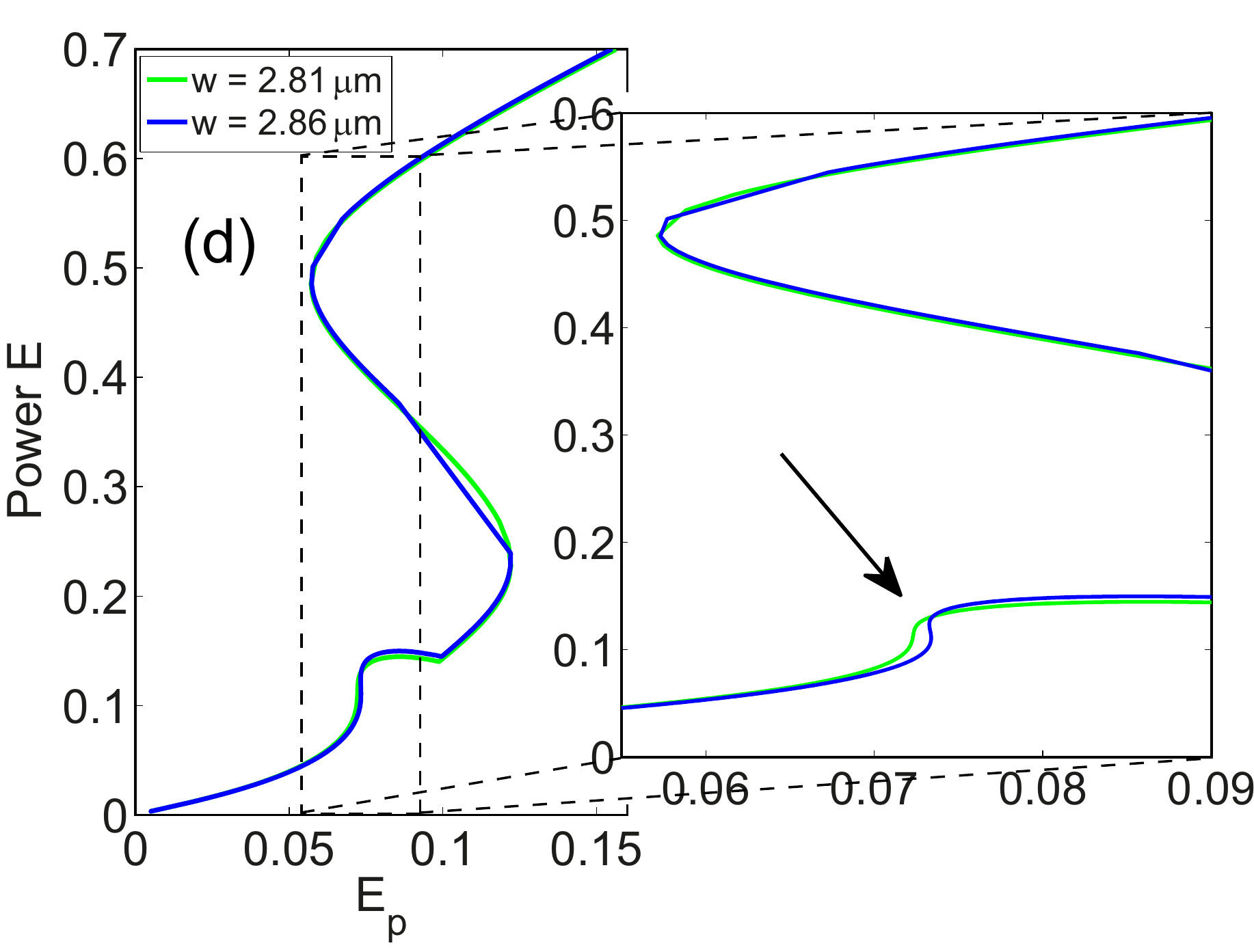}
\\
\includegraphics[width=0.43\textwidth]{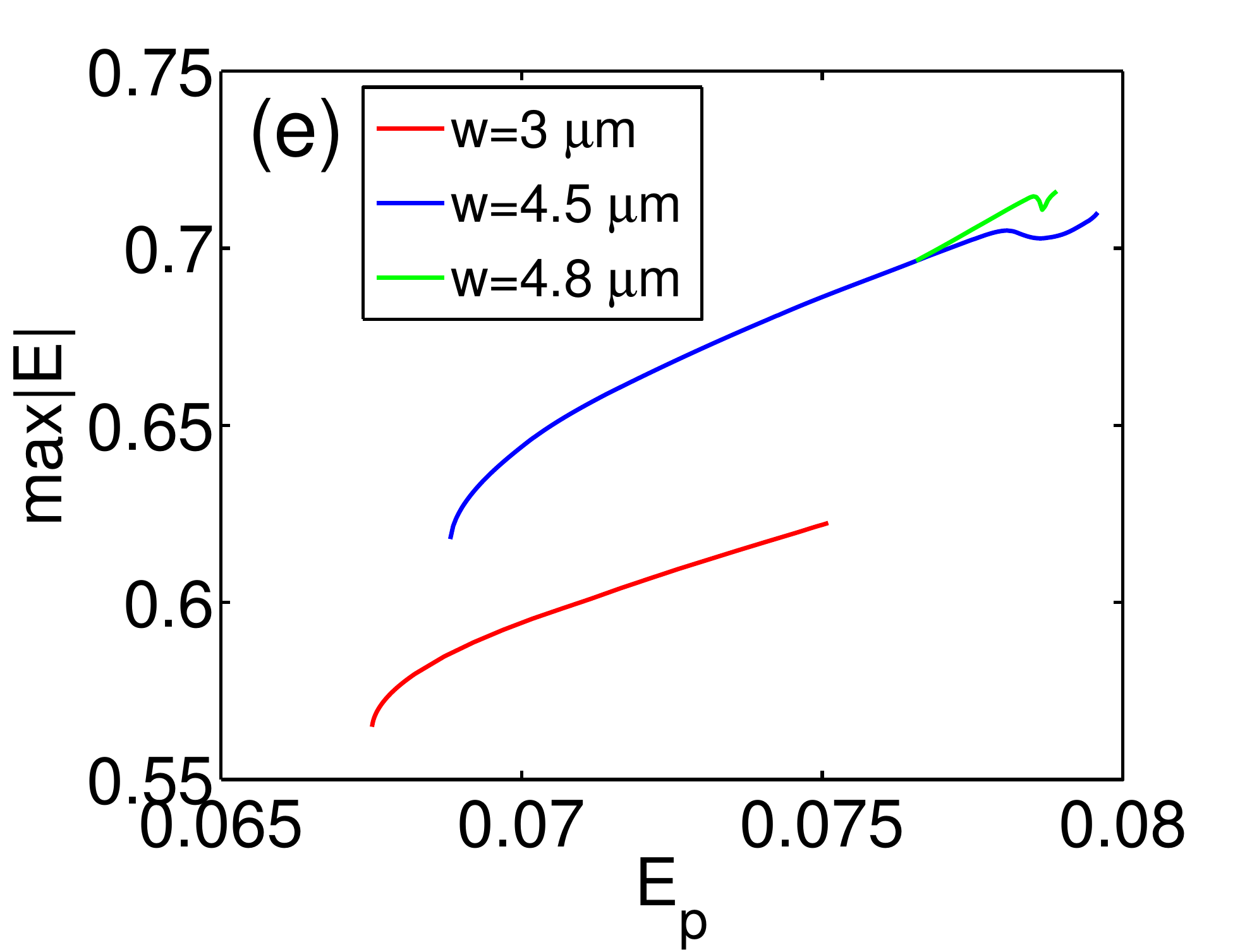}
\includegraphics[width=0.43\textwidth]{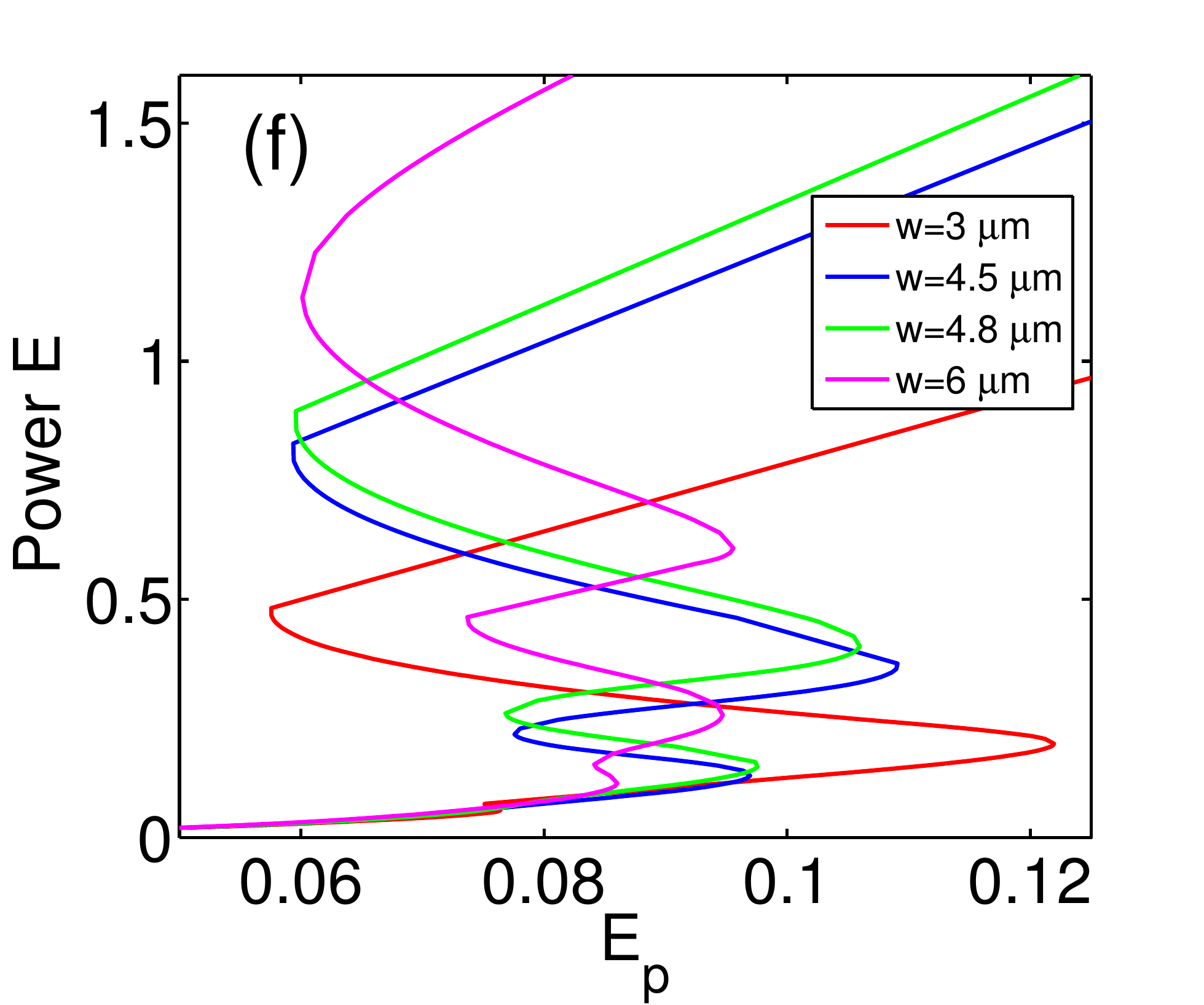}
\caption[Fig 11] {(Color online). (a) Soliton branches, plotted as the maximum of the modulus of the soliton's photon electric field ($E$) against the pump amplitude, $E_p$. This was computed by the 2D Newton-Raphson method. (b) Integrated $E$-field power bistability curves for channel widths $w=1.5$ and $2.5\,\mathrm{\mu m}$. (c) Soliton branches in the vicinity of a bifurcation point, where (d) the $E$-field power bistability curve gains another loop (arrow). Soliton existence over the widest range of pump amplitudes occurs on the interval $w \in [2.5, 2.81] \,\mathrm{\mu m}$. (e) Soliton branches and (f) $E$-field power multistability curves for $w=3, 4.5, 4.8$ and $6 \,\mathrm{\mu m}$. The multistability curve at $w=6 \,\mathrm{\mu m}$ exhibits $3$ loops and $3$ folding points; however, at this channel width no soliton was found using the 2D Newton-Raphson method. Pump detuning, $\Delta=0$ and cavity decay rate, $\gamma=0.04$ throughout this figure.}
\label{fig:soliton_branches_widths}
\end{figure*}

The soliton branches, computed by the 2D Newton method for different channel widths, are shown in Fig.~\ref{fig:soliton_branches_widths}. From inspection of Fig.~\ref{fig:soliton_branches_widths} (a), (c) and (e), as well as data we have not shown, the interval of soliton existence is largest for channel widths in the interval $[2.5,2.81] \,\mathrm{\mu m}$. As the width is increased just beyond this range, in the vicinity of $w=2.86\,\,\mathrm{\mu m}$ a bifurcation point appears, Fig.~\ref{fig:soliton_branches_widths} (c), and the singular soliton splits into two solitons. This point corresponds to the appearance of an additional branch -- or equivalently, a second loop (see arrow in Fig.~\ref{fig:soliton_branches_widths} (d)) -- in the bistability curve, hence becoming a multistability curve. Upon increasing the wire width beyond $w=4.8 \,\,\mathrm{\mu m}$, the multistability curve acquires yet another loop (Fig.~\ref{fig:soliton_branches_widths} (f), curve in magenta). However, note that we found no solitons for $w=6 \,\mathrm{\mu m}$ (the 2D Newton-Raphson method did not converge).

In general, we posit that the occurrence of a bifurcation heralds the appearance of an additional stable light state. We will argue this in the ensuing passage.

\begin{figure*}
\centering
\vspace{10pt}
\includegraphics[width=0.43\textwidth]{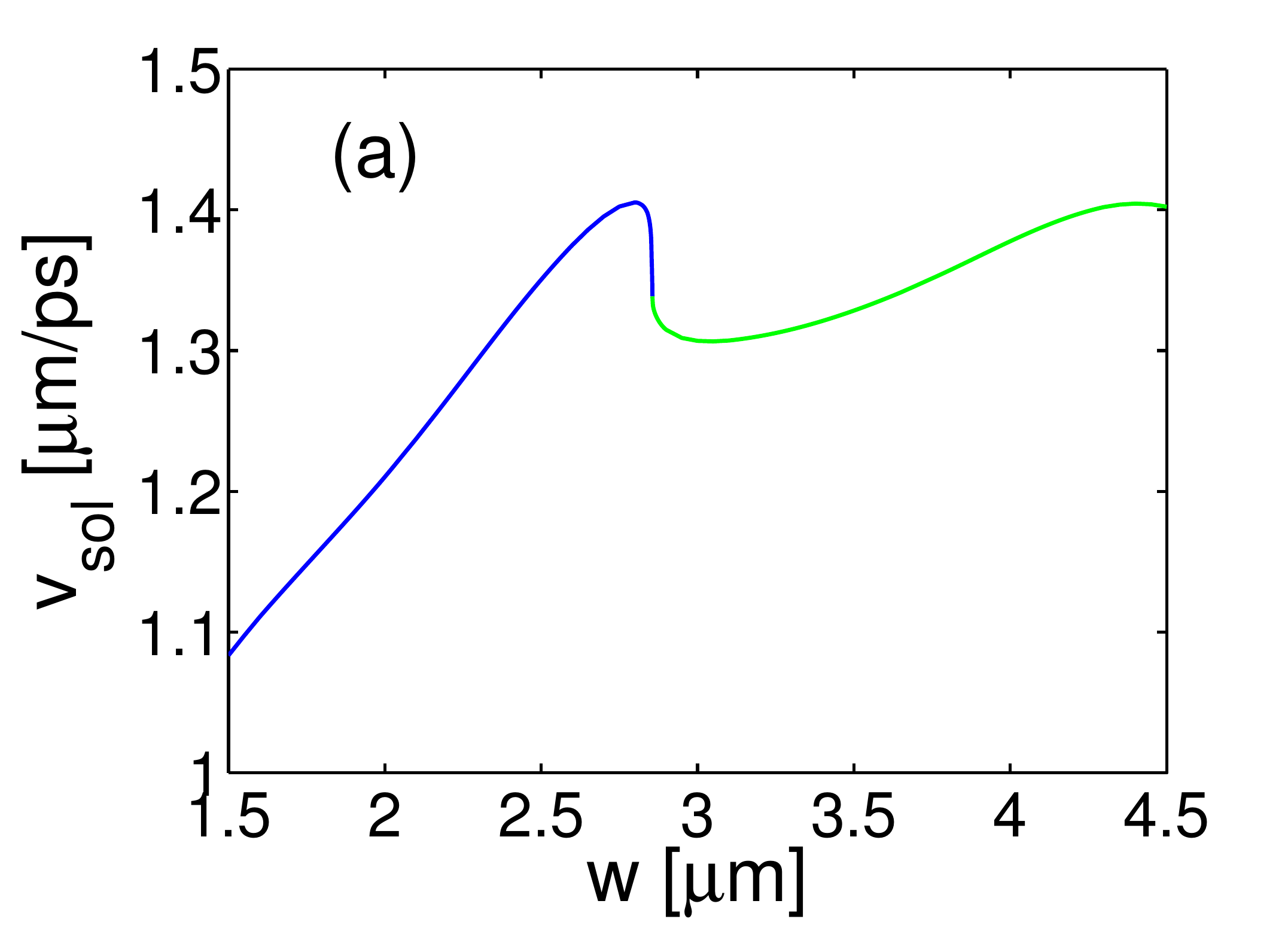}
\includegraphics[width=0.43\textwidth]{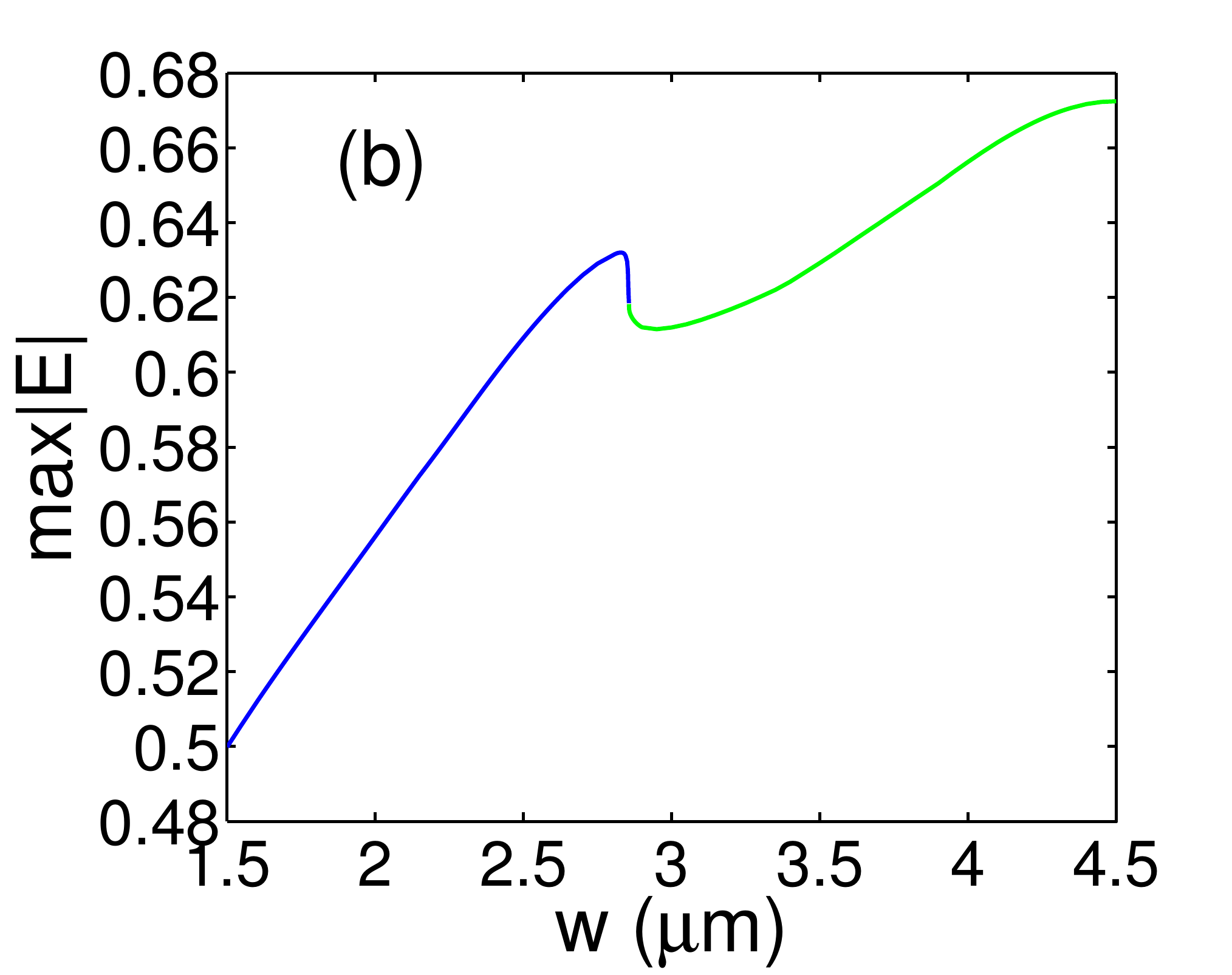}
\caption[Fig 12] {(Color online) (a) Soliton group velocity and (b) maximum soliton amplitude as a function of wire width, for $\Delta=0, \gamma=0.04$. These results were computed using the 2D Newton-Raphson method. The point where the colour of the curve changes ($w = 2.86 \,\,\mathrm{\mu m}$) corresponds to a bifurcation point in the soliton branch.}
\label{fig:soliton_max_velocity_w}
\end{figure*}

The group velocity and the maximum amplitude of the soliton as a function of wire width are shown in Fig.~\ref{fig:soliton_max_velocity_w}. Our model predicts a non-monotonous behaviour of both quantities. From Fig.~\ref{fig:soliton_max_velocity_w} (a), it is clear that the slowest soliton is obtained for the narrowest wire; however, we observe further local maxima and minima in the curve. Here we will attempt to give a qualitative explanation of the non-monotonous behaviour we have predicted.

Starting with $w=1.5\,\mathrm{\mu m}$, as the width is increased, the soliton group velocity increases, reaching a maximum at $w=2.81\,\mathrm{\mu m}$. At this point, we speculate that a second-order mode appears, whose group velocity is necessarily much lower than that of the fundamental mode (nearly a $2/3$ reduction; see end of Sec. \ref{sssec:3.6}). This results in a decrease of the overall soliton group velocity. The soliton can then be viewed as a composite multi-mode entity, consisting of a fundamental and second-order modes. Further increase of the wire width leads to the bifurcation point at $w=2.86 \,\,\mathrm{\mu m}$ in Fig.~\ref{fig:soliton_branches_widths} (c), where we suppose the fourth-order mode appears. With a group velocity of only approximately $1/5$th of the fundamental mode group velocity, composite soliton group velocity is decreased even further, reaching a minimum at $w=3 \,\,\mathrm{\mu m}$. As the wire width is increased beyond this point, the soliton group velocity increases again, as its constituent mode group velocities increase with wire width. This continues until another maximum is reached at around $w=4.4 \,\,\mathrm{\mu m}$. Beyond this point, we expect that a sixth-order mode appears, and the soliton velocity drops again. Following this line of reasoning, we would expect to encounter another bifurcation just below $w=6 \,\,\mathrm{\mu m}$, where a third loop appears in the multistability curve (see curve in magenta in Fig.~\ref{fig:soliton_branches_widths} (f)). We expect that the soliton group velocity drops to zero at $w=6 \,\,\mathrm{\mu m}$, since we were not able to find a soliton for this wire width.

The dependence of the maximum soliton amplitude on the wire width is shown in Fig.~\ref{fig:soliton_max_velocity_w} (b). It is very similar to the relationship between the soliton group velocity and wire width, and hints at a possible link between the two. Although at this stage we cannot comment on a plausible origin of this, it would be interesting to investigate the nature of this correlation in the future. In order to explain quantitatively this peculiar non-monotonous behaviour, we have recently developed a coupled-mode expansion method \cite{PRB2016}. A detailed analysis using this method will be presented in a forthcoming paper.

\subsection{Soliton propagation in tilted microcavity wires}
\begin{figure*}
\centering
\vspace{10pt}
\includegraphics[width=0.43\textwidth]{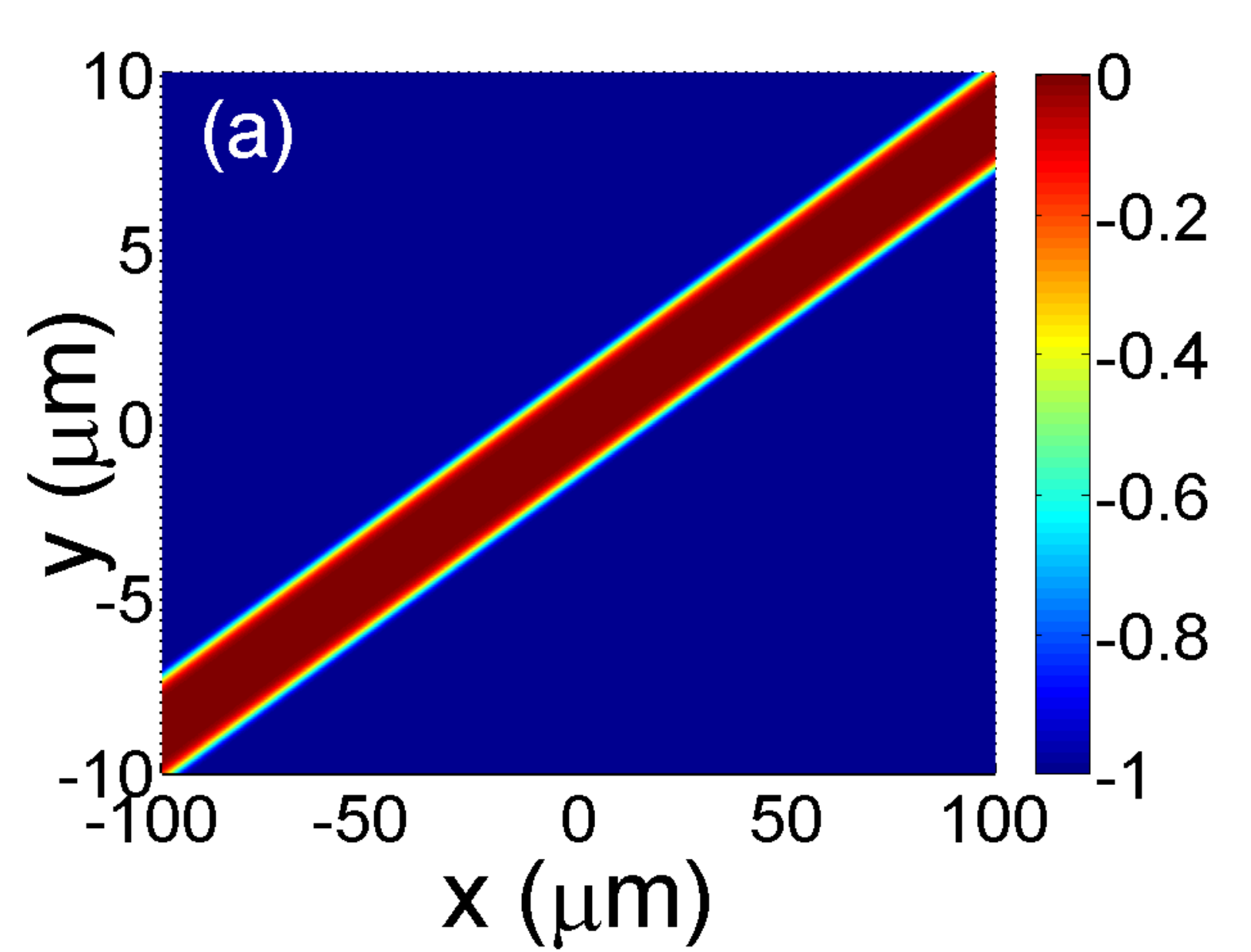}
\includegraphics[width=0.43\textwidth]{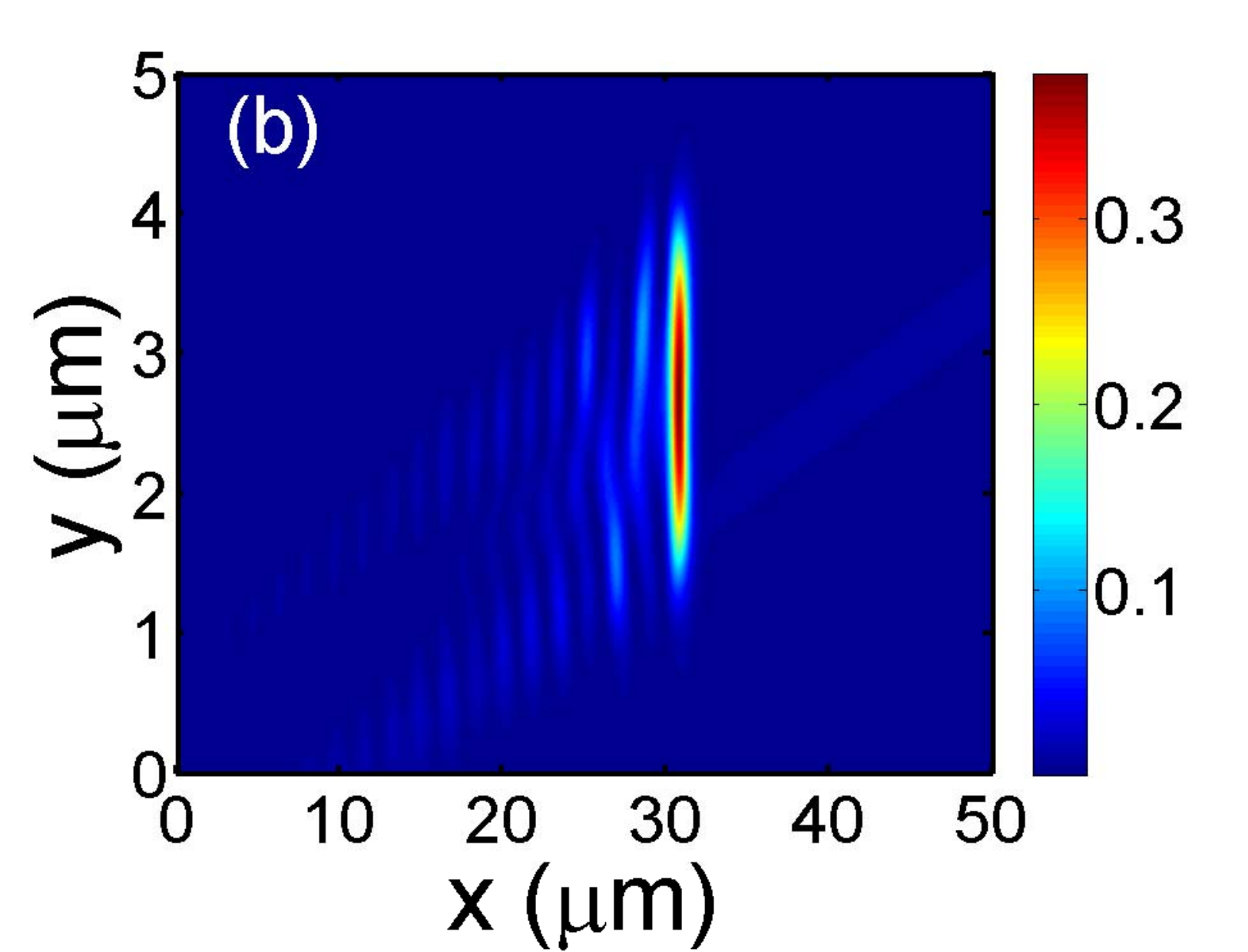}
\caption[Fig 13] {(Color online) (a) Top view of the 3D potential, $U(y)$ (see Eq.~(\ref{eqPot})), of a microcavity wire tilted at an angle $\alpha=5^{\circ}$ with respect to the $x$ axis. (b) Expanded top view of a single frame at $t=24 \,\mathrm{ps}$ of a soliton propagating in the tilted microcavity wire. The squared modulus of the electric field of the photon constituent of the polariton soliton, $\vert E \vert^2$, is plotted in colour. Pump amplitude, $E_p=0.075$; seed amplitude, $E_s=0.5$.}
\label{fig:tilted_waveguide}
\end{figure*}

We have performed a number of dynamical simulations with wires oriented at different tilt angles with respect to the $x$ axis, for $\Delta=0$ and $\gamma=0.04$. In Fig.~\ref{fig:tilted_waveguide} (a), a microcavity wire tilted at an angle $\alpha=5^\circ$ is shown for $w=3 \,\mathrm{\mu m}$, $\theta=20^\circ$ and $E_p=0.075$. We trigger the polariton soliton formation using a seed pulse, as described in the beginning of Sec. \ref{sssec:3.4}. Following brief initial reshaping, the soliton travels undistorted along the tilted microcavity wire (Fig.~\ref{fig:tilted_waveguide} (b) and video in\cite{Supplementary_material}). For these parameters, we found that the soliton is preserved in amplitude up to tilt angles of approximately $6^{\circ}$. Larger tilt angles, however, lead to a gradual decline in amplitude and eventual abrupt destruction of the soliton.

\begin{figure*}
\centering
\vspace{10pt}
\includegraphics[width=0.43\textwidth]{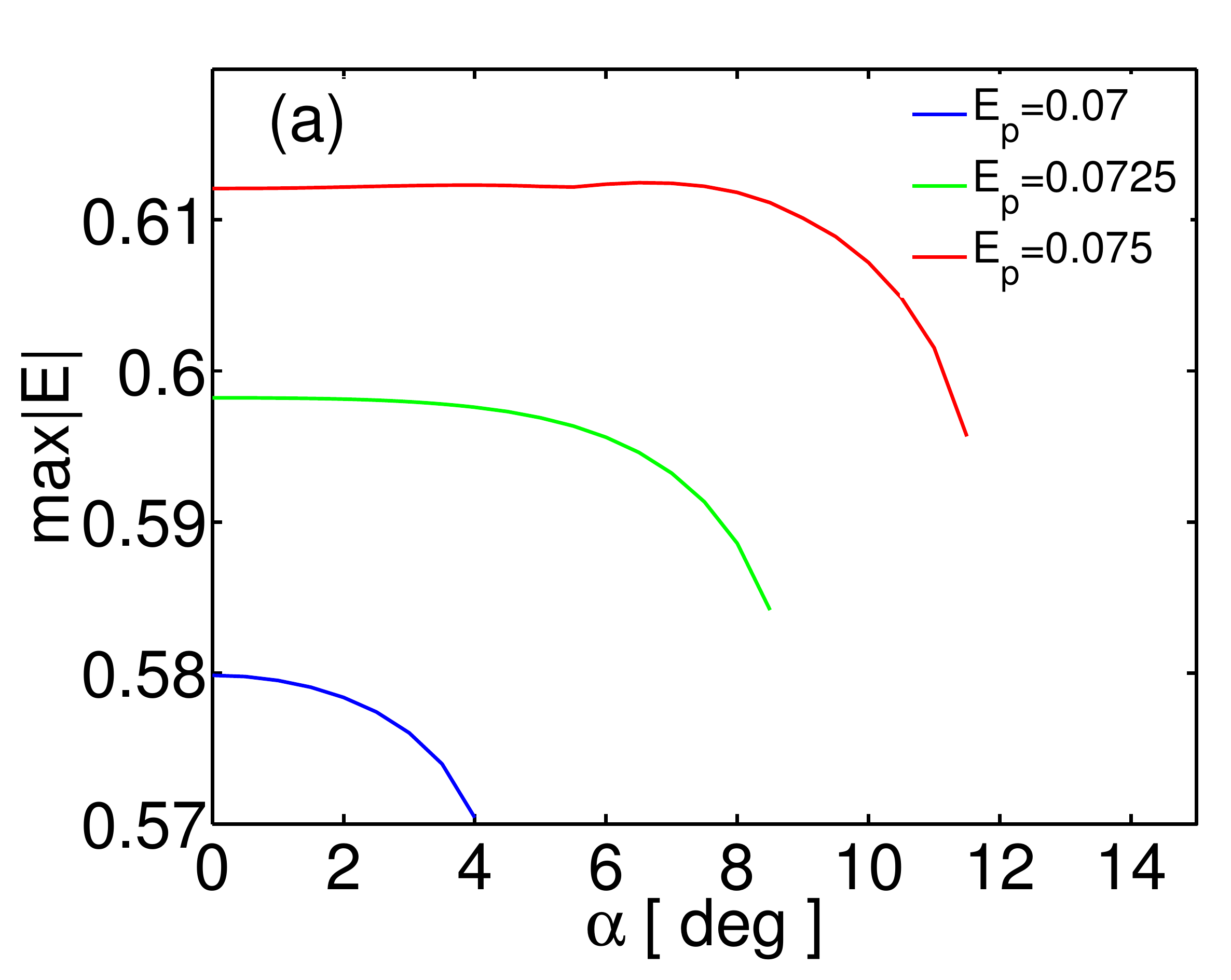}
\includegraphics[width=0.46\textwidth]{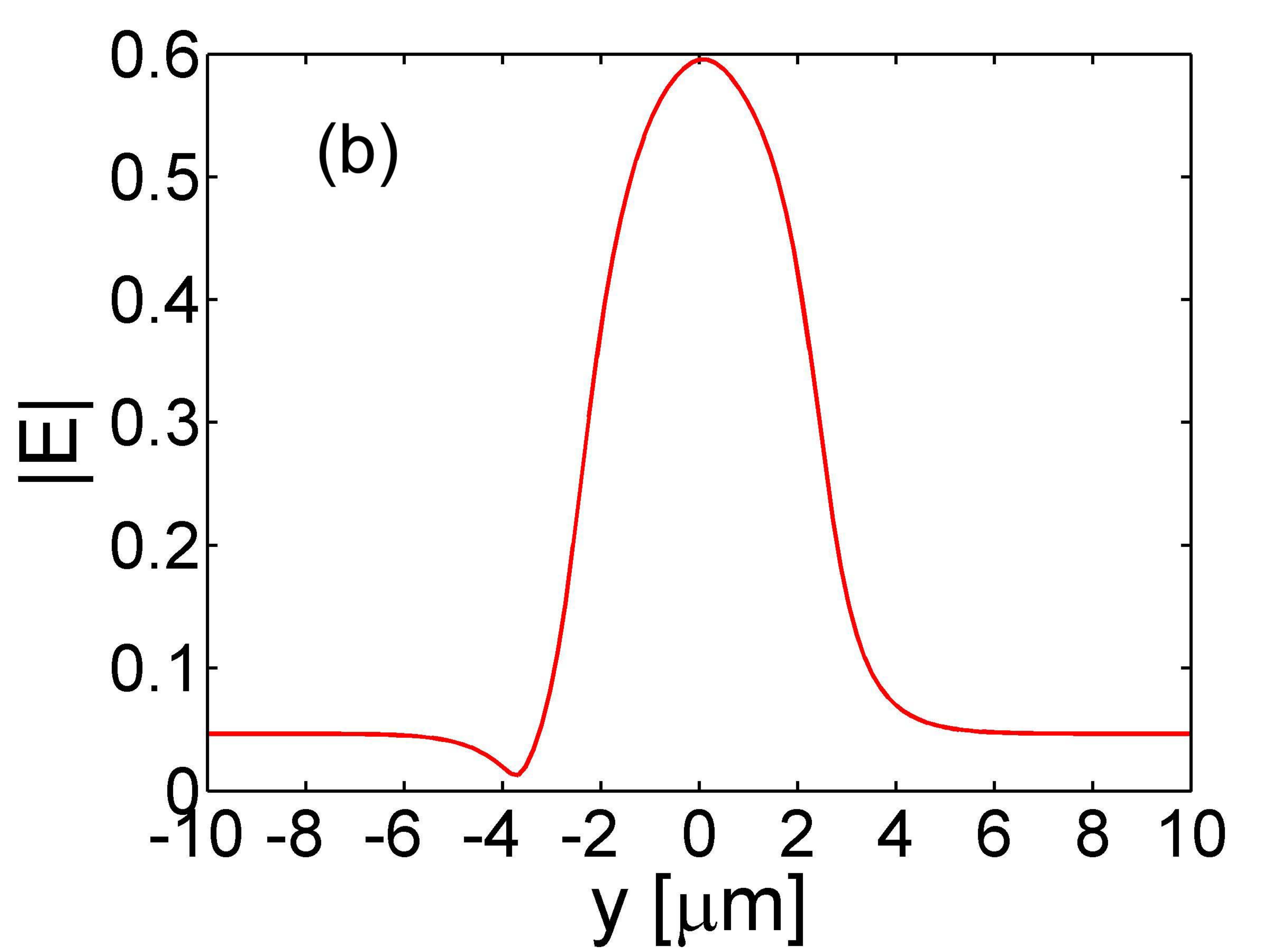}
\caption[Fig 14] {(Color online) (a) Maximum of the modulus of the electric field ($E$) of the photon part of the soliton, as a function of the microcavity wire's tilt angle, $\alpha$. This was calculated using the 2D Newton-Raphson method for three different pump amplitudes. For the case of $E_p=0.075$, the soliton persists with a relatively constant amplitude up to $\alpha=8 ^\circ$. (b) Transverse cross-section through the soliton at $E_p=0.075$ and $\alpha=8 ^\circ$. Note that the soliton profile is asymmetric.}
\label{fig:2DNewton}
\end{figure*}

Our dynamical results above were checked through the 2D Newton-Raphson method. Using the latter, the dependence of the soliton $E$-field maximum on the microcavity wire tilt angle is plotted for different pump amplitudes in Fig.~\ref{fig:2DNewton} (a). The dependence is nonlinear and exhibits typical critical behaviour; the soliton disappears at different critical angles for different $E_p$. The critical angle monotonically increases with increasing $E_p$, reaching a maximum of $\alpha_{max} \approx 12 ^\circ$ at $E_p=0.075$. The soliton amplitude remains virtually unchanged up to $8 ^\circ$. In fact, the soliton can persist up to almost $12^\circ$ with only a small loss in amplitude (about $2.5\%$). A transverse cross-section through the soliton maximum, calculated by the 2D Newton-Raphson method for $\alpha=8 ^\circ$ and $E_p=0.075$, is shown in Fig.~\ref{fig:2DNewton} (b). The soliton exhibits a pronounced asymmetric shape.

The maximum critical angle obtained by the 2D Newton-Raphson method differs by approximately $2$ degrees from the one calculated by the time-dependent split-step method. The two methods are therefore in very good agreement with each other, and give a useful indication of the microcavity wire tilt limitations in designing soliton polaritonic splitters, routers and other functional components of future polaritonic integrated circuits.

\subsection{Soliton propagation in tapered microcavity wires}
\label{Tapered_waveguide}
From Fig.~\ref{fig:soliton_max_velocity_w}, it follows that the soliton amplitude increases with increasing microcavity wire width in the intervals $w \in [1.5, 2.81]$ and $[2.86, 4.5] \,\mathrm{\mu m}$. This effective soliton amplification effect can be exploited in the design of special, tapered microcavity wires, which could be used for a number of applications.

\begin{figure*}
\centering
\vspace{10pt}
\includegraphics[width=0.43\textwidth]{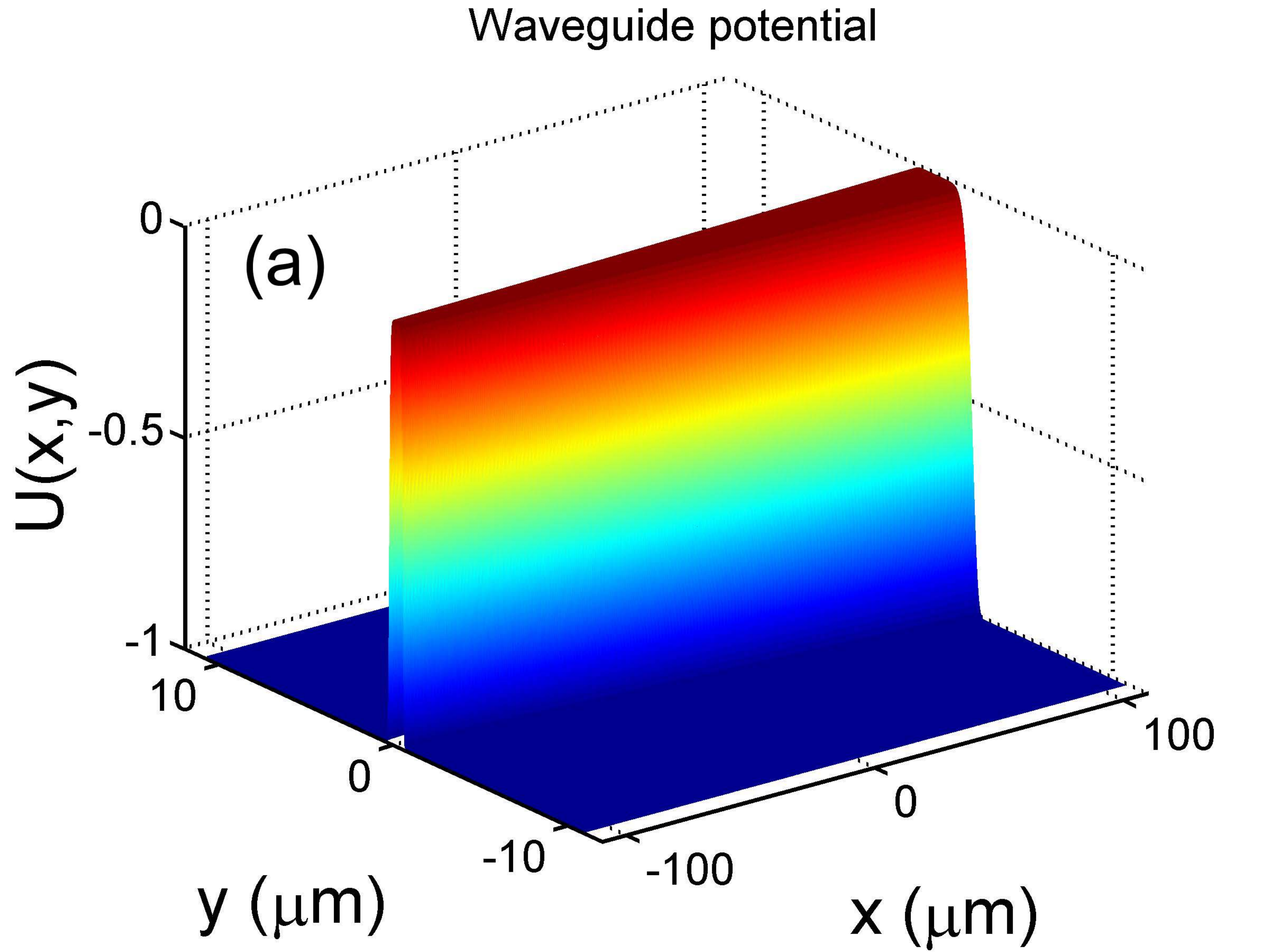}
\includegraphics[width=0.43\textwidth]{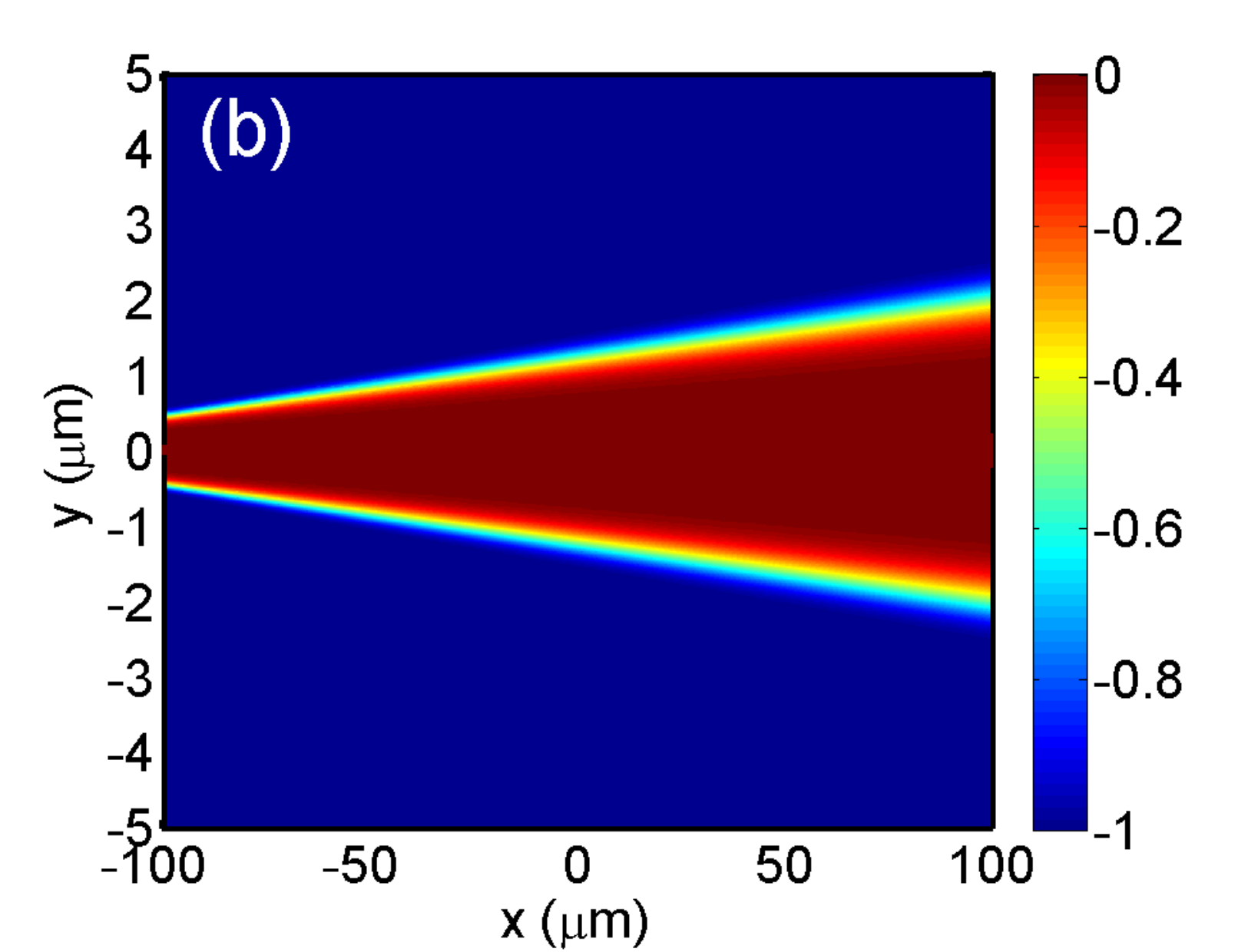}
\includegraphics[width=0.43\textwidth]{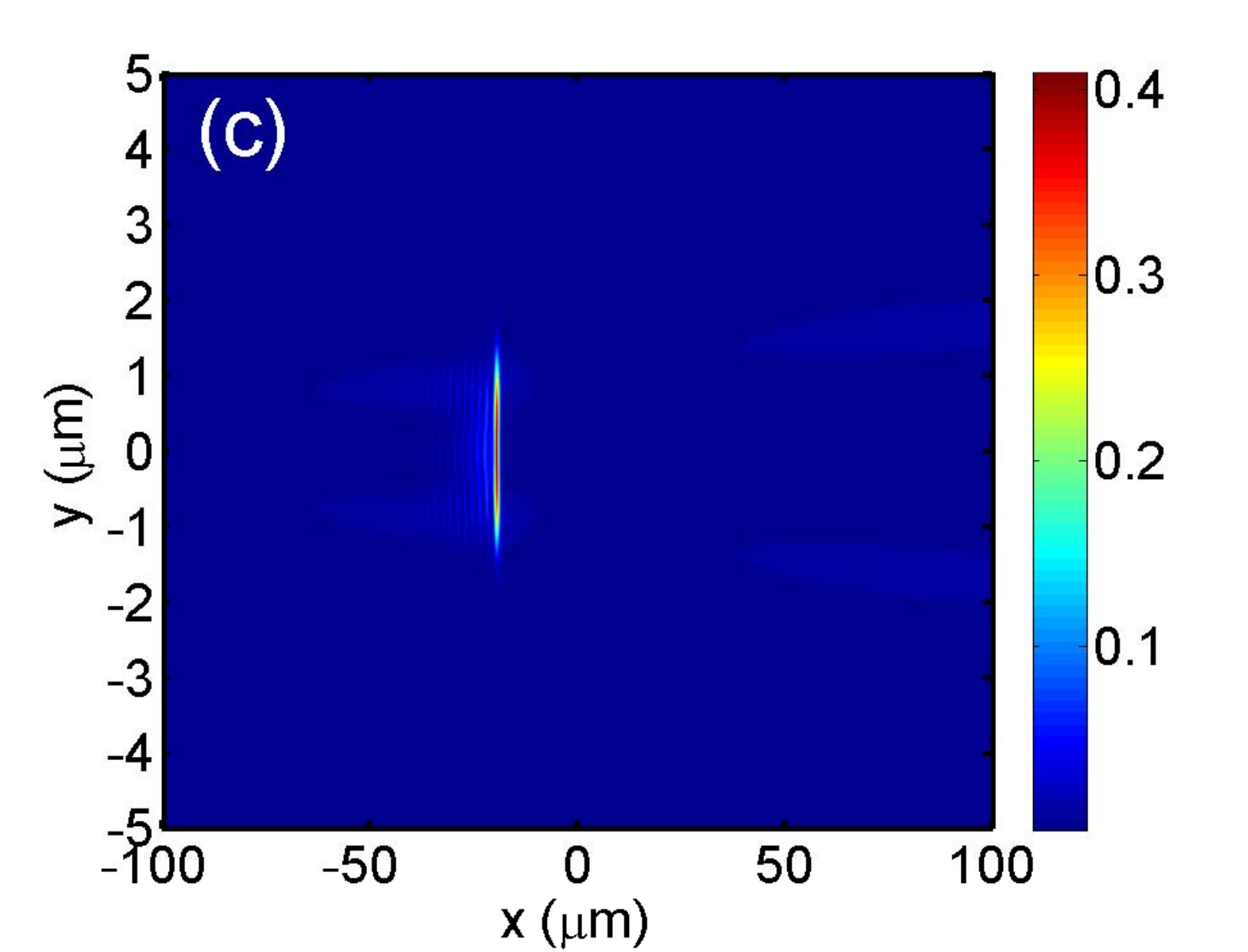}
\includegraphics[width=0.43\textwidth]{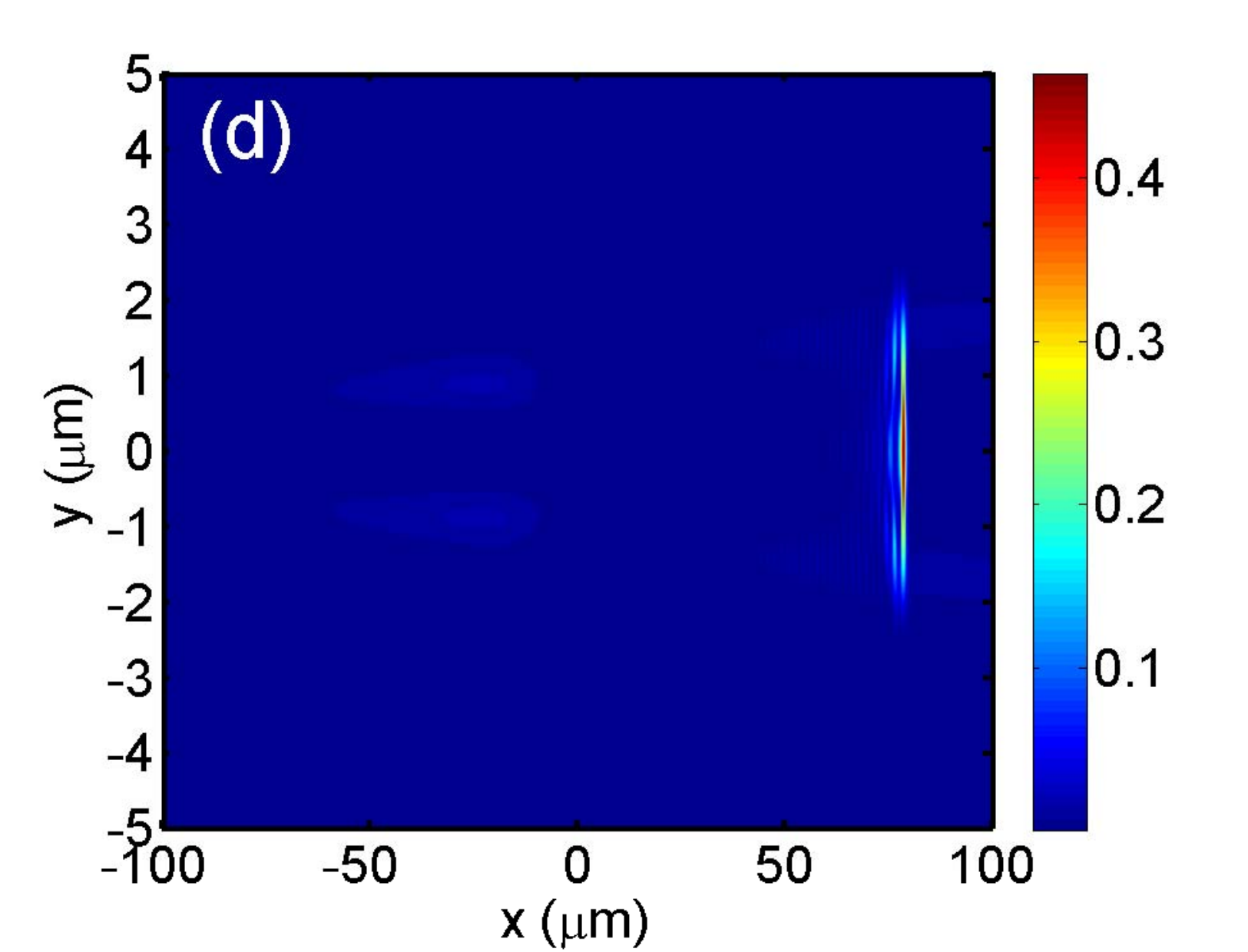}
\includegraphics[width=0.43\textwidth]{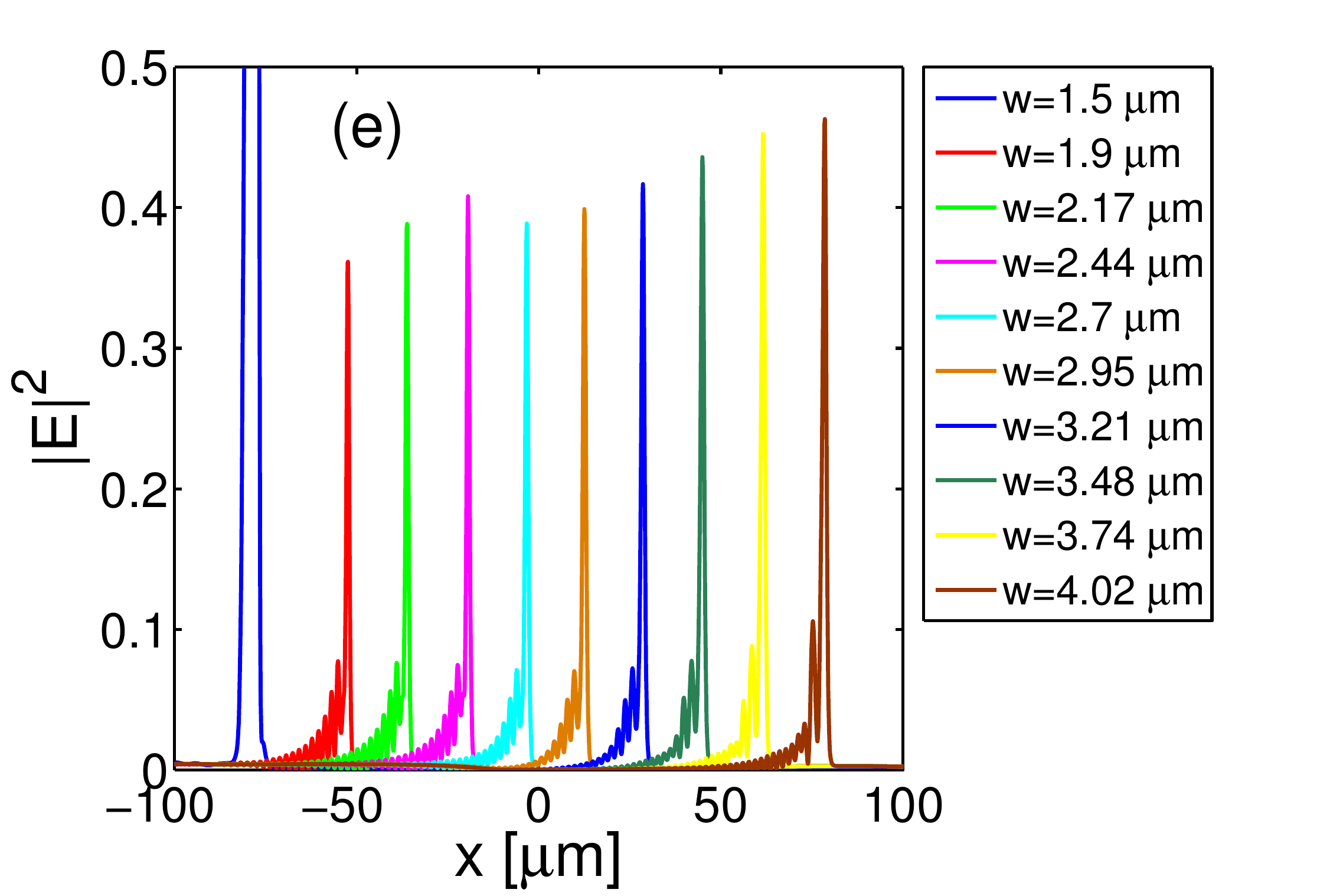}
\caption[Fig 15] {(Color online) (a) 3D confinement potential of a tapered microcavity wire with input facet width, $w=1.00\,\mathrm{\mu m}$, and output facet width, $w=4.02\,\mathrm{\mu m}$. (b) Top view of the potential. (c) and (d): time evolution of a soliton propagating in the tapered microcavity. Modulus squared of the electric field of the photon part of the soliton, $\vert E\vert^2$, at (c) $t=48 \,\mathrm{ps}$ and (d) $t=120 \,\mathrm{ps}$, showing that, as it travels, the soliton broadens in the $y$ direction but remains focused in the $x$. (e) Profiles of the polariton soliton at different moments in time: $t=2.4, 24, 36, 48, 60, 72, 84, 96, 108$ and $120\,\,\mathrm{ps}$, corresponding to tapered microcavity wire widths: $w=1.5, 1.9, 2.17, 2.44. 2.7, 2.95, 3.21, 3.48, 3.74$ and $4.02 \,\mathrm{\mu m}$. The soliton amplitude adiabatically increases, peaks at around the bifurcation point of the soliton branch (see Fig.~\ref{fig:soliton_branches_widths} (c)), then subsequently decreases, reaching a minimum, after which it increases again. The maximum amplitude of the soliton is also plotted in Fig.~\ref{fig:soliton_max_velocity_w} (b).}
\label{fig:tapered_waveguide}
\end{figure*}

In Fig.~\ref{fig:tapered_waveguide} (a) and (b), the confinement potential of a tapered wire with initial and final widths $w=1.00$ and $4.02\,\mathrm{\mu m}$, respectively, is shown. The Rabi frequency potential is constructed in the same way as in Eq. (\ref{eqPot2}), whereby the coupling is $1$ within the channel and zero outside (data not shown). A soliton is launched from the left boundary at time $t=0$. The time evolution of the photon component of the soliton power, $\vert E \vert^2$, is shown in Fig.~\ref{fig:tapered_waveguide} (c) and (d) for two simulation times: $t=48$ and $120 \,\,\mathrm{ps}$, near the beginning and end of the wire, respectively. Snapshots of the evolution of the soliton's longitudinal profile over time are shown in Fig.~\ref{fig:tapered_waveguide} (e). From Fig.~\ref{fig:tapered_waveguide} (c--e) it is clear that, as the soliton traverses the structure, it expands in the $y$ direction, but holds together along $x$ and does not dissipate. A movie of the soliton's travel through the structure is also provided in \cite{Supplementary_material}. In Fig.~\ref{fig:tapered_waveguide} (e), the first (leftmost) profile shown is at a point near the left boundary, where the soliton is being excited by the seed pulse. After some initial reshaping, a soliton is formed. Due to the smooth change in the width of the wire, the soliton power increases adiabatically, reaching a local maximum at $w\approx2.44 \mathrm{\mu m}$. Subsequently, the power drops down sharply, reaching a minimum at $w\approx2.7 \mathrm{\mu m}$, whereupon it gradually increases again. As we shall show below, this non-monotonic behaviour is in very good quantitative agreement with the calculated by the 2D Newton-Raphson method dependence of the soliton amplitude on wire width (Fig.~\ref{fig:soliton_max_velocity_w} (b)).

\begin{figure*}
\centering
\vspace{10pt}
\includegraphics[width=0.43\textwidth]{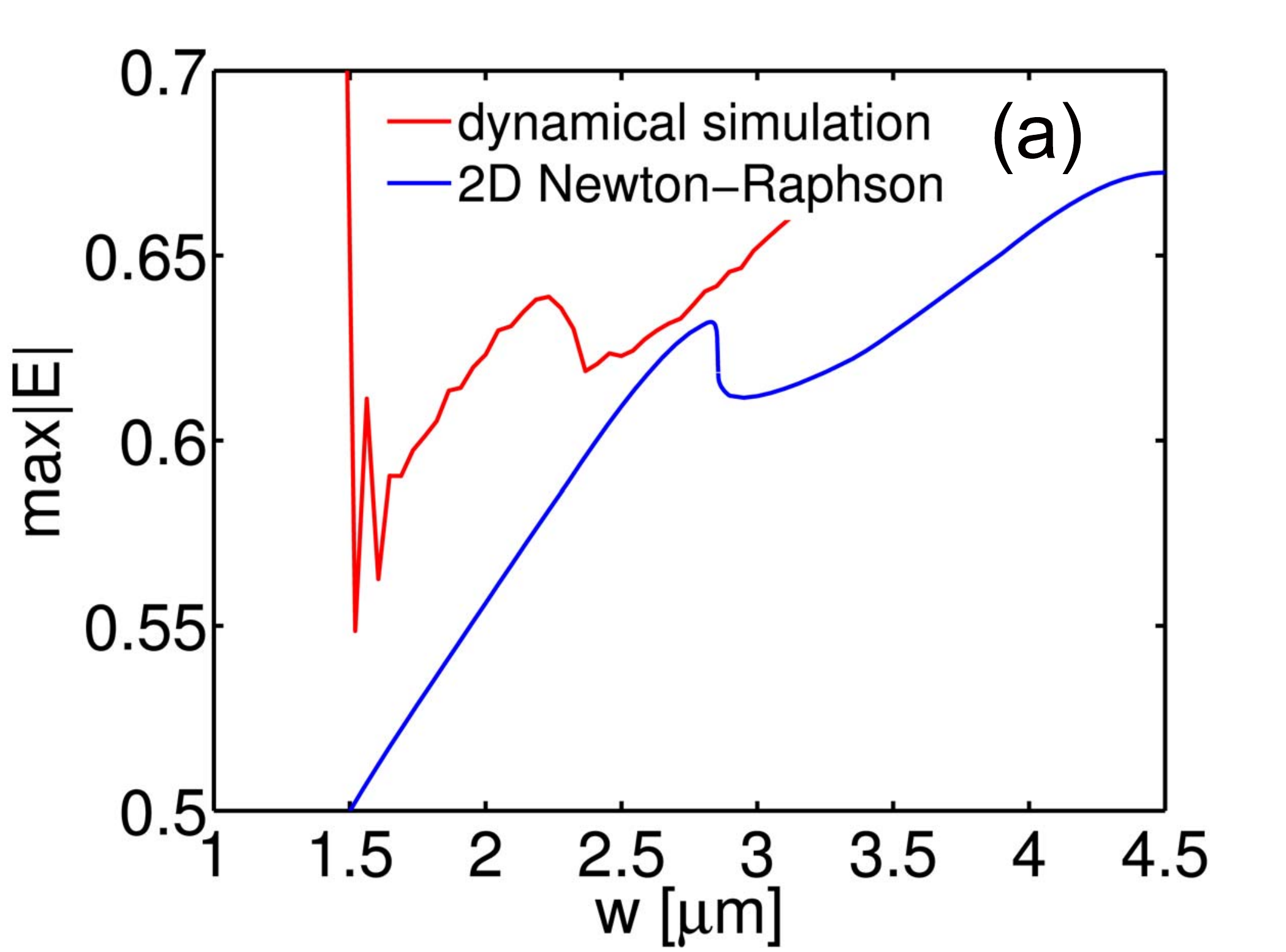}
\includegraphics[width=0.43\textwidth]{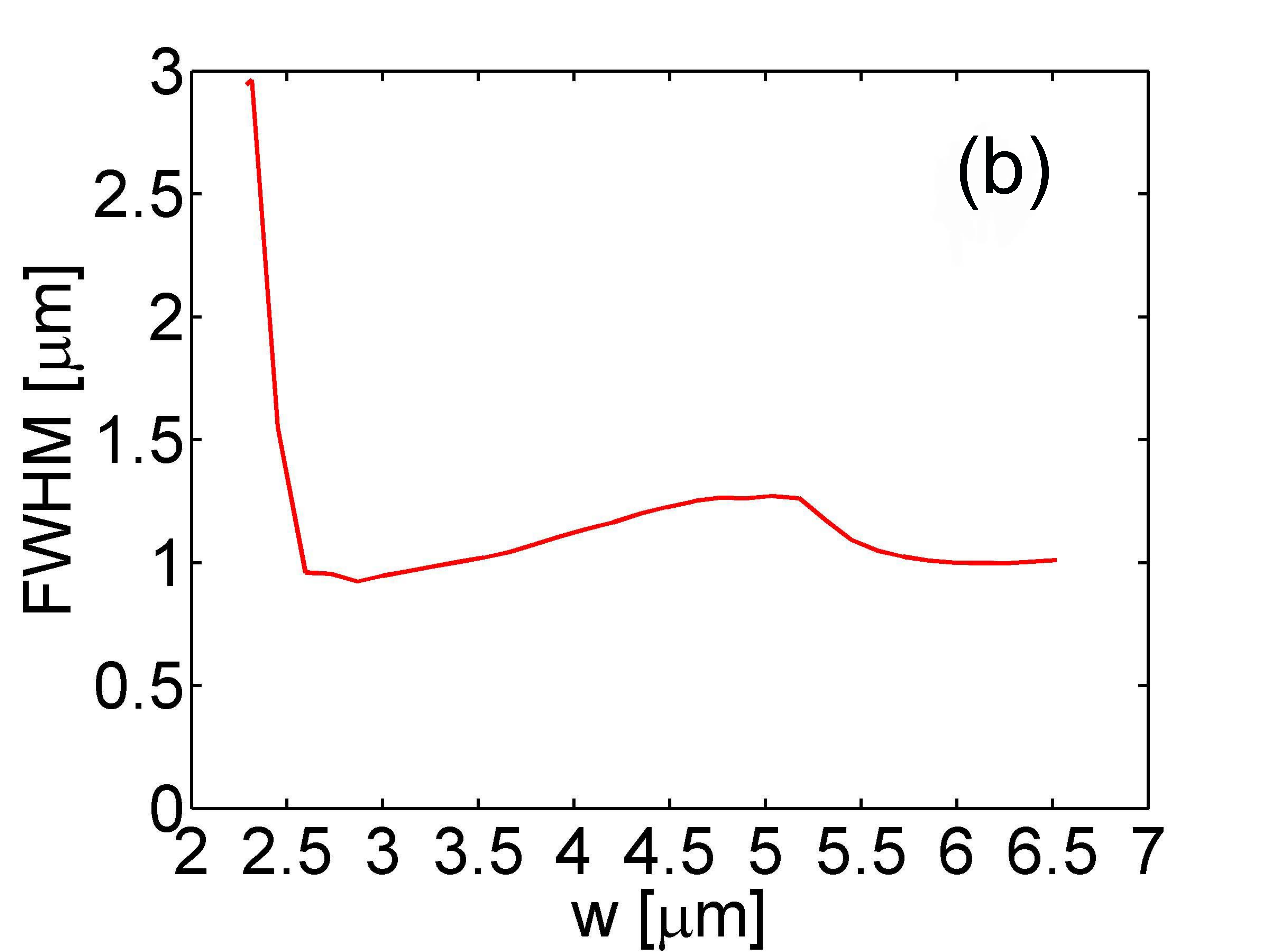}
\caption[Fig 16] {(Color online) (a) Dynamical simulation of the soliton excitation and propagation in a tapered microcavity wire (red curve), along with the dependence of the maximum soliton amplitude on the microcavity wire width, $w$, from Fig.~\ref{fig:2DNewton} (b), calculated with the 2D Newton-Raphson method (blue curve). The pump amplitude is $E_p=0.075$. Save for the initial soliton reshaping (first few points on the red curve), the dynamic simulation yields the same type of non-monotonic dependence on $w$ as the 2D Newton-Raphson one. The shift to the left of the red curve along $w$ is due to the excitation of a soliton at a point to the left of the first calculated point by the 2D Newton-Raphson method (at $w=1.5 \,\,\mathrm{\mu m}$). (b) Longitudinal FWHM of the soliton peak as a function of wire width for the same tapered wire as in (a). After formation, the soliton FWHM is not constant, but deviates by no more than $30 \%$.}
\label{fig:max_power_x_max_power_w}
\end{figure*}

We dynamically simulated the excitation and propagation of a soliton in a tapered wire with initial and final widths, $w=1.00$ and $w=4.02\,\mathrm{\mu m}$, respectively, at $E_p=0.075$. The soliton's maximum amplitude at each point along the wire is displayed in Fig.~\ref{fig:max_power_x_max_power_w} (a). The dependence of the maximum soliton amplitude on wire width, as computed by 2D Newton-Raphson method (see Fig.~\ref{fig:soliton_max_velocity_w} (b)) is plotted on the same graph for comparison. Following initial reshaping, the soliton reaches a maximum at a lower wire width than that predicted by the 2D Newton-Raphson method ($2.81 \,\mathrm{\mu m}$). In spite of this, the overall trend of the two curves is very similar, and the maximum soliton amplitudes computed by the two methods are in a very good agreement. The lateral shift, in fact, is due to the soliton being excited dynamically at a point along the wire, $w < 1.5 \,\,\mathrm{\mu m}$, preceding the first point calculated by the 2D Newton-Raphson method. Hence, the maximum is reached earlier, and the curve is translated to the left.

For $E_p=0.075$, there is no soliton at $w=4.8 \,\,\mathrm{\mu m}$ (Fig.~\ref{fig:soliton_branches_widths} (e), green curve). This is why we have plotted the dependence of the maximum soliton amplitude on wire width in Fig.~\ref{fig:soliton_max_velocity_w} (b) only up to $w=4.5 \,\,\mathrm{\mu m}$.

The soliton FWHM for the same tapered microcavity wire is plotted in Fig.~\ref{fig:max_power_x_max_power_w} (b). After the transient soliton formation process, the FWHM of the soliton peak taken in the $x$ direction is not constant, but changes by no more than $30 \%$ by the time it reaches the end facet of the wire. The driving force behind this FWHM variation is currently unclear; however, this will be a subject of further investigation using our newly developed coupled-mode expansion technique \cite{PRB2016}.

\section{Conclusion}
We have developed a model of the polariton nonlinear dynamics in non-planar microcavity wires, based on driven-dissipative, 2D mean-field Gross-Pitaevskii coupled equations.

For a realistic microcavity wire structure, we found that the typical for planar microcavities conventional bistability of the nonlinear mode power upon variation of the pump amplitude evolves into complex multistability. We have discussed the origin of the multistability in detail, and traced it back to the number of modes arising as a result of the interplay of three distinct factors. Firstly, the number of modes supported by the microcavity confinement potential is determined by the potential depth and width. Secondly, the pump frequency detuning from the reference frequency (blue-shifted with respect to the lower polariton branch resonance frequency for the relevant angle of incidence of the pump) selects a subset of these modes. Thirdly, depending on its parity, the spatial symmetry of the pump further reduces the number of available modes; for example, a symmetric, Gaussian-shaped pump leads to suppression of the odd modes and survival of the even modes.

Further, by employing linear stability analysis and conducting dynamical simulations, we discovered a range of pump amplitudes for which polariton solitons exist in a realistic structure operating in the triggered parametric oscillator regime. We have confirmed these results using the 2D Newton-Raphson method. In contrast to the single-mode polariton solitons in planar microcavities, polariton solitons obtained by our scalar model in microcavity wires exhibit a complex multi-mode structure. This type of spatially localised polariton solitons deserve special attention, since multi-mode solitons are capable of propagating further along an imperfect microcavity than their single-mode counterparts \cite{PRB2016}. In view of this fact, we have developed a modal expansion method \cite{PRB2016} to investigate the inter-modal nonlinear interactions and mechanisms of localisation, and attempt to find a means to leverage these multi-mode interactions and polariton dynamics for the development of polaritonic integrated circuits.

Bearing in mind recent advances in fabrication of microcavities with ultrahigh polariton lifetimes \cite{Snoke1, Snoke2}, we investigated the possibility of exciting solitons in an environment with lower dissipation (i.e., when the exciton and cavity decay rates are lower). For these conditions, we numerically demonstrated an interesting type of radiating polariton soliton exhibiting collapses and revivals during propagation along the microcavity wire.

In order to gauge the plausibility of using microcavity wires for building Y-splitters, we examined the effect of tilted microcavity wires on the solitons' ability to propagate. Using dynamical simulations in tandem with 2D Newton-Raphson calculations, we identified pump amplitude-dependent maximum angles for which the maximum of the soliton amplitude was virtually unaltered ($\ll 1\%$ reduction); the limit appeared to be around $8 ^{\circ}$. We also established the critical tilt angles above which the solitons vanished; the highest such angle was approximately $12 ^{\circ}$.

With a view to creating novel polaritonic devices, we dynamically studied polariton soliton propagation in a tapered microcavity wire. We found that, as the polariton soliton travels over certain ranges of microcavity wire width, both its maximum amplitude and group velocity increase. This microcavity wire geometry could hence be exploited for reshaping and creation of solitons with specific characteristics, or simply to generate soliton amplification. For instance, tapered wire sections can be used as repeaters on a transmission line, compensating for signal loss at junctions or degradation due to structural imperfections. Furthermore, appropriate tapered sections could be selected according to whether fast or slow polariton solitons are required for a particular device application. For instance, fast polariton solitons are necessary for ultrafast polariton switches and modulators with applications in quantum information technologies -- e.g., routing entangled single photons. Conversely, similar to the slow-light effect \cite{Colman2010}, designing geometries supporting slow polariton solitons should lead to an enhancement of the nonlinear response, thereby decreasing the threshold for nonlinear effects, such as polariton parametric scattering and the Kerr effect. These slow polariton solitons may have applications as controllable delay lines and buffers.

In summary, this work lays the foundations for the simulation of the basic building blocks of future polaritonic integrated circuits, such as X- and Y-splitters, couplers and routers, based on soliton logic. Clearly, theoretical studies like these will be of great importance for the physical realisation of low-consumption, low-intensity threshold polariton integrated circuits of the future.

\subsection{Acknowledgments}
The authors wish to thank A. V. Gorbach and D. V. Skryabin for helpful discussions.
G.S. acknowledges funding through the Leverhulme Trust Research Project Grant No. RPG-2012-481. A. P. acknowledges support from SFB 787 of the DFG.


\begin{thebibliography}{99}

\bibitem{Snoke&Sanvitto}
D. W. Snoke, in \textit{Exciton Polaritons in Microcavities}, Eds. V Timofeev and D. Sanvitto, Springer Series in Solid State Sciences, \textbf{172} (Springer, New York, 2011)

\bibitem{Ciuti}
C. Ciuti, P. Schwendimann, B. Deveaud, and A. Quattropani, "Theory of the angle-resonant polariton amplifier," Phys. Rev. B \textbf{62}, R4825--R4828 (2000)

\bibitem{Whittaker}
D. M. Whittaker, "Classical treatment of parametric processes in a strong-coupling planar microcavity," Phys. Rev. B \textbf{63}, 193305-1 -- 193305-1(2001)

\bibitem{Kwong}
N.H. Kwong, R. Takayama, I. Rumyantsev, M. J. Kuwata-Gonokami, and R. Binder, "Third-order exciton-correlation and nonlinear cavity-polariton effects in semiconductor microcavities," Phys. Rev. B \textbf{64}, 045316 (2001)

\bibitem{Amo1}
A. Amo, D. Sanvitto, F. P. Laussy, D. Ballarini, E. del Valle, M. D. Martin, A. Lema\^{\i}tre, J. Bloch,
D. N. Krizhanovskii, M. S. Skolnick, C. Tejedor, and L. Vi\~{n}a, "Collective fluid dynamics of a polariton condensate in a semiconductor microcavity," Nature \textbf{457}, 291--296 (2009)

\bibitem{Amo2}
A. Amo, S. Pigeon, D. Sanvitto, V. G. Sala, R. Hivet, I. Carusotto, F. Pisanello, G. Leménager, R. Houdré, E Giacobino, C. Ciuti, A. Bramati, "Polariton Superfluids Reveal Quantum Hydrodynamic Solitons," Science \textbf{332}, 1167 -- 1169 (2011)

\bibitem{Snoke1}
B. Nelsen, G. Liu, M. Steger, D. W. Snoke, R. Ballili, K. West, and L. Pfeiffer, Phys. Rev. X \textbf{3}, 041015 (2013)

\bibitem{Snoke2}
M. Steger, G. Liu, B. Nelsen, C. Gautham, D. W. Snoke, R. Ballili, L. Pfeiffer, and K. West, Phys. Rev. B \textbf{88}, 235314 (2013)

\bibitem{Snoke3}
M. Steger, C. Gautham, D. W. Snoke, L. Pfeffer, and K. West, Optica \textbf{2}, 1 (2015)

\bibitem{Deng}
H. Deng, H. Haug, and Y. Yamamoto, \textit{Exciton-polariton Bose-Einstein Condensation}, Rev. Mod. Phys. \textbf{82}, 1489 (2010)

\bibitem{Egorov&Dmitry}
O. A. Egorov, D.V. Skryabin, A.V. Yulin, and F. Lederer, "Bright Cavity Polariton Soliton," Phys. Rev. Lett. \textbf{102}, 153904-1--153904-4 (2009)

\bibitem{Sich_NaturePhotonics}
M. Sich, D. N. Krizhanovskii, M. S. Skolnick, A. V. Gorbach, R. Hartley, D. V. Skryabin, E. A. Cerda-M\'{e}ndez, K. Biermann, R. Hey and P. V. Santos, "Observation of bright polariton solitons in a semiconductor cavity," Nature Photonics, \textbf{6}, 50--55 (2012)

\bibitem{Wertz_APL}
E. Wertz, L. Ferrier, D. D. Solnyshkov, P. Senellart, D. Bajoni et al., "Spontaneous formation of a polariton condensate in a planar GaAs microcavity," Appl. Phys. Lett., \textbf{95}, 051108-1--051108-3 (2009)

\bibitem{Wertz_Nature}
E. Wertz, L. Ferrier, D. D. Solnyshkov, R. Johne, D. Sanvitto, A. Lema\^{\i}tre, I. Sagnes, R. Grousson, A. V. Kavokin, P. Senellart, G. Malpuech and J. Bloch, "Spontaneous formation and optical manipulation of extended polariton condensates", Nature Physics, \textbf{6}, 860--864 (2010)

\bibitem{our_OL}
G. Slavcheva, A. V. Gorbach, A. Pimenov, A. G. Vladimirov and D. V. Skryabin, Opt. Lett. \textbf{40}, 1787 (2015)

\bibitem{PRB2016}
G. Slavcheva, A. V. Gorbach, and A. Pimenov, Phys. Rev. B \textbf{94}, 245432 (2016)

\bibitem{McCall&Hahn}
S. L. McCall and E. L. Hahn, Phys. Rev. \textbf{183}, 457 (1969)

\bibitem{Scully&Zubairy}
M. O. Scully and M. S. Zubairy, \textit{Quantum Optics}, (Cambridge Univeristy Press, Cambridge, 1997)

\bibitem{Supplementary_material}
Supplementary material

\bibitem{Colman2010}
P. Colman, C. Husko, S. Combrie, I. Sagnes, C. W. Wong, and A. De Rossi, Nature Photonics, \textbf{4}, 862 (2010)

\end{thebibliography}
\end{document}